%% file: main.tex
\documentclass[a4paper,11pt]{article}
\pdfoutput=1 
\usepackage{jheppub} 
\usepackage{lineno}
\usepackage{slashed}
\usepackage{subcaption}
\usepackage{mathtools}
\usepackage{commath}
\usepackage{multirow}
\usepackage{hyperref}
\usepackage{cleveref}
\usepackage{booktabs}
\usepackage{nicefrac} 
\usepackage{comment}
\usepackage{graphicx}
\usepackage{float}
\usepackage{enumitem}
\usepackage{physics}

\title{Electroweak Symmetry Restoration in the N2HDM 
via Domain Walls}

\author[a]{Mohamed Younes Sassi,}
\author[a,b]{Gudrid Moortgat-Pick}

\affiliation[a]{II. Institut für Theoretische Physik,
University of Hamburg,\\Luruper Chaussee 149, 22761 Hamburg, Germany}
\affiliation[b]{Deutsches Elektronen-Synchrotron DESY, Notkestr. 85, 22607 Hamburg, Germany}

\emailAdd{mohamed.younes.sassi@desy.de}
\emailAdd{gudrid.moortgat-pick@desy.de}

\preprint{DESY-24-109}

\abstract{Domain walls are a type of topological defects that can arise in the
early universe after the spontaneous breaking of a discrete symmetry. They can form in several beyond the Standard Model theories with an
extended Higgs sector such as the Next-to-Two-Higgs-Doublet model
(N2HDM). In this work, we discuss the domain wall solution related
to the singlet scalar of the N2HDM and demonstrate the possibility of restoring the electroweak symmetry inside and in the vicinity of the
domain wall. Such symmetry restoration can have profound implications on early universe cosmology as the weak sphaleron rate inside the domain wall would, in principle, be unsuppressed compared to the rate outside the wall. We also discuss the possibility of generating CP-violating vacua localized in the vicinity of the domain wall. Our work is a first step towards the realization of electroweak baryogenesis mediated by domain walls in the N2HDM.}

\begin{document} 

\maketitle
\flushbottom

\input{intro}

\input{N2HDM}

\input{DWinN2HDM}

\input{PhenoScenarios}

\input{DoubleDWSol}

\input{Summary_and_conclusions}

\appendix

\acknowledgments
This work is funded by the Deutsche Forschungsgemeinschaft (DFG) through Germany’s Excellence Strategy – EXC 2121 “Quantum Universe” — 390833306. Figures presented in this work were generated using \texttt{MatPlotLib} \cite{Hunter:2007} and \texttt{NumPy} \cite{2020NumPy-Array}.

\appendix

\bibliography{references.bib}

\end{document}

%% file: intro.tex
\section{Introduction}

Extended Higgs sectors are well-motivated extensions of the standard model of particle physics (SM). These extensions can be used to solve several shortcomings of the SM. For instance, electroweak baryogenesis, a mechanism providing an elegant solution to the problem of matter-antimatter asymmetry of the universe, requires a first-order electroweak phase transition (EWPT) in order to satisfy the Sakharov condition of departure from thermal equilibrium. It is well known that the EWPT in the SM is a cross-over, which keeps the particle plasma in the early universe in thermal equilibrium during the phase transition. However, one can obtain a first-order phase transition in several models with extended Higgs sectors such as singlet extensions of the SM \cite{Zhang:2023jvh, Carena:2022yvx} the Two-Higgs-Doublet-Models (2HDM) \cite{Biekotter:2023eil, Goncalves:2021egx} and the next-to-2HDM (N2HDM) \cite{Biekotter:2021ysx, Chaudhuri:2024vrd}. In addition, a scalar sector with more than one Higgs doublet is a crucial ingredient in several models tackling the hierarchy problem \cite{Quevedo:2010ui}, the strong-CP problem \cite{Dutta:2023lbw, Ringwald:2024uds, DiLuzio:2020wdo}, dark matter \cite{Dutta:2023cig,Engeln:2020fld, Bringmann:2023iuz, Cruz:2023xxg} and neutrino masses \cite{Mohapatra:2004zh, King:2003jb, Aoki:2009mb, Antipin:2017wiz}. 

In this work, we focus on the N2HDM where the SM Higgs sector is extended with another Higgs doublet and an extra real singlet scalar. In order to avoid some shortcomings of this model such as flavor changing neutral currents (FCNCs) at tree level, one usually imposes discrete symmetries on the doublet scalars to force up and down-type fermions to couple to only one doublet scalar at a time \cite{Branco:2011iw, Glashow:1976nt}. The invariance of the model under these discrete symmetries leads to the presence of several disconnected degenerate minima for the scalar potential. When the scalar fields acquire a vacuum expectation value (VEV) in the early universe, these discrete symmetries get spontaneously broken, leading to the formation of domain walls interpolating between the degenerate disconnected minima of the vacuum manifold. These domain walls are a type of topologically protected cosmic defects, where the discrete symmetry is restored inside their core. The presence of stable domain walls in a model is, however, a serious constraint given that these defects tend to dominate the energy density of the universe shortly after their formation \cite{Zeldovich:1974uw, Kibble:1976sj}. However, it is easy to circumvent this problem by making the discrete symmetry approximate, leading to a bias in the potential between the minima related by the discrete symmetry \cite{Gelmini:1988sf, Saikawa:2017hiv, PhysRevD.28.1419}. In such a case the region of the true minimum will expand in the region of the false minimum and the domain wall network annihilates. Other mechanisms to annihilate the domain wall networks include the possibility of having initial conditions that lead to vacua being favored over the others \cite{Larsson:1996sp}, a latter symmetry restoration of the discrete symmetry or the possibility that the spontaneously broken discrete symmetry does not get restored in the early universe and therefore domain walls networks do not form in the first place \cite{Dvali:1995cc}. 

Even though domain walls can be problematic from a cosmological point of view, it was shown that they can lead to very interesting phenomena when coupled with other scalar fields. For instance, it was recently shown that after EWSB in the 2HDM \cite{Law:2021ing, Sassi:2023cqp}, domain walls  can form in several classes with different properties such as exhibiting CP-violating or/and electric charge breaking vacua inside their core, which can provide a very rich phenomenology at the time of their formation in the early universe such as the CP-violating scattering of fermions off the wall or transforming fermions of a given SU(2) doublet into each other via an electric charge breaking scattering. This includes phenomena such as e.g. top quarks turning into bottom quarks \cite{Sassi:2023cqp}. It was also shown in \cite{Blasi:2022woz, Agrawal:2023cgp} that domain walls in the real singlet extension of the SM can be used in order to facilitate the occurrence of the EWPT, since the domain walls related to the real singlet scalar act as impurities catalyzing the phase transition. Such a scenario can be used to overcome the problem of vacuum trapping \cite{Biekotter:2021ysx} where the universe gets stuck in the symmetric phase due to the nucleation probability of the true SM vacuum being very small. 

In this work, we consider another interesting phenomenon induced by the domain wall of the real singlet scalar in the N2HDM, namely the possibility of electroweak symmetry restoration (EWSR) in the vicinity of the domain wall. In such a case, the sphaleron rate is much less suppressed inside and in the vicinity of the wall than outside of it. Therefore, this effect combined with a source for CP-violation, can lead, a priori to the generation of a matter-antimatter asymmetry in the early universe induced by domain walls. Such a mechanism was examined in previous works \cite{Brandenberger:1991dr, Brandenberger:1992ys, Brandenberger:1994bx, Brandenberger:1994mq, Schroder:2024gsi, Cline:1998rc, Dasgupta:1996ys, Davis:1992fm} in the framework of general topological defects such as cosmic strings and domain walls. One significant advantage of such a mechanism compared to conventional electroweak baryogenesis is that the need for a first order phase transition can be avoided, given that the topological defect will provide the separation in the regions with drastically different sphaleron rates, ensuring the out of thermal equilibrium condition. For the case of cosmic strings, it was shown in \cite{Cline:1998rc} that any matter-antimatter asymmetry produced by this mechanism is orders of magnitudes smaller than the observed asymmetry. This is mainly due to cosmic strings being one dimensional defects which renders the volume in space where the mechanism is active to be very small. Such volume suppression is, however, not present in the case of moving domain walls as they are two-dimensional objects and therefore this mechanism can be effective in a large volume \cite{Brandenberger:1994mq, Schroder:2024gsi}.        

The phenomenon of EWSR inside the wall is directly related to the effective mass terms in the potential for the doublet fields. Far from the wall, the scalar potential of the Higgs doublets is in the broken phase with the minima of the Higgs doublets corresponding to those satisfying $v_{sm} = \sqrt{v^2_1 + v^2_2 } \approx 246 \text{ GeV}$. In the vicinity of the wall, the value for the singlet scalar field's vacuum $v_s$ responsible for the domain wall solution goes to zero, leading to a change in the effective mass terms for the Higgs doublets. Depending on the parameter point and in the case of negative couplings between the singlet and doublet scalar fields, the effective mass term of the Higgs doublets can get a huge positive boost inside the wall and the potential of the 2HDM part of the model transitions to the symmetric phase where the minima of the doublets vanish and the electroweak symmetry gets restored. 

In our work, we provide a detailed analysis of the behavior of the Higgs doublets in the background of the singlet domain wall. This includes the phenomenon of EWSR as well as the possibility of the Higgs doublet VEVs getting larger inside the wall. We also discuss the width of the region of symmetry restoration in the vicinity of the wall as a function of the model parameters and the amount of EWSR inside the wall. As a further aspect of our investigation, we discuss the phenomenology of EWSR inside the wall for different scenarios of parameter points, demonstrating the practical viability of this mechanism in providing a way to achieve unsuppressed sphaleron rates for electroweak baryogenesis without the need for a first order phase transition. We also discuss the generation of CP-violating vacua localized in the vicinity of the wall, providing a CP-violating source needed for electroweak baryogenesis while naturally evading experimental constraints from electron dipole moment. Our work is a proof of principle for the different ingredients needed for generating a matter-antimatter asymmetry in the early universe using domain walls. A complete calculation of the amount of baryogenesis generated using this mechanism is subject of a future publication.

Our paper is organized as follows: in section \ref{section2} we briefly introduce the N2HDM and the used notation. In section \ref{section3}, we discuss the theoretical and experimental constraints that we impose on the parameter points of the N2HDM that we consider and discuss in detail the domain wall solutions of the model, providing a detailed analysis of the different behaviors of the doublet fields in the vicinity of the singlet wall. In section \ref{section4}, we discuss different phenomenological scenarios of the model and the possibility of EWSR inside the wall in each of them. In section \ref{section5}, we briefly discuss the possibility of generating CP-violating condensates localized in the vicinity of the walls. We summarize and conclude our work in section \ref{section6}.

%% file: N2HDM.tex
\section{The Next-to-Two-Higgs-Doublet Model}\label{section2}
In this section, we briefly introduce the Next-to-Two-Higgs-Doublet model and the needed notation used in the work. For a comprehensive review of this model, the reader is referred to \cite{Muhlleitner:2016mzt,Muhlleitner:2017dkd, Chen:2013jvg}. \\
In the N2HDM, the standard model Higgs sector is extended with an extra $SU(2)_L \times U(1)_Y$ doublet $\Phi_2$ and a real singlet $\Phi_s$. The Higgs sector potential is given by:
\begin{align}
   \notag V_{N2HDM} &= m^2_{11}\Phi^{\dagger}_1\Phi_1 + m^2_{22}\Phi^{\dagger}_2\Phi_2 +  m^2_{12}(\Phi^{\dagger}_1\Phi_2 + h.c.) + \frac{\lambda_1}{2}\bigl(\Phi^{\dagger}_1\Phi_1\bigr)^2 + \frac{\lambda_2}{2}\bigl(\Phi^{\dagger}_2\Phi_2\bigr)^2 \\ \notag
    &   + \lambda_3\bigl(\Phi^{\dagger}_1\Phi_1\bigr)\bigl(\Phi^{\dagger}_2\Phi_2\bigr)  + \lambda_4\bigl(\Phi_1^{\dagger} \Phi_2\bigr)\bigl(\Phi_2^{\dagger} \Phi_1\bigr) 
    +\biggl[\frac{\lambda_5}{2}\bigl(\Phi_1^{\dagger} \Phi_2\bigr)^2 + h.c\biggr] \\ \notag
    & + \frac{m^2_S}{2}\Phi^2_s + \frac{\lambda_6}{8}\Phi^4_s + \frac{\lambda_7}{2}\Phi^2_s(\Phi^{\dagger}_1\Phi_1) + \frac{\lambda_8}{2}\Phi^2_s(\Phi^{\dagger}_2\Phi_2) \\ & + \biggl[a_1\Phi_s + a_3 \Phi^3_s + b_1(\Phi^{\dagger}_1\Phi_1)\Phi_s + b_2(\Phi^{\dagger}_2\Phi_2)\Phi_s + c_1 (\Phi^{\dagger}_1\Phi_2\Phi_s + h.c.) \biggr] .
\label{eq:treepot}    
\end{align}
In order to avoid flavor changing neutral currents, one imposes a $Z_2$ symmetry that acts on the scalar fields in the following way:
\begin{align}
    \Phi_1 \rightarrow \Phi_1, && \Phi_2 \rightarrow -\Phi_2, && \Phi_s \rightarrow \Phi_s.
\end{align}
This symmetry is softly broken by the term $m_{12}(\Phi^{\dagger}_1\Phi_2 + h.c)$. When the parameters $a_1$, $a_3$, $b_1$, $b_2$ and $c_1$ are zero, the potential also allows for an accidental discrete symmetry $Z'_2$, which only acts on the singlet:
\begin{equation}
    \Phi_s \rightarrow -\Phi_s.
\end{equation}
As we are mainly interested in studying the behavior of the doublet scalar fields $\Phi_1$ and $\Phi_2$ in the background of the singlet domain walls, we limit ourselves in this work to study the case when the $Z'_2$ symmetry is not explicitly broken, i.e. to make all the terms in the last line of (\ref{eq:treepot}) vanishing. This choice is motivated in order to simplify the calculation and discussion of the parameter dependence and the fact that only very small values for the symmetry breaking terms are needed in order to avoid domain walls dominating the energy budget of the universe. Incorporating terms that break the $Z'_2$ symmetry is subject of future work in which we also discuss the matter-antimatter asymmetry generated by these domain walls and the possibility of detecting gravitational waves emitted by the annihilation of the biased network of domain walls \footnote{In that context, these parameters will be crucial, as they will determine the time interval in the early universe during which the mechanism of baryogenesis via domain walls is active until the annihilation of the biased domain wall network.}. 

After electroweak and $Z'_2$ symmetries breaking, the scalar doublets and singlet acquire a vacuum expectation value. The most general vacuum can be written as:
\begin{align}
   \langle \Phi_1 \rangle = \text{U} \langle \tilde{\Phi}_1 \rangle = \text{U} \dfrac{1}{\sqrt{2}}
    \begin{pmatrix}
          0 \\      v_1
     \end{pmatrix},      
&& \langle \Phi_2 \rangle = \text{U} \langle \tilde{\Phi}_2 \rangle = \text{U} \dfrac{1}{\sqrt{2}}
      \begin{pmatrix}
     v_+ \\
     v_2e^{i\xi}
      \end{pmatrix} , && \langle \Phi_s \rangle = v_s,
\label{eq:vacuumform}      
\end{align}
where U is an element of the $\text{SU(2)}_L\times\text{U(1)}_Y$ group that is given by:
\begin{equation}
        \text{U} = e^{i\theta} \text{exp}\biggl(i\dfrac{\tilde{g}_i\sigma_i}{2v_{sm}}\biggl),
\label{eq:EWmatrix}        
\end{equation}
with $\theta$ and $\tilde{g}_i$ denoting the Goldstone modes of the scalar doublets, $\sigma_i$ the Pauli matrices and $v_{sm} \approx 246 \text{ GeV} $ the standard model vacuum expectation value.

The scalar doublets admit three possible types of vacua. The most general one, where $v_+ \neq 0$, breaks the electromagnetism symmetry $U(1)_{em}$ and gives a mass to the photon. Consequently, such vacua are physically not allowed at present time. The second type occurs when the phase between the two scalar doublets $\xi$ does not vanish. Such vacuum is CP-violating as it generates an imaginary mass to the fermions via the Yukawa sector. Due to constraints from electron dipole moment experiments, such CP-violating vacua should have very small values for $\xi$ to be realized in nature. The third type is the neutral vacuum, occurring when $v_+ = 0$ and $\xi = 0$. In this work, we consider the case when the singlet scalar acquires a vacuum expectation value $v_s \neq 0$, which breaks $Z'_2$ spontaneously and gives rise to domain walls in the early universe. As for the doublet vacua, we only limit ourselves to neutral vacua as they lead to SM-like behavior. However, it was shown recently in \cite{Sassi:2023cqp, Law:2021ing} that domain walls in the 2HDM (related to the spontaneous breaking of the $Z_2$ symmetry) exhibit different classes of domain wall solutions including electric charge breaking and/or CP-violating domain wall solutions even when the vacua on both regions are neutral (the electric charge breaking and CP-violating vacua are only localized inside or in the vicinity of the wall). This behavior is obtained when the Goldstone modes in (\ref{eq:EWmatrix}) are different in both domains. In the context of the N2HDM, we will discuss in section \ref{section5} that it is also possible to get non-vanishing CP-violating condensates $\xi(x)$ in the vicinity of the singlet domain wall even when the doublet VEVs on both domains have the same sign but different Goldstone modes.

The particle spectrum of the N2HDM includes 3 CP-even Higgs particles with masses denoted as $m_{h_1}$, $m_{h_2}$ and $m_{h_3}$, one CP-odd particle with mass $m_A$ and two charged Higgs bosons $m_{H^{\pm}}$. It is more advantageous to express the potential parameters in terms of physical quantities such as the masses of the physical particles and $\tan(\beta) = v_2/v_1$. This is achieved by diagonalizing the mass matrix $M^2_\rho$ (see (\ref{eq:massmatrix})) given in the interaction basis $(\rho_1, \rho_2, \rho_3)$, where $\rho_{1,2,3}$ correspond to field expansions around the vacua $v_{1,2,s}$ in (\ref{eq:vacuumform}).
\begin{equation}
M_{\rho}^{2}=\left(\begin{array}{ccc}
v^{2}\lambda_{1}\cos(\beta)^{2}+m_{12}^{2}\, tan(\beta) \,\, &
v^{2}\lambda_{345}\,\cos(\beta)\,\sin(\beta)-m_{12}^{2}\,\, &
v\,v_{s}\lambda_{7}\,\cos(\beta)\\
v^{2}\lambda_{345}\,\cos(\beta)\,\sin(\beta)-m_{12}^{2} \,\, &
v^{2}\lambda_{2}\,\sin(\beta)^{2}+m_{12}^{2}/tan(\beta)\,\,&
v\,v_{s}\lambda_{8}\,\sin(\beta)\\
v\,v_{s}\lambda_{7}\,\cos(\beta)\,\, & v\,v_{s}\lambda_{8}\,\sin(\beta)\, \,&
v_{s}^{2}\,\lambda_{6}
\end{array}\right),
\label{eq:massmatrix}
\end{equation}
where $v^2 = v^2_1 + v^2_2$. This mass matrix is diagonalized using a rotation matrix R which fulfills the requirement $RM^2_{\rho}R^{T} = diag(m^2_{h_1}, m^2_{h_2}, m^2_{h_3})$, where the masses $m_{h_{1,2,3}}$ correspond to the masses of the CP-even Higgs bosons in the physical mass basis $(h_1, h_2, h_3)$. The diagonalizing matrix R is parametrized using the mixing angles $\alpha_1$, $\alpha_2$ and $\alpha_3$ as:
\begin{equation}
R = 
\begin{pmatrix}
c(\alpha_1)c(\alpha_2)  & s(\alpha_1)c(\alpha_2) & s(\alpha_2)\\
-\bigl(c(\alpha_1)s(\alpha_2)s(\alpha_3) + s(\alpha_1)c(\alpha_3)\bigr) & c(\alpha_1)c(\alpha_3) - s(\alpha_1)s(\alpha_2)s(\alpha_3) & c(\alpha_2)s(\alpha_3) \\
-c(\alpha_1)s(\alpha_2)c(\alpha_3) + s(\alpha_1)s(\alpha_3) & -\bigl(c(\alpha_1)s(\alpha_3) + s(\alpha_1)s(\alpha_2)c(\alpha_3) \bigr) & c(\alpha_2)c(\alpha_3)
\end{pmatrix},
\label{eq:Rmatrix}
\end{equation}
where $c(\alpha_i)$ denotes $\cos(\alpha_i)$ and $s(\alpha_i)$ denotes $\sin(\alpha_i)$. The values of the mixing angles are constrained between $-\pi/2$ and $\pi/2$. We adopt the conventional mass hierarchy $m_{h_1} < m_{h_2} < m_{h_3}$. Note that the interaction basis $(\rho_1, \rho_2, \rho_3)$ is related to the physical mass basis $(h_1, h_2, h_3)$ by:
\begin{equation}
    \begin{pmatrix} h_1 \\ h_2  \\  h_3  \end{pmatrix} = R \begin{pmatrix} \rho_1 \\ \rho_2  \\  \rho_3  \end{pmatrix}.
\end{equation}
One can then relate the potential parameters and the masses of the scalars in the N2HDM using the following formulas: 
\begin{align}
\lambda_1 &= \frac{1}{v^2_1}\biggl(-m^2_{12}\tan(\beta) + \sum_i m^2_{h_i}R^2_{i1}\biggr), \\
\lambda_2 &= \frac{1}{v^2_2}\biggl(-\frac{m^2_{12}}{\tan(\beta)} + \sum_i m^2_{h_i}R^2_{i2}\biggr), \\
\lambda_3 &= \frac{1}{v_1v_2}\biggl(m^2_{12} + \sum_i R_{i2}R_{i1}m^2_{h_i}\biggr) - \lambda_4 - \lambda_5, \\
\lambda_4 &= \frac{m_{12}}{v_1v_2} - 2 \frac{m^2_{H^{\pm}}}{v^2} + \frac{m^2_{A}}{v^2}, \\
\lambda_5 &= \frac{m^2_{12}}{v_1v_2} - \frac{m^2_A}{v^2}, \\
\lambda_6 &= \frac{1}{v^2_s}\biggl( R^2_{i3}m^2_{H_i}\biggr), \\
\lambda_7 &= \frac{1}{v_1v_s}\biggl( R_{i3}R_{i1}m^2_{h_i} \biggr), \label{eq:lambda7} \\
\lambda_8 &= \frac{1}{v_2v_s}\biggl( R_{i3}R_{i2}m^2_{h_i} \biggr), \label{eq:lambda8} \\
m^2_{11} &= m^2_{12}\tan(\beta) - \frac{\lambda_1}{2}v^2_1 - \biggl(\frac{\lambda_3 + \lambda_4 + \lambda_5}{2}\biggr)v^2_2 - \frac{\lambda_7}{2}v^2_s, \\
m^2_{22} &= \frac{m^2_{12}}{\tan(\beta)} - \frac{\lambda_2}{2}v^2_2 - \biggl(\frac{\lambda_3 + \lambda_4 + \lambda_5}{2}\biggr)v^2_1 - \frac{\lambda_8}{2}v^2_s, \\
m^2_s &= -\frac{\lambda_6}{2}v^2_s - \frac{\lambda_7}{2}v^2_1 - \frac{\lambda_8}{2}v^2_2.
\end{align}
In the next chapter, we discuss the domain wall solutions that arise after spontaneous symmetry breaking of the $Z'_2$ discrete symmetry. We investigate the influence of the domain wall profile of the singlet scalar on the doublet scalars and in particular the possibility of restoring the electroweak symmetry inside the domain wall.  

%% file: DWinN2HDM.tex
\section{Domain walls in the N2HDM}\label{section3}
Domain walls are a type of topological defects that arise after spontaneous symmetry breaking of a discrete symmetry. In our model, different types of domain wall solutions can be found depending on which discrete symmetry gets broken in the early universe.\\
Domain wall solutions are constructed by imposing vacua related by a discrete symmetry at the boundaries $\pm \infty$. In the case of a spontaneously broken $Z'_2$ symmetry, DW solutions interpolate between regions with vacua $\langle \phi_s \rangle = -v_s$ and $\langle \phi_s \rangle = v_s$ and therefore, necessarily cross $\langle \phi_s \rangle = 0$ inside the core of the wall. In the case when the $Z_2$-discrete symmetry gets spontaneously broken (alongside the electroweak symmetry), possible domain wall solutions interpolate between the vacua located on two disconnected 3-spheres of the vacuum manifold: 
\begin{align}
  \langle \phi_1 \rangle &= v_1, && \langle \phi_2 \rangle = -v_2, && \langle \phi_s \rangle = v_s, &&  U = U_1 && \text{ at } -\infty, \\
    \langle \phi_1 \rangle &= v_1, &&  \langle \phi_2 \rangle = v_2, && \langle \phi_s \rangle  = v_s, && U = U_2 && \text{ at } +\infty, 
\end{align}
with $U_1$ and $U_2$ corresponding to different Goldstone modes $\theta$ and $g_i$ of $SU(2)_L \times U(1)_Y$, leading to the creation of different classes of DW solutions as was recently found in \cite{Sassi:2023cqp, Law:2021ing}.\\

In the following sections, we focus on the domain wall solutions obtained in the N2HDM after the spontaneous symmetry breaking of $Z'_2$ leading to a non-zero $v_s$ as well as describing the effects of the domain wall solution for $\phi_s(x)$ on the field configurations of $\phi_1(x)$ and $\phi_2(x)$, demonstrating the possibility of restoring the EW symmetry inside the wall. 

\subsection{Symmetry restoration in the early universe}

Before we start discussing the domain wall solutions in the N2HDM, we first briefly consider the thermal evolution of the N2HDM scalar potential in the early universe. One crucial condition for the validity of our analysis is the restoration of the $Z'_2$ symmetry in the early universe at high temperatures. This requirement is important because in the case of non-restoration of the $Z'_2$ symmetry, domain walls wouldn't have formed in the first place and the singlet would be in the broken phase already at very high temperatures. 

To check whether a parameter point features electroweak and/or $Z'_2$ symmetry restoration in the early universe, one needs to follow the evolution of the effective thermal potential of the N2HDM at high temperatures given by \cite{Biekotter:2021ysx}:
\begin{align}
    V_{N2HDM}(T,\Phi_1,\Phi_2,\Phi_s) = V^{tree}_{N2HDM} + V_{CW} + V^T_{N2HDM},
\end{align}
where $V^{tree}_{N2HDM}$ denotes the tree-level potential defined in (\ref{eq:treepot}), $V_{CW}$ the Coleman-Weinberg one-loop correction:
\begin{align}
    V_{CW}(\phi_i) = \sum_j \dfrac{n_j}{64\pi^2} (-1)^{2s_i} m_j^4(\phi_i)\biggl[ \ln (\dfrac{|m^2_j(\phi_i)|}{\mu^2}) - c_j \biggr],
\end{align}
where $n_j$ denotes the multiplicities of the thermal bath particles given by index j, $m_j(\phi_i)$ the mass formulas of the particle as a function of the scalar field $\phi_i$ and $c_j$ the $\overline{\text{MS}}$ renormalization constants with $c_j = 3/2$ for fermions and scalars and $c_j = 5/6$ for gauge bosons \cite{Biekotter:2021ysx}. 
$V^T_{N2HDM}$ denotes the one-loop thermal correction to the scalar potential generated by the interaction of the scalar sector with the thermal bath in the early universe \cite{LeBellac_1996}:
\begin{align}
    &V^T_{N2HDM} = \sum_j \dfrac{n_jT^4}{(2\pi)^2}J_{\pm}(\dfrac{m^2_j(\phi_i)}{T^2}) ,\\
    &J_{\pm}(\dfrac{m^2_j(\phi_i)}{T^2}) = \mp \int^{+\infty}_0 dx x^2 \log\biggl[ 1 \pm exp\biggl( -\sqrt{x^2 + \dfrac{m^2_j(\phi_i)}{T^2}} \biggr) \biggr],
\end{align}
where $n_j$ denotes the multiplicities of the thermal bath particles given by index j, $m_j(\phi_i)$ the mass formulas of the particle as a function of the scalar field $\phi_i$ and $J_{\pm}$ the thermal functions for fermions (+) and bosons (-).\\
In the high-temperature limit, the thermal functions reduce to \cite{LeBellac_1996}:
\begin{align}
    J_-(y) & \approx -\dfrac{\pi^4}{45} + \dfrac{\pi^2}{12}y - \dfrac{\pi}{6}y^{\frac{3}{2}} - \dfrac{1}{32}y^2\log\bigl(\dfrac{|y|}{a_b}\bigr) + O(y^3), \\
    J_+(y) & \approx -\dfrac{7\pi^4}{360} + \dfrac{\pi^2}{24}y + \dfrac{1}{32}y^2\log\bigl(\dfrac{|y|}{a_f}\bigr) + O(y^3),
\end{align}
where $a_b = \pi^2\text{exp}(3/2 - 2\gamma_E)$ and $a_f = 16\pi^2\text{exp}(3/2 - 2\gamma_E)$. This has the effect of rendering the effective mass terms to be temperature-dependent: 
\begin{align}
    V_{N2HDM}(T, \Phi_1, \Phi_2, \Phi_s) = m^2_{11}(T) |\Phi_1|^2 + m^2_{22}(T) |\Phi_2|^2 + m^2_S(T) \Phi^2_s +  m^2_{12}\Phi_1\Phi_2 + ... .
\end{align}
It is therefore possible that for a non-zero temperature T, the effective mass terms turn positive, leading the minimum of the potential to be at the origin of field space and therefore to the restoration of the symmetry.
In the case when the potential has its minimum at the origin of the field space $(\Phi_1, \Phi_2, \Phi_s)_T = (0, 0, 0)$ at some temperature T, then the symmetries are restored. To investigate this, one can use an analytical or numerical approach. The analytical approach discussed in \cite{Biekotter:2021ysx} calculates the Hessian matrix of the potential at its origin (0,0,0) at high temperatures T, which gives us information about the curvature of the potential around the origin. This involves calculating the principal minors of the Hessian matrix $H^0_{i,j} = \partial^2V/\partial\phi_i\partial\phi_j|_{(0,0,0)}$. One can then define the quantities\footnote{These coefficients also incorporate terms from the daisy resummation of infrared-divergent contributions in the thermal potential. The authors of \cite{Biekotter:2021ysx} use the Arnold-Espinosa method \cite{Arnold:1992rz} for the derivation of the coefficients $c_{ii}$. } $c_{ii} \equiv
\lim\limits_{\,T\rightarrow\infty}\,
H_{ii}^{0}/ T^2 $ \cite{Biekotter:2021ysx}:
\begin{align}
c_{11} & \simeq
-0.025+c_{1}-\frac{1}{2\pi}\left(\frac{3}{2}\lambda_{1}\sqrt{c_1}+\lambda_{3}\sqrt{c_{2}}+\frac{1}{2}\lambda_{4}\sqrt{c_{2}}+\frac{1}{4}\lambda_{7}\sqrt{c_{3}}\right)
\ ,
\label{coeff_1}
\\
c_{22} & \simeq
-0.025+c_{2}-\frac{1}{2\pi}\left(\frac{3}{2}\lambda_{2}\sqrt{c_2}+\lambda_{3}\sqrt{c_{1}}+\frac{1}{2}\lambda_{4}\sqrt{c_{1}}+\frac{1}{4}\lambda_{8}\sqrt{c_{3}}\right)
\ ,
\label{coeff_2}
\\
c_{33} & =
c_{3}-\frac{1}{2\pi}\left(\lambda_{7}\sqrt{c_1}+\lambda_{8}\sqrt{c_{2}}+\frac{3}{4}\lambda_{6}\sqrt{c_{3}}\right)\, ,
\label{coeff_3}
\end{align}
where the coefficients $c_{i}$ are defined as \cite{Biekotter:2021ysx}:
\begin{align}
\label{c1} 
c_{1} &=
\frac{1}{16}({g'}^{2}+3g^{2})+\frac{\lambda_{1}}{4}+\frac{\lambda_{3}}{6}+\frac{\lambda_{4}}{12}+\frac{\lambda_{7}}{24}\,
,\\
\label{c2}
c_{2} &=
\frac{1}{16}({g'}^{2}+3g^{2})+\frac{\lambda_{2}}{4}+\frac{\lambda_{3}}{6}+\frac{\lambda_{4}}{12}+\frac{\lambda_{8}}{24}+\frac{1}{4}y_{t}^{2}
\, ,\\
\label{c3}
c_{3} &= \frac{1}{6}(\lambda_{7}+\lambda_{8})+\frac{1}{8}\lambda_{6} \, ,
\end{align} 
with g and $g'$ denoting the weak gauge couplings and $y_t$ the Yukawa coupling to the top quark. For positive $c_{11}$ and $c_{22}$, the electroweak symmetry is restored at high T and in case $c_{33}>0$, the $Z'_2$ symmetry is restored. In this work, we focus on the restoration of the $Z'_2$ symmetry at higher temperatures to ensure the formation of the singlet domain walls. In case when $c_{11,22} < 0$, the doublets have a temperature-dependent VEV, and the required Hessian matrix has to be evaluated at $(v_1(T), v_2(T), 0)$ to reliably determine whether the $Z'_2$ symmetry is restored. Such a calculation is more complicated and can only be done numerically for the N2HDM. However, we observed that for several parameter points satisfying only the condition $c_{33} > 0$, it is possible to restore the $Z'_2$. Considering the leading order in $T^2$ for the thermal potential, the effective mass term for the singlet field at a given temperature T can be approximated by $$M_S(T) \approx m^2_s + c_{33}T^2 + \dfrac{\lambda_7}{4} v^2_1(T) +  \dfrac{\lambda_8}{4} v^2_2(T).  $$ 

In case $M_S(T)$ is positive, $v_s(T) = 0$ minimizes the thermal potential. Such a scenario can be obtained for $\lambda_{7,8} > 0$ or small negative $\lambda_{7,8}$. It is also possible to obtain a thermal history where the EW and/or the $Z'_2$ symmetry are restored for an intermediate temperature interval as was found in \cite{Biekotter:2021ysx}. Such a scenario can only be examined by a full numerical approach and therefore we limit ourselves to the more conservative scenario where $c_{11}$, $c_{22}$ and $c_{33}$ are positive (unless otherwise specified for some parameter scans) ensuring that all symmetries get restored at some high-temperature T. This constraint was included in our implementation of \texttt{ScannerS} \cite{Coimbra:2013qq, Ferreira:2014dya, Muhlleitner:2016mzt, Costa:2015llh, Muhlleitner:2020wwk} in order to only generate parameter points for our scans where the EW and $Z'_2$ symmetries get restored at some stage in the early universe. Note that the expressions for $c_{11}$, $c_{22}$ and $c_{33}$ derived in \cite{Biekotter:2021ysx} use the Arnold-Espinosa resummation scheme \cite{Arnold:1992rz}. It is known that different resummation schemes can lead to different outcomes for the thermal evolution of the scalar fields in the early universe (see e.g. page 23 in \cite{Biekotter:2021ysx}). A detailed discussion of these aspects is beyond the scope of this work and therefore, we use these formulas in the framework of the Arnold-Espinosa method only as an attempt to verify the restoration of the $Z'_2$ symmetry in the early universe.  

\subsection{$Z'_2$ Domain Walls}
We discuss the singlet domain wall solution in the N2HDM and its effects on the VEVs of the doublets $v_1$ and $v_2$. Here, we only consider the zero-temperature potential for the calculation of the field configuration. We therefore assume that the change in the scalar potential at small temperatures after EWSB is negligible. This choice was made in order to simplify obtaining results for large sets of parameter scans, as otherwise, one would need to calculate the exact VEVs at a particular temperature $T<T_{ew}$ for every parameter point, with $T_{ew}$ the temperature at which the EW symmetry gets spontaneously broken, making the computation lengthy and more complicated. 

In order to get domain wall solutions, we require the singlet vacua $v_s$ to have a different sign at both spatial asymptotics $\pm \infty$ depicting two regions of the universe with different signs for $v_s$. Outside the wall, the electroweak symmetry is broken and the doublets $\Phi_1$ and $\Phi_2$ acquire VEVs $(v_1(\pm \infty), v_2(\pm \infty)) \neq (0,0)$. In this work, we only consider neutral CP-conserving vacua, therefore: $v_+(\pm \infty) = 0$ and $\xi(\pm \infty) = 0$.\\
The spatially varying field configuration is then obtained by solving the equations of motion for the scalar fields using the latter boundary conditions:
\begin{align}
    \dfrac{\partial^2v_i}{\partial t^2} - \dfrac{\partial^2v_i}{\partial x^2} + \dfrac{\partial V_{N2HDM}}{\partial v_i} = 0, 
\end{align}
where \textit{i} denotes the different scalar fields. 
Due to the complicated nature of these non-linear differential equations, we find the solutions numerically using the gradient flow method \cite{Law:2021ing} which minimizes the total energy per unit area of the static vacuum configuration given by:
\begin{equation}
    \sigma_{dw} = \int^{+\infty}_{-\infty} \dd x \text{ } \dfrac{\dd \Phi_i}{\dd x} \dfrac{\dd \Phi_i^{\dagger}}{\dd x} + V_{N2HDM}(x),
\label{eq:energy}    
\end{equation}
where the first term denotes the kinetic energy contribution from each scalar field and the second term gives the potential energy of the vacuum configuration.
\begin{figure}[H]
\centering
\includegraphics[width=0.65\textwidth]{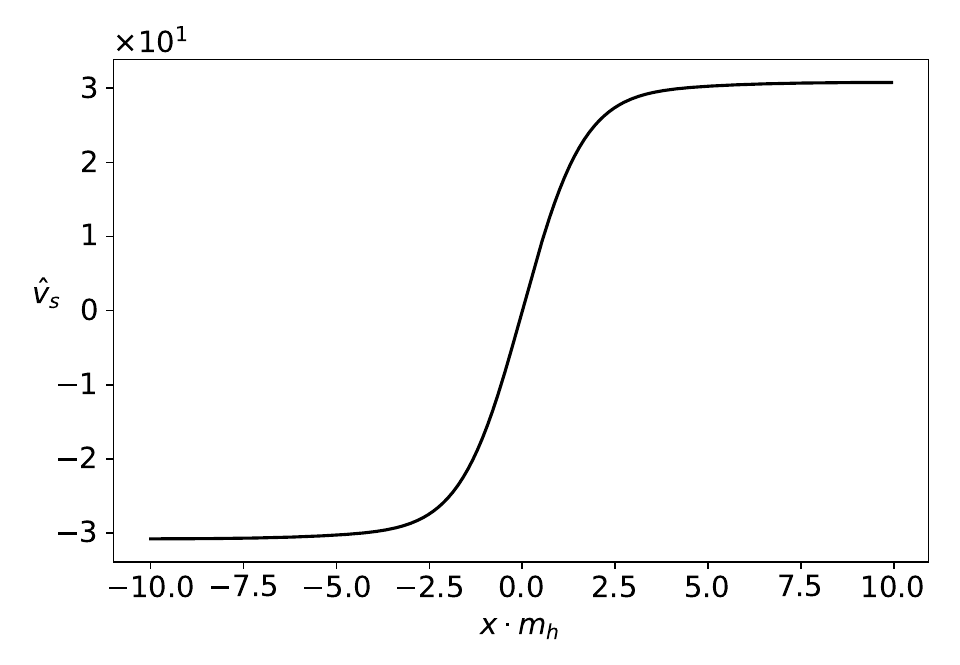}
\caption{Domain wall solution for the singlet scalar field $\phi_s$. The result is shown in terms of the rescaled VEV $\hat{v}_s = v_s/v_{sm}$.}
\label{fig:dwsinglet}
\end{figure}
The spatial profile for the rescaled VEV $\hat{v}_s(x) = v_s(x)/v_{sm}$ is shown in Figure \ref{fig:dwsinglet}. In the simpler case of a model with only a real singlet scalar field, the width of the wall can be well approximated by $\delta_s = (\sqrt{\dfrac{\lambda_6}{4}}v_s)^{-1}$ \cite{Saikawa:2017hiv}. This expression is, however, only applicable to our domain wall solution when the coupling terms between the doublets and the singlet are very small or zero ($\lambda_{7,8} \approx 0$). For non-vanishing values of $\lambda_{7,8}$ the back-reaction of the doublet fields will lead to a deviation of the wall's width from $\delta_s$.

In the background of the domain wall of the singlet $v_s(x)$, the potential for $(v_1,v_2)$ is x-dependent. We rewrite the potential $V_{N2HDM}$ (\ref{eq:treepot}) as:
\begin{align}
   \notag V_{N2HDM}(x) &= -m^2_{12}v_1(x)v_2(x)\cos{\xi(x)} +   \biggl(\dfrac{m^2_{11}}{2} + \dfrac{\lambda_7}{4}v^2_s(x)\biggr)v^2_1(x) + \biggl(\dfrac{m^2_{22}}{2} + \dfrac{\lambda_8}{4}v^2_s(x)\biggr)v^2_2(x) \\ 
 & \notag + M_+(x)v^2_+(x)  + \dfrac{m^2_s}{2}v^2_s(x) + \biggl(\dfrac{\lambda_3 + \lambda_4 + \lambda_5\cos{(2\xi(x))}}{4}\biggr)v^2_1(x)v^2_2(x) \\& + \dfrac{\lambda_1}{8}v^4_1(x) + \dfrac{\lambda_2}{8}v^4_2(x) + \dfrac{\lambda_2}{8}v^4_+(x) + \dfrac{\lambda_6}{8}v^4_s(x),  
\end{align}
where,
\begin{equation}
    M_+(x) = \dfrac{m^2_{22}}{2} + \dfrac{\lambda_2}{4}v^2_2(x) + \dfrac{\lambda_3}{4}v^2_1(x) + \dfrac{\lambda_8}{4}v^2_s(x). 
\end{equation}

Due to the coupling terms between the doublet scalar fields $\Phi_1$ and $\Phi_2$ with the singlet scalar field $\Phi_s$, the profile of the vacuum configuration for the doublets in the background of the singlet domain wall will not be homogeneous in space. In the vicinity of the domain wall's core, $v_1(x)$ and $v_2(x)$ can depart considerably from their asymptotic values. The behavior of $v_1(x)$ and $v_2(x)$ inside the wall is largely influenced by the effective mass of the doublets, which we define as:
\begin{align}
M_{1}(x) &=  \dfrac{m^2_{11}}{2} + \dfrac{\lambda_{345}}{4}v^2_2(x) + \dfrac{\lambda_7}{4}v^2_s(x), 
\label{eq:meff1}\\
M_{2}(x) &=  \dfrac{m^2_{22}}{2} + \dfrac{\lambda_{345}}{4}v^2_1(x) + \dfrac{\lambda_8}{4}v^2_s(x).
\label{eq:meff2}
\end{align}
\begin{figure}[t]
     \centering
     \begin{subfigure}[b]{0.49\textwidth}
         \centering
         \includegraphics[width=\textwidth]{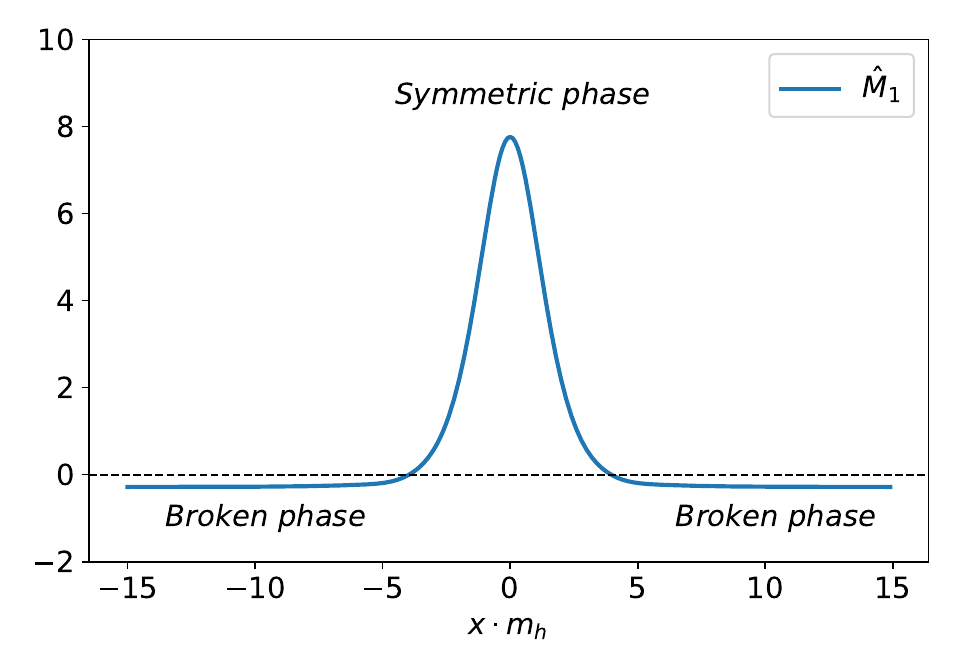}
       \subcaption{$\hat{M}_{1}$ for $P_1$}  \label{subfig:m1pp1}
     \end{subfigure}
     \hfill
     \begin{subfigure}[b]{0.49\textwidth}
         \centering
         \includegraphics[width=\textwidth]{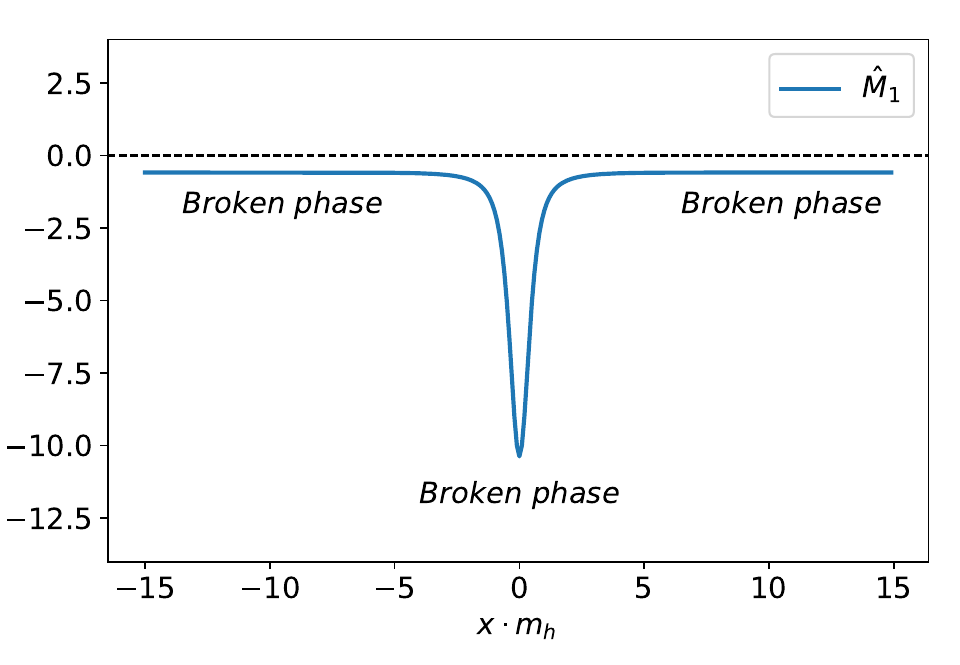}
        \subcaption{$\hat{M}_{1}$ for $P_2$} \label{subfig:m1pp2}
     \end{subfigure}
\caption{Rescaled spatial profile of the effective mass term $\hat{M}_1 = M_1(x)/m^2_{h}$ (with $m_h = 125.09 \text{ GeV}$) for different parameter points $P_1$ and $P_2$. (a) Spatial profile of $\hat{M}_1$ for $P_1$. Far from the wall, $\hat{M}_1$ is negative and the potential of the doublet is in the broken phase. Inside the wall, $\hat{M}_1$ turns positive due to the term $\lambda_7v^2_s$ vanishing, and the potential of the doublet is in the symmetric phase. (b) Spatial profile of $\hat{M}_1$ for $P_2$. Inside of the wall, $\hat{M}_1$ is even more negative and we remain in the broken phase.} 
\label{fig:effm1}
\end{figure}
In the case when $m^2_{12}$ is small or vanishing, the effective masses $M_{1,2}$ far from the wall are negative (see Figure \ref{subfig:m1pp1} ). This is required for the potential to develop non-vanishing vacuum expectation values. For $\lambda_{7,8}<0$, it is possible to get $M_{1,2}(\pm \infty)<0$ even if $m^2_{11}$ and $m^2_{22}$ are positive. The effective mass terms inside the wall are reduced to:
\begin{equation}
M_{1,2}(0) = \dfrac{m^2_{11,22}}{2} + \dfrac{\lambda_{345}}{4}v^2_{2,1}(0).
\label{eq:effinsidewall}
\end{equation}
It is therefore possible to turn the effective mass terms inside the wall positive, which leads the potential $V_{N2HDM}(\Phi_1, \Phi_2, 0)$, effectively describing a 2HDM model, to be in the symmetric phase where the minima of the scalar doublets are zero. 

\begin{figure}[h]
     \centering
     \begin{subfigure}[b]{0.49\textwidth}
         \centering
         \includegraphics[width=\textwidth]{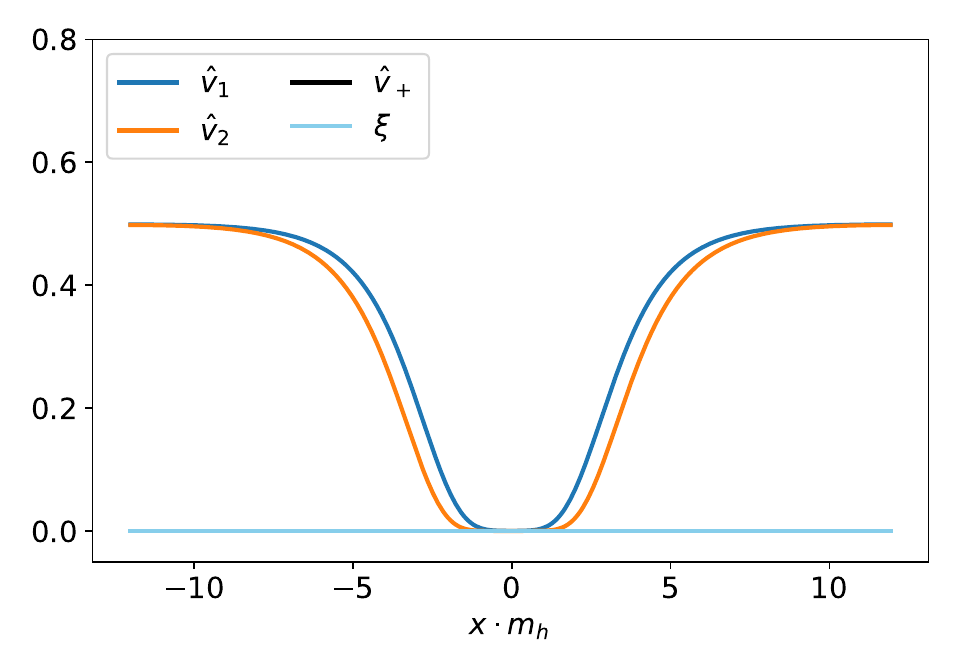}
       \subcaption{}  \label{subfig:ewsr}
     \end{subfigure}
     \hfill
     \begin{subfigure}[b]{0.49\textwidth}
         \centering
         \includegraphics[width=\textwidth]{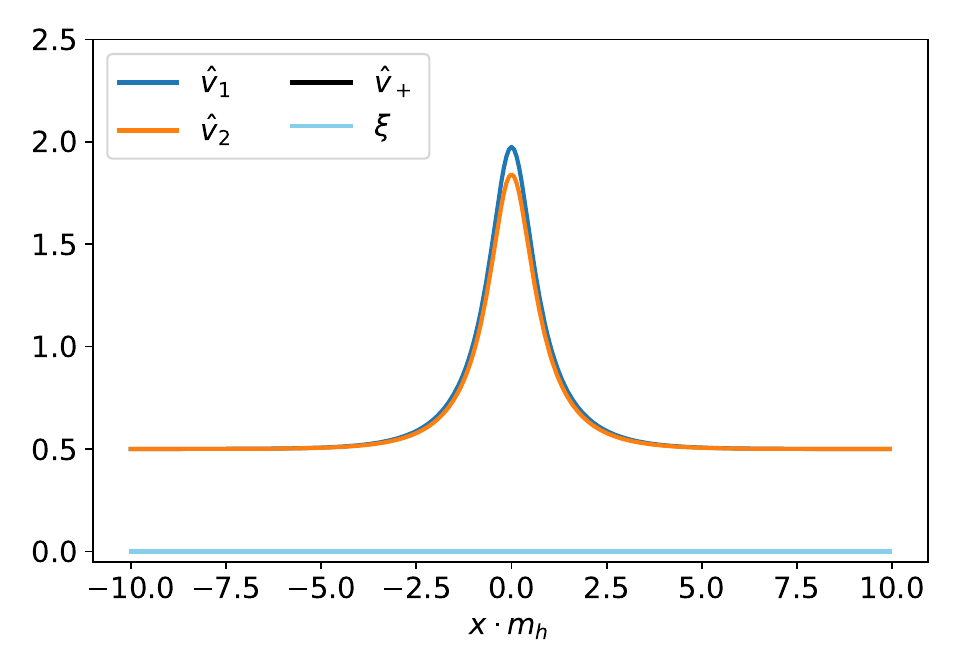}
        \subcaption{} \label{subfig:bigger}
     \end{subfigure}
\caption{Rescaled VEV profiles for the doublet fields $\hat{v}_{1,2}(x) = v_{1,2}/v_{sm} $ for different parameter points. (a) For parameter point $P_1$, $\hat{v}_{1,2}(x)$ inside the wall are much smaller than outside the wall. (b) For parameter point $P_2$, $\hat{v}_{1,2}(x)$ inside the wall gets higher than outside the wall, due to the effective mass term being more negative in the DW.} 
\label{fig:DWSolutions}
\end{figure}

We show the behavior of the doublets inside the domain wall of the singlets for 2 different parameter points $P_1$ and $P_2$ (see Table \ref{tab:benchamarks}). In case $M_{1,2}(0)$ are positive inside the wall (e.g. for $\lambda_{7,8}<0$), the doublet scalar fields can have a vanishingly small vacuum expectation values $v_{1,2}(0) = 0$ (see Figure \ref{subfig:ewsr}).
\begin{table}
    \centering
    \begin{tabular}{ccccccccccc}
         & $m_{h_1}$  & $m_{h_2}$  & $m_{h_3}$   & $\tan(\beta)$ & $v_s$  & $\alpha_1$ & $\alpha_2$ & $\alpha_3$ & $m^2_{12}$ \\
         \hline
         \hline
       $P_1$  & 95.81 & 125.09 & 420.18 & 1 & 7599.26 &  0.88& 1.01 & 0.47 & 0 \\
       $P_2$  & 125.09 & 248.80 &  828.88   & 1 & 1755.75 & 0.70 & -0.07 & -0.001 & 0 \\
        \hline
       $P_3$  & 125.09 & 242.27 & 1698.10  & 1 & 1041.23 &  0.797 & -0.049 & -0.176 & 0 \\
       $P_4$  & 125.09 & 392.9 &  1141.1 &  1 & 1009.4 & 0.77 & 0.11 & -0.14 & 0  \\
        \hline
       $P_5$  & 125.09 & 391.31 & 693.66 &  1 & 2868.37 &  0.73& 0.33& 1.39 & 198916 \\
       $P_6$  & 125.09 & 242.59 &  622.04   & 1 & 798.66 & 0.877 & -0.55 & -1.48 & 179776 \\
       \hline
       \hline
    \end{tabular}
    \caption{Benchmark parameter points demonstrating different behavior for the doublet VEVs inside the singlet domain wall. The mass parameters $m_{h_1}$, $m_{h_2}$, $m_{h_3}$ as well as $v_s$ are given in GeV while $m^2_{12}$ is given in $\text{GeV}^2$.}
    \label{tab:benchamarks}
\end{table}
Even when the doublets scalar potential for $v_s = 0$ has its global minimum at $v_1 = 0$ and $v_2 = 0$ (as is the case when the effective mass terms are positive and $m_{12} = 0$), one does not always achieve $v_{1,2}(0)=0$ inside the wall whenever the effective masses are positive. This is due to the interplay between the kinetic and potential energy contributions of the domain wall's field configuration. When the energy barrier between the asymptotic vacua $(v_1, v_2, \pm v_s)$ and the extremum at $(v_1=0, v_2 = 0, v_s = 0)$ is large, the contribution of the potential energy to the total energy of the solution will be large. According to the Bogomolnyi method for static kink solutions \cite{vachaspati_2023}, the minimal energy solution of the kink field configuration requires the contribution to the total energy of the domain wall from the potential part and the kinetic part to be equal. This leads the field configuration to have a high contribution from the kinetic energy and therefore the fields inside the wall will have a rapidly changing profile without having to pass through $v_{1,2} = 0$. In the case of a small potential barrier, the potential energy and therefore also the kinetic energy will be both small leading to a thicker wall, and therefore the VEVs of the doublets can have enough space to converge to the actual minimum of the potential at $x=0$.
\begin{figure}[t]
     \centering
     \begin{subfigure}[b]{0.49\textwidth}
         \centering
         \includegraphics[width=\textwidth]{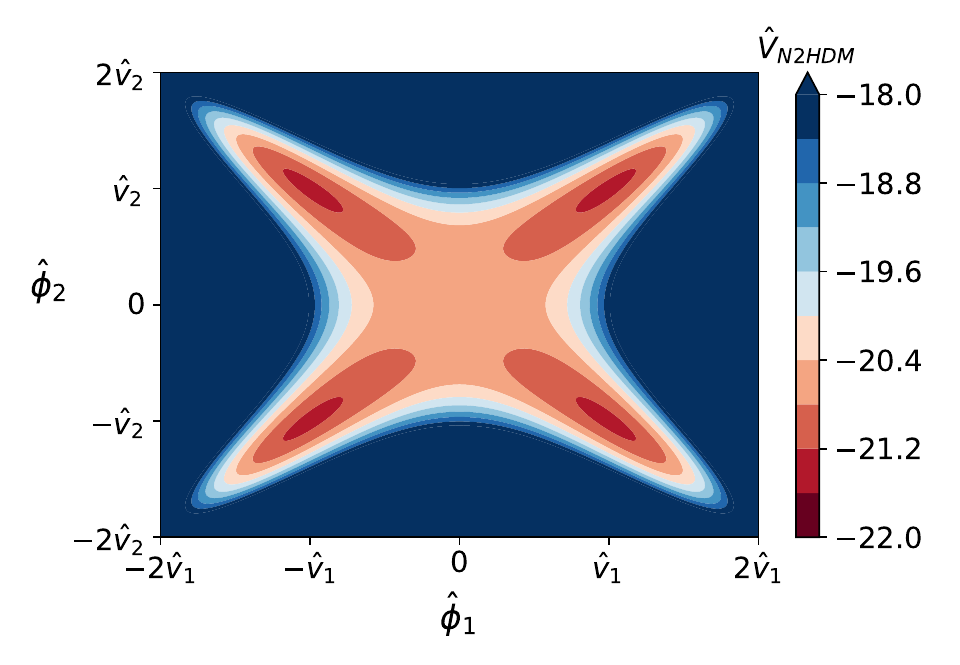}
       \subcaption{}  \label{subfig:potoutsidelower}
     \end{subfigure}
     \hfill
     \begin{subfigure}[b]{0.49\textwidth}
         \centering
         \includegraphics[width=\textwidth]{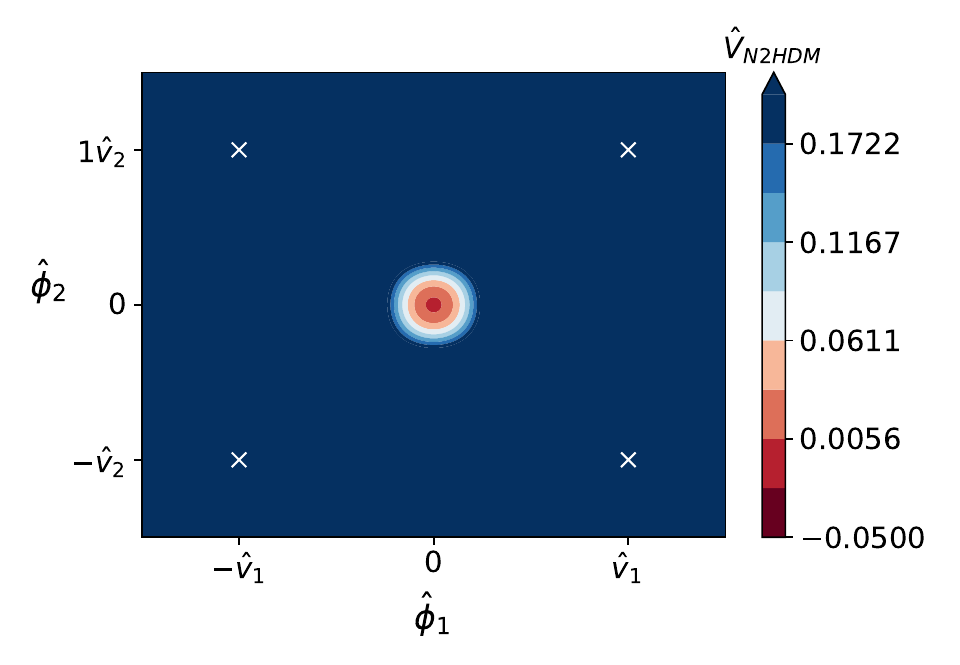}
        \subcaption{} \label{subfig:potinsidelower}
     \end{subfigure}
     \begin{subfigure}[b]{0.49\textwidth}
         \centering
         \includegraphics[width=\textwidth]{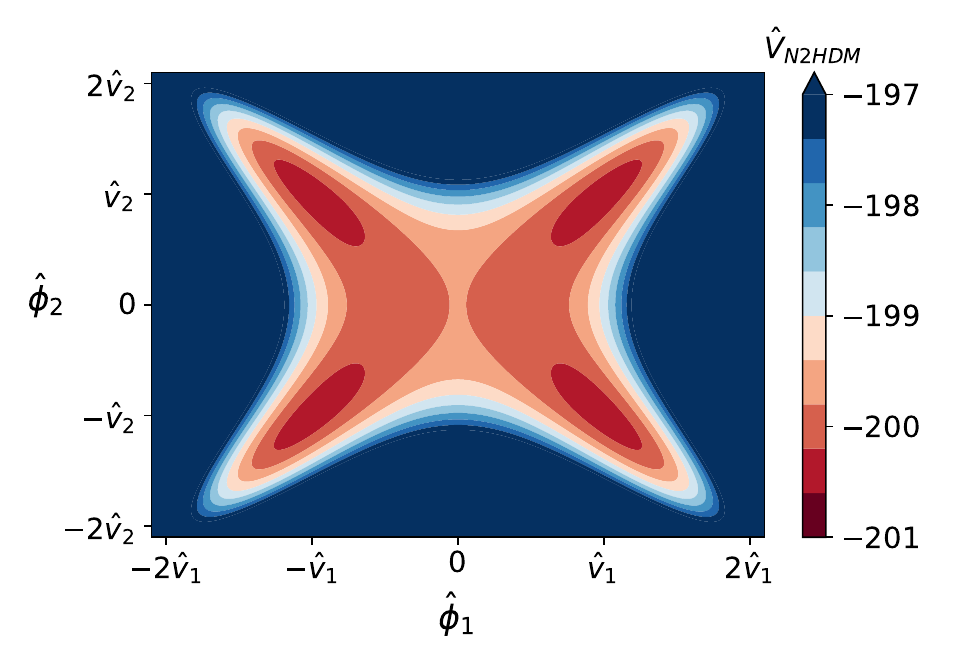}
       \subcaption{}  \label{subfig:potoutsidebigger}
     \end{subfigure}
     \hfill
     \begin{subfigure}[b]{0.49\textwidth}
         \centering
         \includegraphics[width=\textwidth]{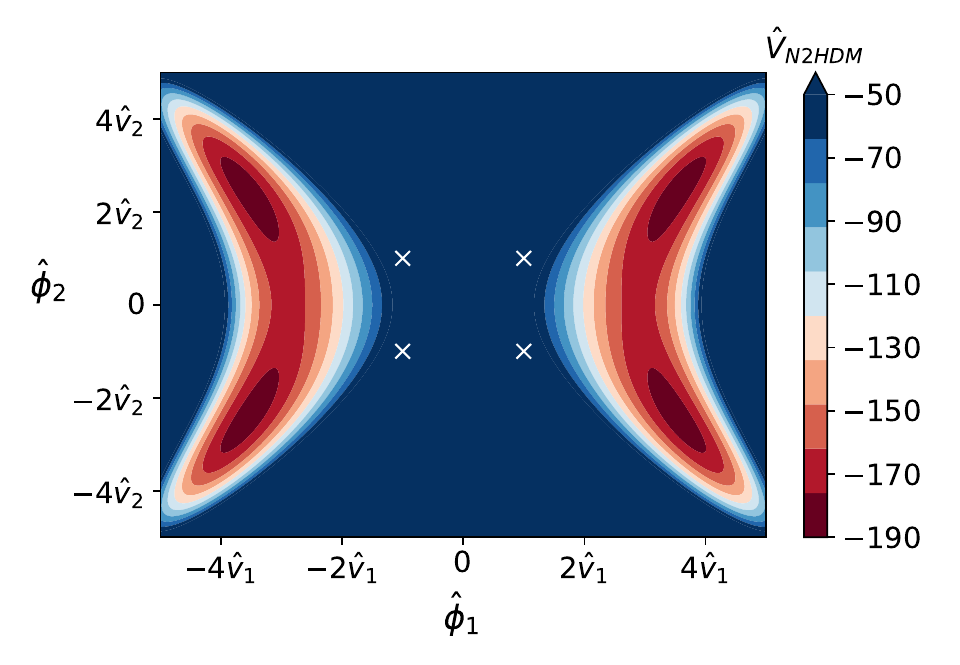}
        \subcaption{} \label{subfig:potinsidebigger}
     \end{subfigure}
\caption{The potential of the N2HDM: (b) inside ($\phi_s = 0$) and (a) outside ($\phi_s = v_s$) the domain wall for 2 parameter points $P_1$ (a,b) and $P_2$ (c,d).  } 
\label{fig:potentials}
\end{figure}

The opposite effect can happen when $\lambda_{7,8}>0$ (e.g. parameter point $P_2$). In this case the effective mass terms in (\ref{eq:meff1}) and (\ref{eq:meff2}) receive a negative contribution inside the wall (see Figure \ref{subfig:m1pp2}) and $v_{1,2}$ grow bigger (see Figure \ref{subfig:bigger}).

In order to explain these two distinct behaviors, we show in Figure \ref{fig:potentials} the potential of the Higgs doublets $\Phi_1$ and $\Phi_2$ inside and outside the singlet domain wall. For $P_2$, due to the quadratic effective masses being smaller inside the wall, the potential gets "stretched" and the minima of the potential have a higher value than those outside of the wall (depicted as a white cross sign in Figure $\ref{subfig:potinsidebigger}$). For $P_1$, the effective mass terms are higher and positive inside the wall. This leads to the 2HDM potential being in the symmetric phase just like the potential before EWSB.
We observe in Figure \ref{subfig:potinsidelower} that the minimum of the potential in this case is zero for both fields $\Phi_1$ and $\Phi_2$. Therefore, the vacuum expectation values for $v_1$ and $v_2$ inside the domain wall become very small in order to minimize the energy of the scalar fields configuration. 

In the following, we perform scans of random N2HDM parameter points to determine the different behaviors that can occur for the doublet fields inside the wall. The samples of parameter points were obtained using \texttt{ScannerS} \cite{Coimbra:2013qq, Ferreira:2014dya, Muhlleitner:2016mzt, Costa:2015llh, Muhlleitner:2020wwk}. We impose the theoretical constraints of boundedness from below, perturbative unitarity and vacuum stability, as well as experimental constraints from flavor physics and precision observables S, T and U. The boundedness from below condition is used to make sure that the potential does not tend to minus infinity at some direction. Perturbative unitarity ensures that the eigenvalues of the tree level S-matrix for $2\rightarrow 2$ scattering processes are smaller than $8\pi$ \cite{Muhlleitner:2016mzt}. We also impose the constraint of $Z'_2$ symmetry restoration at higher temperatures to ensure that all used parameter points lead to the formation of domain walls after the spontaneous breaking of that discrete symmetry at some point in the early universe.

We solve the differential equations describing the scalar field configuration for each generated parameter point satisfying the constraints. The results are quantified using the quantities:
\begin{align}
 r_1 = \dfrac{v_1(0)}{v_1(\pm \infty)}, &&
 r_2 = \dfrac{v_2(0)}{v_2(\pm \infty)}, &&
 \hat{v}_{ew}(0) = \dfrac{\sqrt{v^2_1(0)+v^2_2(0)}}{v_{sm}},
\end{align}
which gives a measure for the restoration of the electroweak symmetry inside the wall. 
We analyze the correlations between the results of the parameter scans and the difference in the effective mass inside and outside the wall $\Delta_{1,2}$:
\begin{align}
\Delta_{1,2} &= M_{1,2}(0) - M_{1,2}(\pm \infty) = \dfrac{1}{4}\lambda_{345}(v^2_{2,1}(0) - v^2_{2,1}(\pm \infty)) - \dfrac{1}{4}\lambda_{7,8}v^2_{s}. 
\label{eq:deltas}
\end{align}
Using the quantities $\Delta_{1,2}$ is motivated by the observation in the results of $P_1$ and $P_2$ that when the effective masses become less negative inside the wall ($\Delta_{1,2} > 0$) we obtain $r_{1,2}<1$ and when the effective masses inside the wall become more negative than outside of it ($\Delta_{1,2} < 0$) we obtain $r_{1,2}>1$. 
\subsubsection{General parameter scan with $\mathbf{m^2_{12} = 0}$}

\begin{table}[H]
\centering
{\renewcommand{\arraystretch}{1.0}
\footnotesize
\begin{tabular}{cccccc}
$m_{h_{a}}$  & $m_{h_{b}}$  & $m_{h_{c}}$ & $m_{A}$ &
$m_{H^{\pm}}$  & $\text{tan}\beta$ \\
\hline
\hline
$125.09$  & $[125.09,700]$  & $[125.09,1400]$ & $[400, 700]$ &
$[650,700]$  & $1$ \\
\hline
\hline
$C^{2}_{h_{a}t\bar{t}}$ & $C^{2}_{h_{a}VV}$  & $R_{b3}$  & $m_{12}^{2}$ & $v_{S}$ & $type$ \\
\hline
\hline
$[0.6, 1]$ & $ [0.8, 1.2]$ & $[-1,1]$ & $0$ & $[200,1500]$ & $2$ 
\end{tabular}
}
\caption{\small Set of input parameters for our \texttt{ScannerS} scan. We focus first on parameter points with $m^2_{12} =0$. $C_{h_{a}t\bar{t}}$ and $C_{h_{a}VV}$ are defined respectively as the coupling factors of the CP-even Higgs boson $h_a$ to the SM gauge bosons and the top quark and are defined as $C_{h_aVV} = \cos{(\beta)}R_{a1} + \sin{(\beta)}R_{a2} $ and $C_{h_at\bar{t}}=R_{a2}/\sin{(\beta)}$ (see \cite{Muhlleitner:2020wwk}). }
\label{tab:input_parameters_Scan}
\end{table}

\begin{figure}[h]
     \centering
     \begin{subfigure}[b]{0.49\textwidth}
         \centering
         \includegraphics[width=\textwidth]{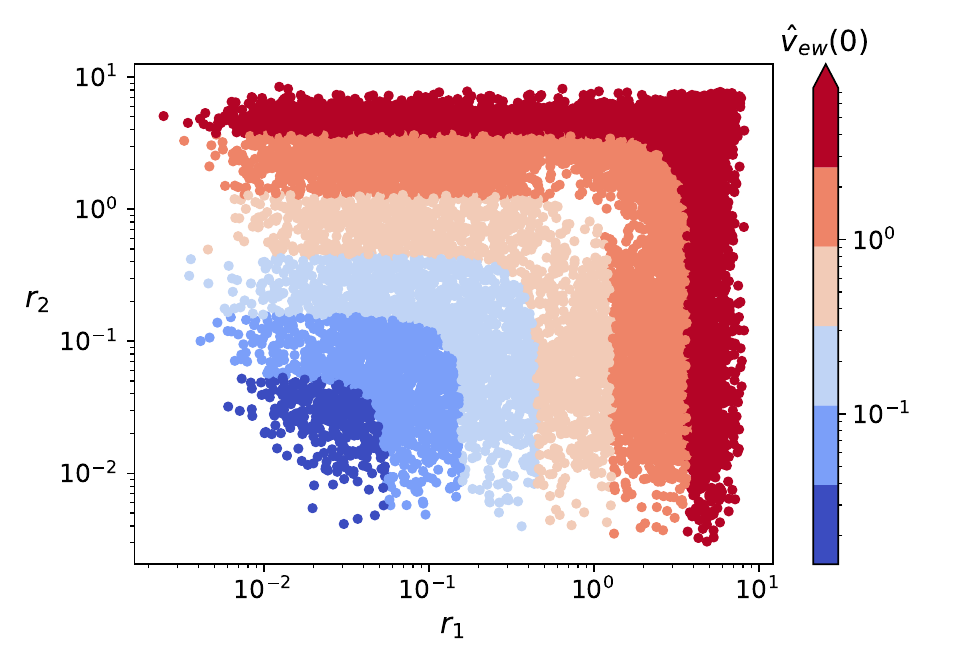}
        \subcaption{} \label{subfig:scan1vsq}
     \end{subfigure}
     \begin{subfigure}[b]{0.49\textwidth}
         \centering
         \includegraphics[width=\textwidth]{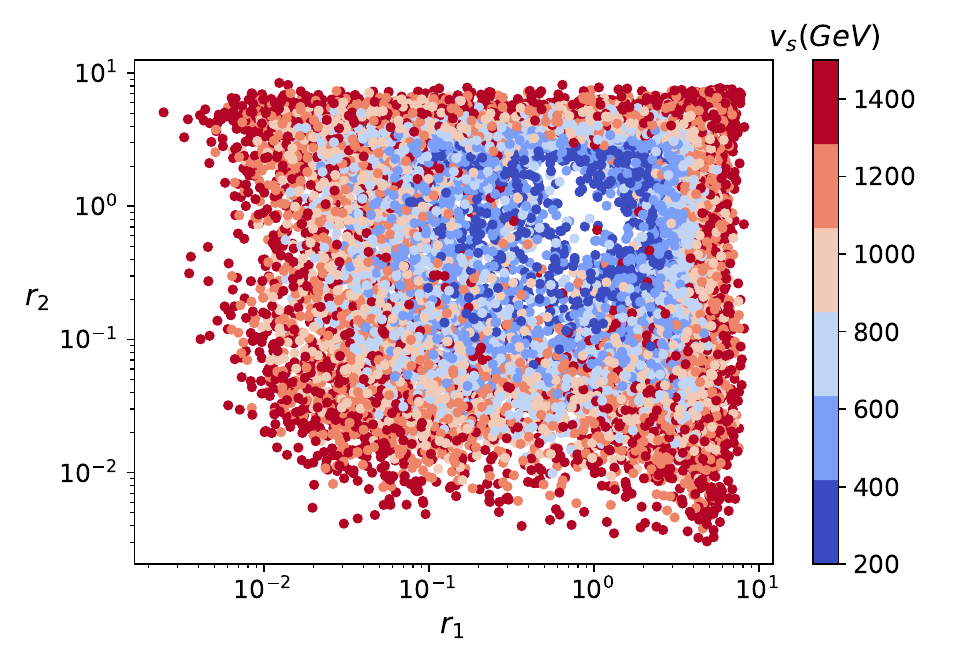}
       \subcaption{}  \label{subfig:scan1vs}
        \end{subfigure}
       \begin{subfigure}[b]{0.49\textwidth}
         \centering
         \includegraphics[width=\textwidth]{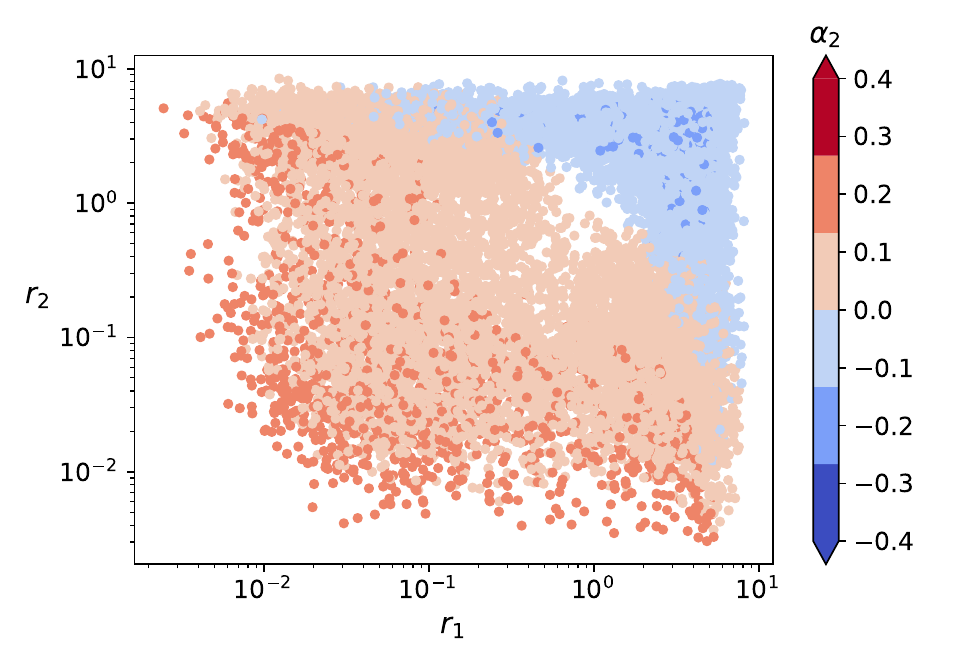}
        \subcaption{} \label{subfig:scan1alpha2}
     \end{subfigure}
      \begin{subfigure}[b]{0.49\textwidth}
         \centering
         \includegraphics[width=\textwidth]{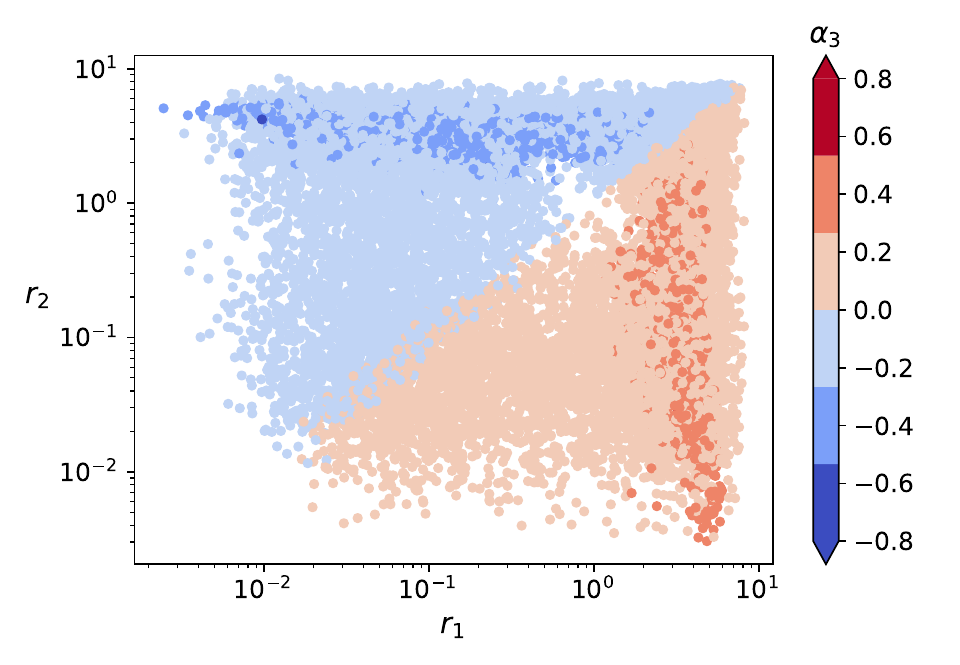}
       \subcaption{}  \label{subfig:scan1alpha3}
     \end{subfigure}
\caption{Result of the scan in terms of the ratios $r_1$ and $r_2$. (a) Results for $r_1$ and $r_2$ as a function of the electroweak symmetry VEV $v_{ew}(0)$ inside the wall. (b) shows the correlations as a function of $v_s$.  (c) shows the correlations as a function of $\alpha_2$. (d) shows the correlations as a function of $\alpha_3$. } 
\label{fig:ratios}
\end{figure}
We start with a general parameter scan of the N2HDM using a sample of 20000 parameter points (see Table \ref{tab:input_parameters_Scan}). For this particular scan, we take $m^2_{12} = 0$ and we only require $Z'_2$ symmetry restoration as we found that generating parameter points with \texttt{ScannerS} that also satisfy the constraint of EW symmetry restoration was much easier using non-zero values for $m_{12}$. Electroweak symmetry non-restoration at higher temperatures can constrain conventional models of electroweak baryogenesis since the non-restoration of the symmetry leads to the suppression of sphaleron transitions up to very high temperatures. However, the mechanism of electroweak baryogenesis using domain walls relies on the symmetry restoration (or on the partial symmetry restoration for intermediate values of $r_{1,2}$) only in the vicinity of the wall, where the sphaleron rate will be less suppressed than in the regions far from the wall and therefore EW symmetry non-restoration would not disadvantage such a mechanism for baryogenesis. 

The results of the scan are shown in Figure \ref{fig:ratios}. The ratios $r_{1,2}$ inside the wall can achieve low values up to $r_{1,2} \approx 0.001$ but also very high values corresponding to a much higher VEV inside the wall. Notice that the requirement that the effective mass terms turn positive inside the wall is not enough to induce a total electroweak symmetry restoration inside the wall (even if the potential is symmetric in that region). This is the case because the region in space with positive effective masses is not large enough for $v_{1,2}$ to converge to zero.

Due to the complexity of the model parameters, it is hard to obtain correlations between the physical variables of the model and the ratios $r_1$ and $r_2$. For this particular scan of the parameter points, we obtain some dependence between the values of the mixing angles $\alpha_2$, $\alpha_3$ and the ratios $r_1$ and $r_2$. As can be seen in Figure \ref{subfig:scan1alpha2}, one can achieve both ratios being small only for positive $\alpha_2$, while for $\alpha_2 < 0$, the ratios are bigger than one. This happens because we obtain both negative $\lambda_{7}$ and $\lambda_{8}$ only when $\alpha_2 > 0$. Concerning the dependence on the singlet VEV $v_s$ we find (see Figure \ref{subfig:scan1vs}) that the smallest ratios $r_{1,2}$ are obtained for larger $v_s$.
\begin{figure}[h]
     \centering
     \begin{subfigure}[b]{0.49\textwidth}
         \centering
         \includegraphics[width=\textwidth]{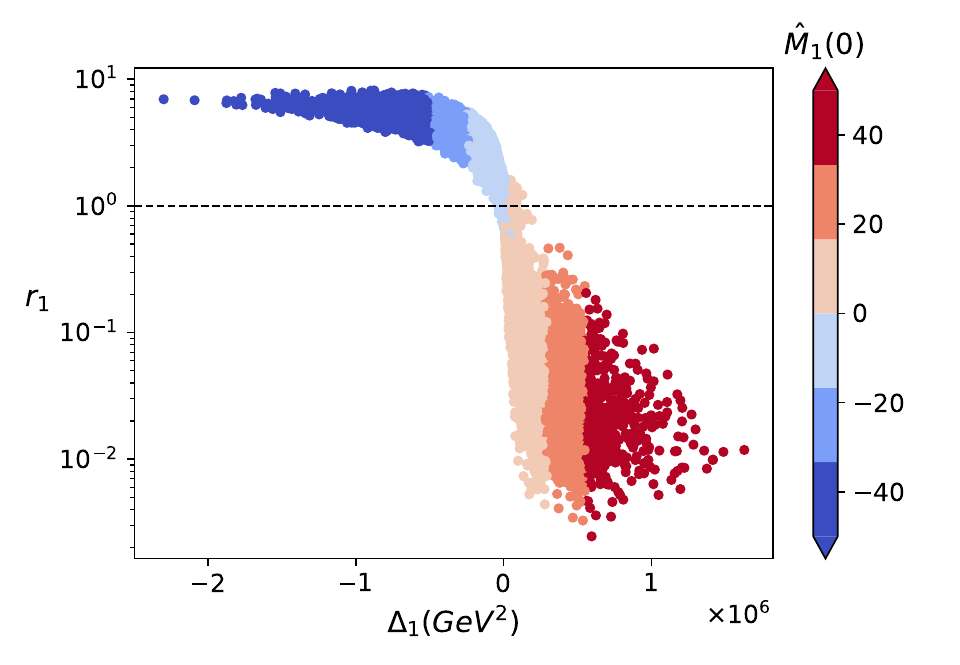}
        \subcaption{} \label{subfig:delta1r1}
     \end{subfigure}
     \begin{subfigure}[b]{0.49\textwidth}
         \centering
         \includegraphics[width=\textwidth]{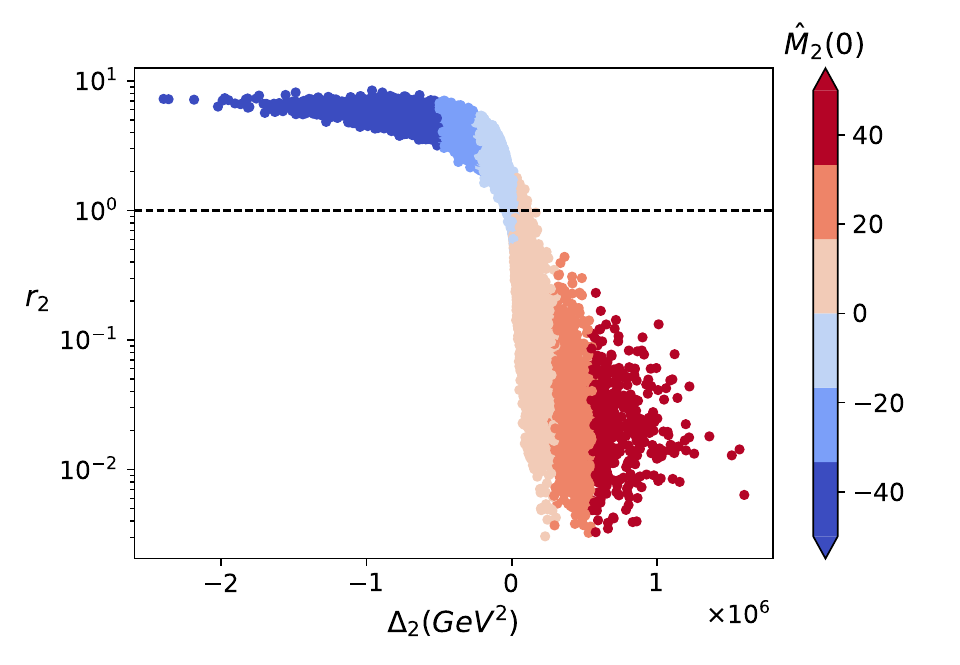}
       \subcaption{}  \label{subfig:delta2r2}
        \end{subfigure}
\caption{Correlations between the ratios $r_{1,2}$ and the difference in the effective mass inside and outside the wall $\Delta_{1,2}$.} 
\label{fig:deltasratios}
\end{figure}

In Figure \ref{fig:deltasratios}, we verify the validity of our assumption concerning the correlations between the sign of $\Delta_{1,2}$ and the ratios $r_{1,2}$. The anticipated behavior that $r_{1,2}>1$ for negative $\Delta_{1,2}$ and $r_{1,2}<1$ for positive $\Delta_{1,2}$ holds for most of the parameter points. However, one can see that some parameter points can have ratios $r_{1,2}>1$ even when $\Delta_{1,2} > 0$. 

\begin{figure}[h]
     \centering
     \begin{subfigure}[b]{0.49\textwidth}
         \centering
         \includegraphics[width=\textwidth]{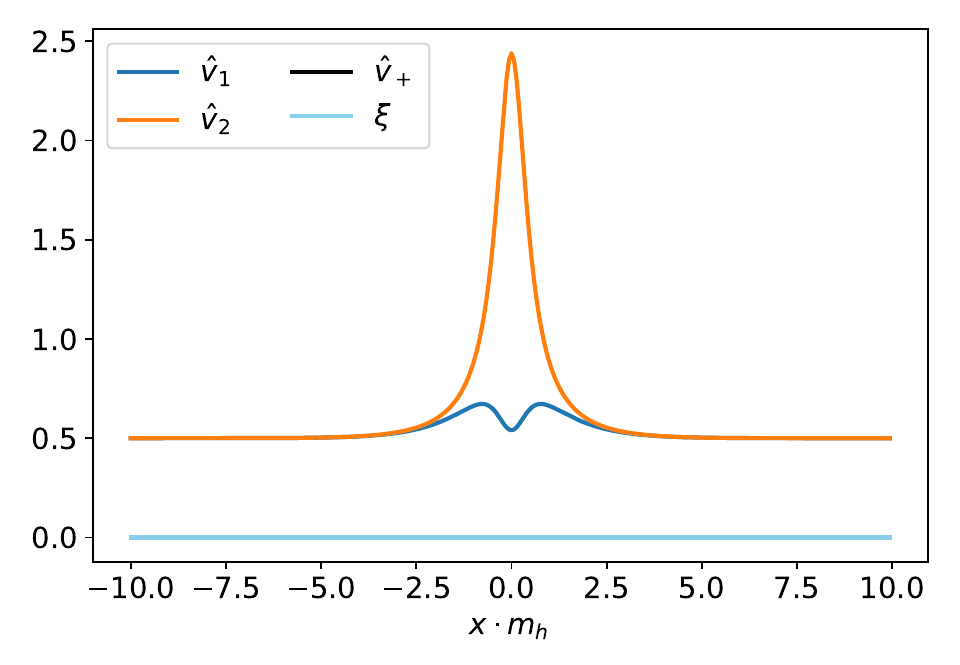}
        \subcaption{} \label{subfig:pp3}
     \end{subfigure}
     \begin{subfigure}[b]{0.49\textwidth}
         \centering
         \includegraphics[width=\textwidth]{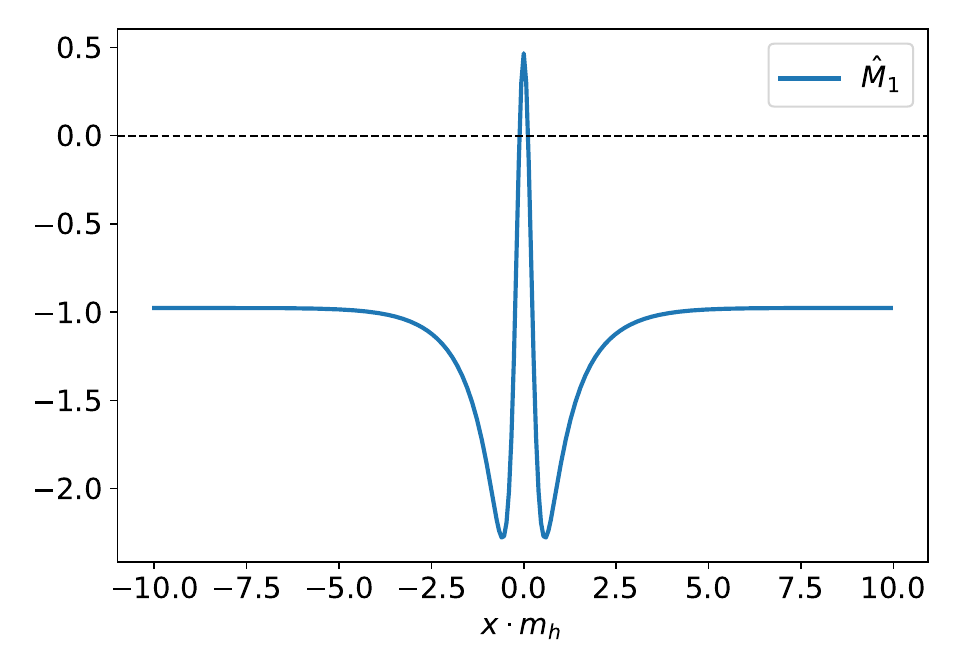}
       \subcaption{}  \label{subfig:pp3effmass}
        \end{subfigure}
\caption{Anomalous behavior where a positive $\Delta_1$ leads to $r_1>1$. (a) Electroweak vacuum configuration for the parameter point $P_3$ (see Table \ref{tab:benchamarks}). (b) Effective mass $\hat{M}_{1}$(x) for the parameter point $P_3$.} 
\label{fig:pp3}
\end{figure}
As an example for this anomalous behavior, consider the benchmark point $P_3$ (see Table \ref{tab:benchamarks}).
The domain wall solution for this parameter point is shown in Figure \ref{subfig:pp3}. The profile of $v_1(x)$ initially grows as we approach the wall then gets a sharp drop near $x=0$ with $v_1(0)$ still bigger than its value outside the wall leading to $r_1>1$. This behavior is explained by the profile of the effective mass (shown in Figure \ref{subfig:pp3effmass}). The effective mass $M_{1}$ is initially smaller (more negative) in the vicinity of the wall due to the term $\lambda_{345}v^2_2(x)$ in (\ref{eq:meff1}) being negative and sizable. This leads $v_1(x)$ to grow in that region. However, as we approach the core of the wall at $x=0$, the large positive contribution from $\lambda_7v^2_s$ leads the effective mass term to be bigger than its value outside the wall ($\Delta_1>0$). This sharp positive contribution is, however, only localized in a very small region around $x=0$. It is therefore energetically more favorable for $v_1(x)$ to not decrease substantially in that small region and we end with $v_1(0)>v_1(\pm \infty)$. The same behavior can also happen for $r_2>1$ and $\Delta_2>0$. This scenario can occur for parameter points where $\lambda_7$ is negative, while $\lambda_8$ is positive and large, leading to $v_2(0)$ being large inside the wall which in turn leads to the contribution $\lambda_{345}v^2_2(x)$ inside the effective mass term $M_{1}$ to be sizable.   
\begin{figure}[h]
     \centering
     \begin{subfigure}[b]{0.49\textwidth}
         \centering
         \includegraphics[width=\textwidth]{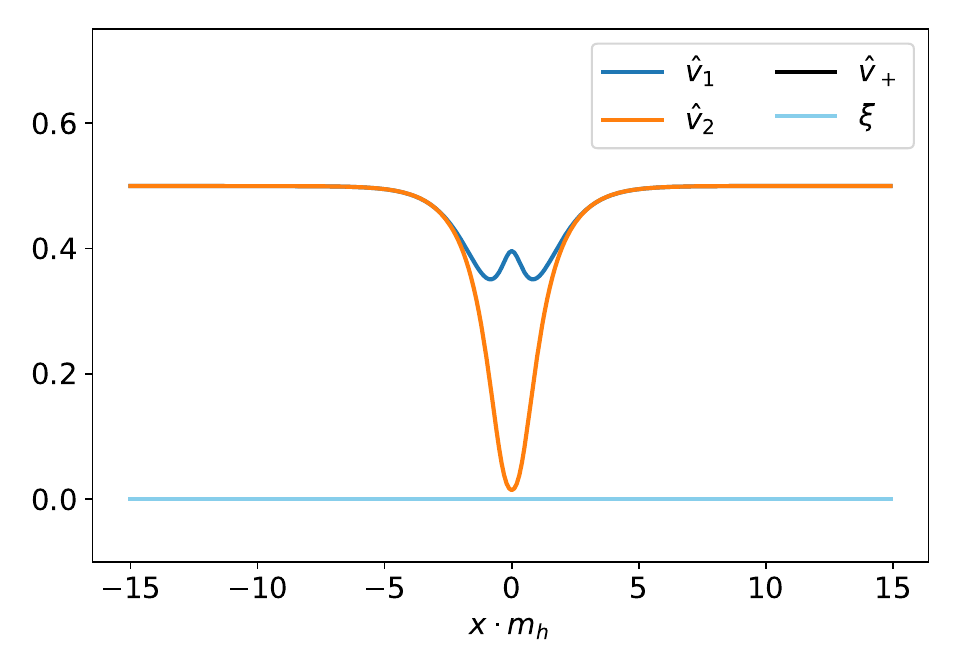}
        \subcaption{} \label{subfig:pp4}
     \end{subfigure}
     \begin{subfigure}[b]{0.49\textwidth}
         \centering
         \includegraphics[width=\textwidth]{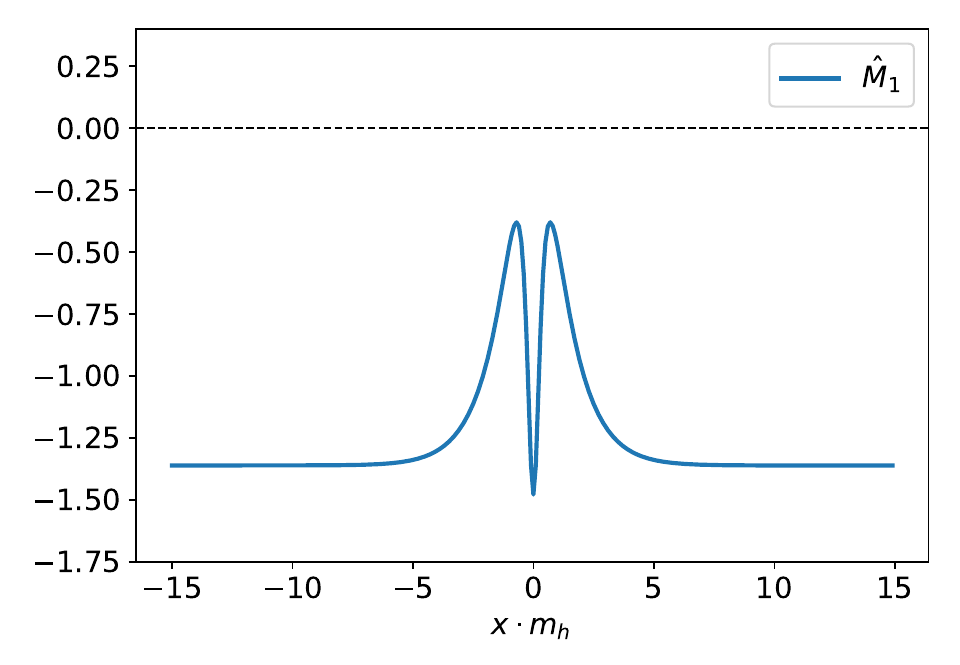}
       \subcaption{}  \label{subfig:pp4effmass}
        \end{subfigure}
\caption{Anomalous behavior where a negative $\Delta_1$ leads to $r_1<1$. (a) Electroweak vacuum configuration for the parameter point $P_4$ (see Table \ref{tab:benchamarks}). (b) Effective mass $\hat{M}_{1}$(x) for the parameter point $P_4$.} 
\label{fig:pp4}
\end{figure}

We also find a few parameter points where $r_1$ is slightly smaller than 1 even for $\Delta_1 < 0$. As an example for this scenario, we take the benchmark point $P_4$ (see Table \ref{tab:benchamarks}). The profile of the doublets is shown in Figure \ref{subfig:pp4}, where we observe the opposite behavior of the previous anomalous scenario, namely that $v_1(x)$ initially decreases as we approach the wall and grows inside the core of the wall. This behavior is explained by the profile of the effective mass $M_{1}$ (see Figure \ref{subfig:pp4effmass}) which initially grows in the vicinity of the wall leading to smaller $v_1$. However, due to $\lambda_7$ being positive, the effective mass obtains a sharp large negative contribution at $x=0$ which makes $v_1(x)$ grow again inside the wall. This negative contribution is, however, only limited to a very small region in space which is not enough to make $v_1(0)$ grow higher than its asymptotic value. This scenario happens especially for parameter points where $\lambda_7>0$ while $\lambda_8<0$ and sizable, leading to $v_2$ being very small inside the wall and as a consequence, the term $\lambda_{345}v^2_2$ in (\ref{eq:meff1}) gives a large positive boost to $M_{1}$ (for $\lambda_{345}<0$) in the vicinity of the wall. 
\subsubsection{Parameter scan with $\mathbf{m^2_{12} \neq 0}$}
\begin{figure}[h]
     \centering
     \begin{subfigure}[b]{0.49\textwidth}
         \centering
         \includegraphics[width=\textwidth]{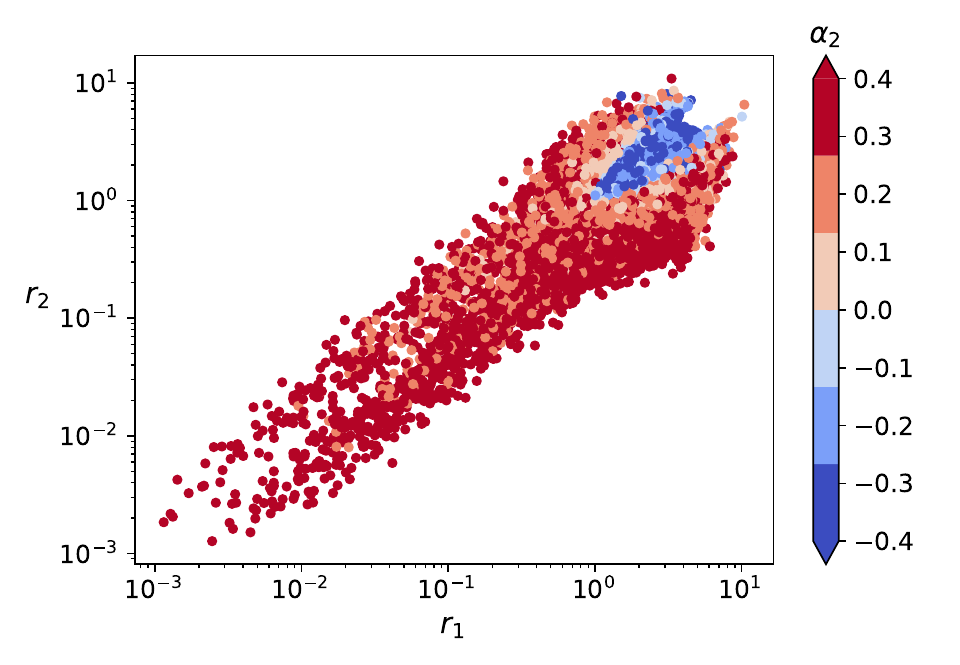}
        \subcaption{} \label{subfig:scanm12a2}
     \end{subfigure}
     \begin{subfigure}[b]{0.49\textwidth}
         \centering
         \includegraphics[width=\textwidth]{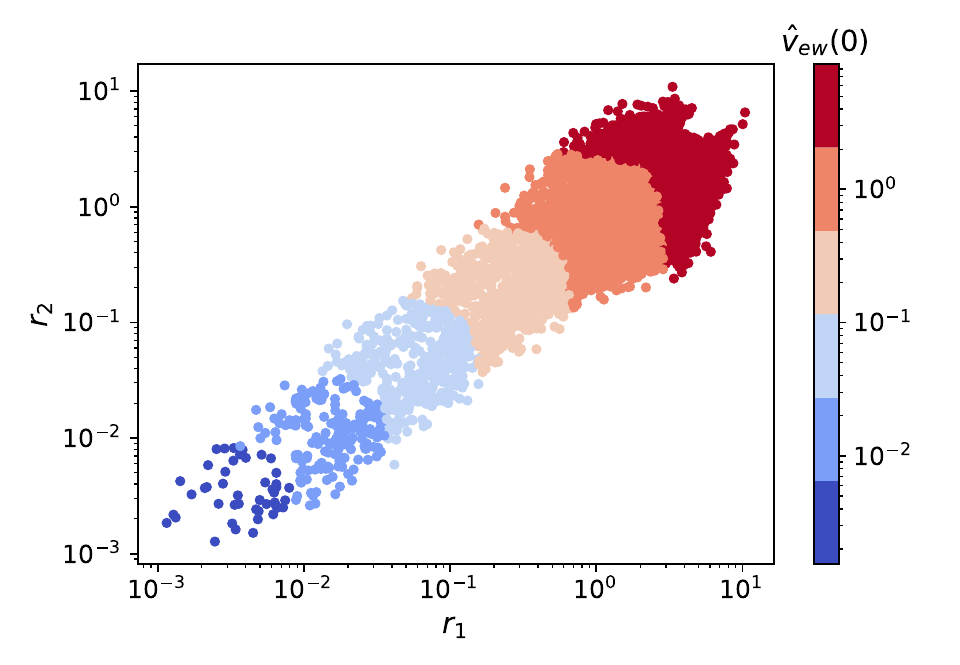}
       \subcaption{}  \label{subfig:smallm12}
        \end{subfigure}
\caption{Results of the parameter scan for non-zero $m^2_{12}$. It is possible to achieve smaller values for $r_{1,2}$ inside the wall compared to the previous parameter scan.} 
\label{fig:m12nonzero}
\end{figure}
To see the effects of a non-zero $m^2_{12}$ term, we perform a parameter scan for 20000 points taking the same constraints used as in the previous scan (see Table \ref{tab:input_parameters_Scan}) but with $0 < m^2_{12} < 10^5 \text{ GeV}^2$. Notice that, in contrast to the previous case, even for positive values of the effective masses $M_{1,2}$ outside the wall, the potential can get a VEV due to the non-zero $-m^2_{12}\Phi_1\Phi_2$-term that can lead to a dominant negative contribution to $V_{N2HDM}$. For this parameter scan, we impose the condition of symmetry restoration for both the $Z'_2$ and the EW symmetry at high temperatures.

In contrast to the case with $m^2_{12} = 0$, we do not observe the possibility of having one ratio $r_i$ being very small while the other ratio $r_j$ is big (see Figure \ref{subfig:scanm12a2}). Note that such behavior in the previous case was obtained for parameter points that lead to one doublet field $\Phi_i$ having $M_{i}(0)>0$ while $M_j(0)<0$. Therefore, the potential inside the wall had its minimum at $(v_i = 0, v_j \neq 0)$. For the case $m^2_{12} \neq 0$, and for parameter points where the two doublets have a different sign for their effective masses inside the wall, the term $-m^2_{12}\Phi_1\Phi_2$ shifts the minimum inside the wall to $(v_i \neq 0, v_j \neq 0)$ which reduces the difference between $r_i$ and $r_j$.     
\begin{figure}[h]
     \centering
     \begin{subfigure}[b]{0.49\textwidth}
         \centering
        \includegraphics[width=\textwidth]{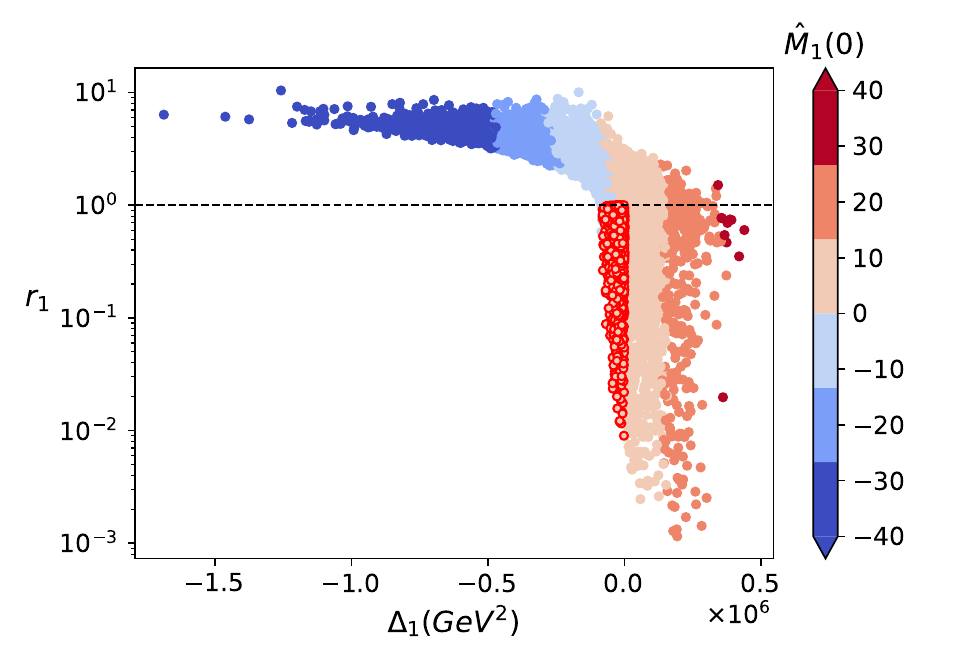}
        \subcaption{} \label{subfig:delta1r1m12}
     \end{subfigure}
     \begin{subfigure}[b]{0.49\textwidth}
         \centering
         \includegraphics[width=\textwidth]{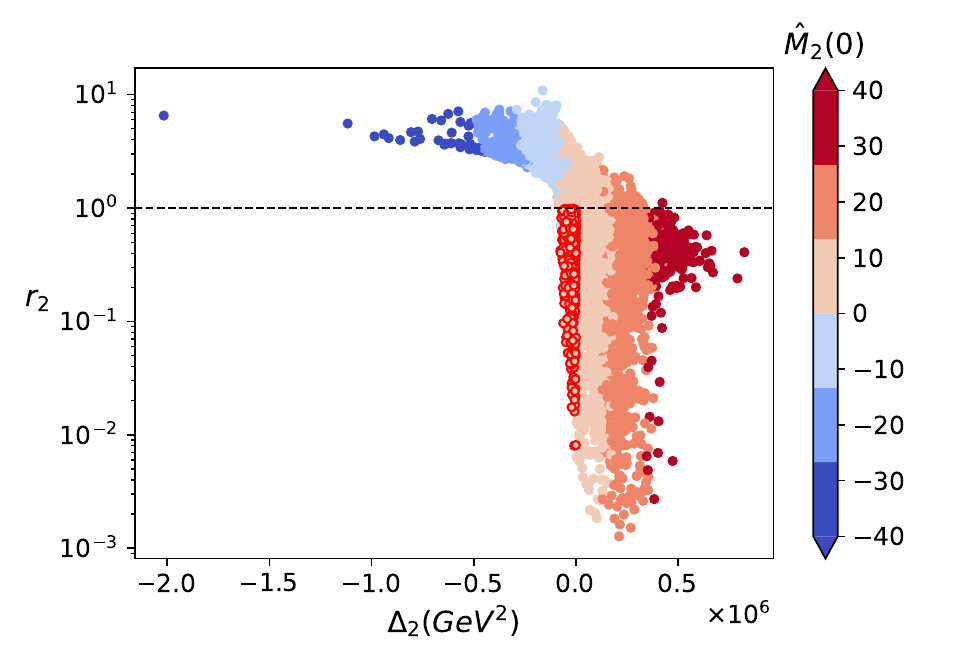}
       \subcaption{}  \label{subfig:delta2r2m12}
        \end{subfigure}
\caption{Correlations between the ratios $r_{1,2}$ and the difference in the effective mass inside and outside the wall $\Delta_{1,2}$ for the case when $m_{12} \neq 0$.} 
\label{fig:deltasratiosm12}
\end{figure} 

\begin{figure}[t]
     \centering
     \begin{subfigure}[b]{0.49\textwidth}
         \centering
        \includegraphics[width=\textwidth]{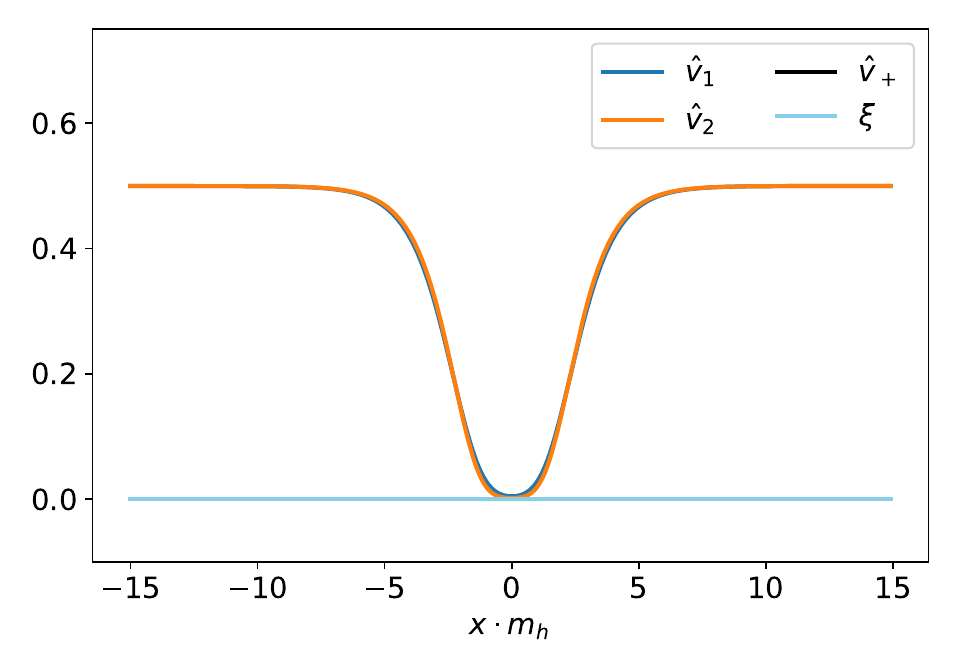}
        \subcaption{} \label{subfig:v1v2p5}
     \end{subfigure}
     \begin{subfigure}[b]{0.49\textwidth}
         \centering
         \includegraphics[width=\textwidth]{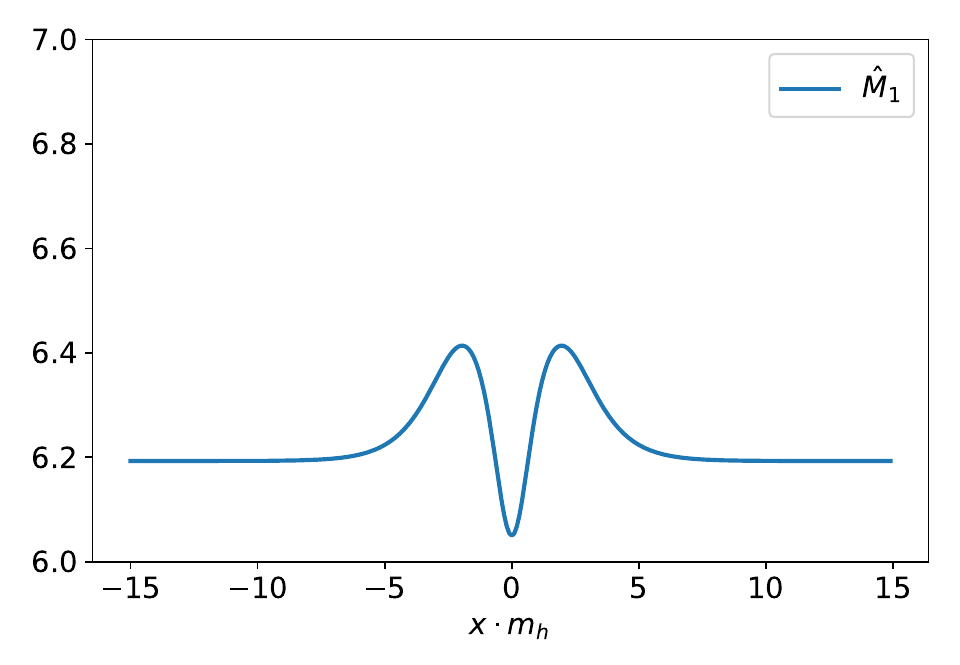}
       \subcaption{}  \label{subfig:m1p5}
        \end{subfigure}
\caption{(a) Profiles of the doublet VEVs $v_{1,2}$ for parameter point $P_5$. (b) Profile of the effective mass term $M_{1}$ for parameter point $P_5$.} 
\label{fig:p5}
\end{figure}

Another key difference is that we observe an anomaly for some parameter points satisfying $\Delta_{1,2}<0$, but result in $r_{1,2}<1$ (shown in bubbles with red edges in Figure \ref{fig:deltasratiosm12}). This happens for parameter points that have positive effective masses $M_{1,2}(x)$ everywhere in space and $\Delta_{1,2}$ being negative\footnote{Recall that for the case $m^2_{12} = 0$, such a scenario is not possible because the effective masses outside the wall are always negative.}. For these parameter points, we found that the potential inside the domain wall can still have its minimum at $(v_1,v_2) = (0,0)$ even if the effective mass terms get smaller (but are still positive). Therefore, the ratios $r_{1,2}$ will be small because the profile of $v_{1,2}(x)$ will converge to zero inside the wall. As an example for this scenario, we choose the benchmark point $P_5$ (see Table \ref{tab:benchamarks}). The results for the profile of $v_{1,2}(x)$ and $M_{1}(x)$ are shown in Figure \ref{fig:p5}.
For this parameter point, $\lambda_8$ is negative and $M_{2}$ grows substantially inside the wall. This leads the potential of the 2HDM in the direction $v_2$ to have its minimum at a small value near 0, making contributions from $-m_{12}v_1(0)v_2(0)$ vanishing or being small. For this parameter point, $\lambda_7$ is small and positive. Therefore $M_{1}(0)$ is smaller than outside the wall but stays positive. The minimum for $v_1$ inside the wall will then be 0, as the term $-m^2_{12}v_1v_2$ is negligible compared to the effective mass term $M_{1}$.  

\begin{figure}[h]
     \centering
     \begin{subfigure}[b]{0.49\textwidth}
         \centering
        \includegraphics[width=\textwidth]{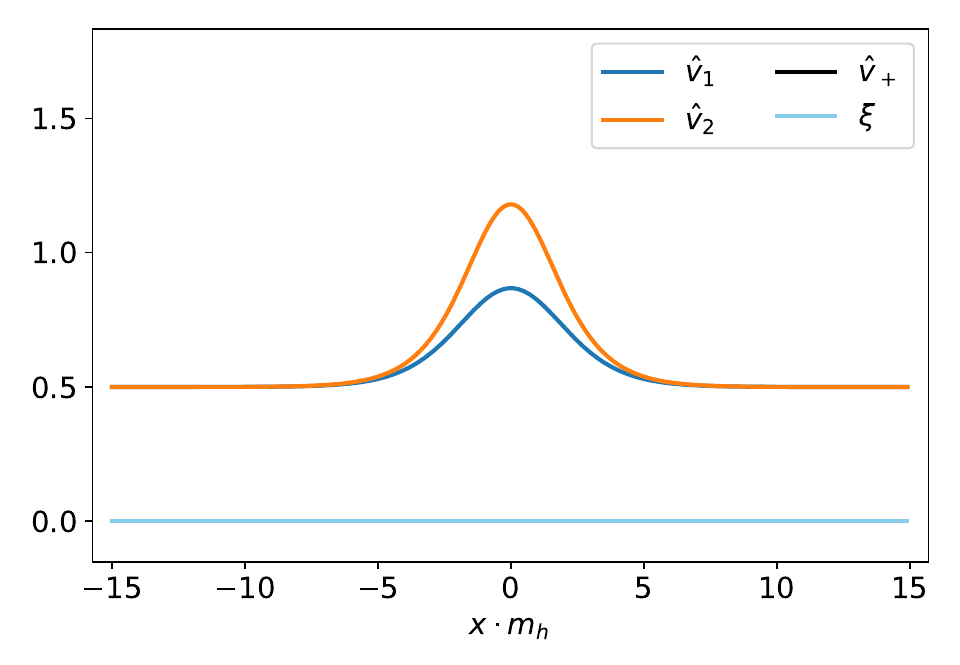}
        \subcaption{} \label{subfig:v1v2p6}
     \end{subfigure}
     \begin{subfigure}[b]{0.49\textwidth}
         \centering
         \includegraphics[width=\textwidth]{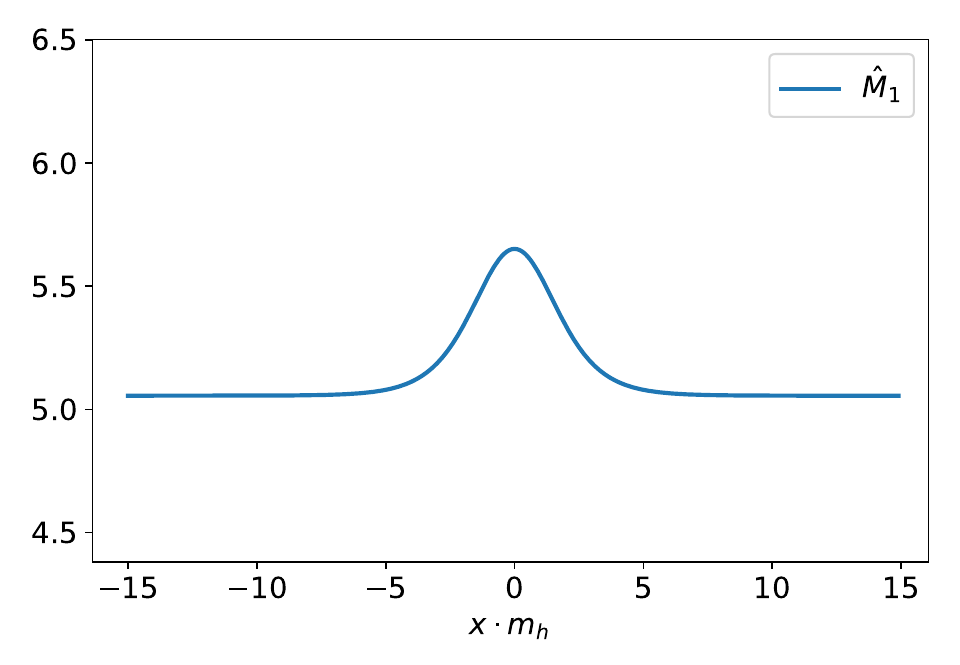}
       \subcaption{}  \label{subfig:m1p6}
        \end{subfigure}
\caption{(a) Profiles of the doublet VEVs $v_{1,2}(x)$ for parameter point $P_6$. (b) Profile of the effective mass term $M_{1}(x)$ for parameter point $P_6$.} 
\label{fig:p6}
\end{figure}
Even though we obtain mostly similar correlations between the ratios $r_{1,2}$ and the difference in the effective masses $\Delta_{1,2}$ for the case of $m_{12} \neq 0$, more parameter points can have $r_{1,2}>1$ for $\Delta_{1,2}>0$ (see Figure \ref{fig:deltasratiosm12}). This happens due to the minimum of the potential $V_{N2HDM}(v_1, v_2, 0)$ not being at the origin $(v_1, v_2) = (0,0)$  but at a finite value for the VEVs and even for positive effective masses. This happens when contributions from the term $-m^2_{12}\Phi_1\Phi_2$ are substantial compared to the effective mass terms.

As an example of this behavior we take the benchmark parameter point $P_6$ (see Table \ref{tab:benchamarks}). The profiles of $v_{1,2}(x)$ as well as $M_{1}(x)$ are shown in Figure \ref{fig:p6}. We find that $v_1(x)$ grows inside the wall even though the effective mass term $M_{1}(x)$ gets larger. For this parameter point, we obtain a large positive value for $\lambda_8 = 0.22$ compared to a negative $\lambda_7 = -0.04$. This has the effect that $M_{2}(x)$ sharply decreases inside the wall. This in turn "stretches" the 2HDM part of the potential in the $v_2$ direction, leading to a growing $v_2$ inside the wall. Because of the term $-m^2_{12}\Phi_1\Phi_2$, the minimum of the potential inside the wall in the direction of $v_1$ can be different than zero, even when the effective mass $M_{1}(0)$ is positive. Indeed, because $v_2(0)$ does not vanish inside the wall for this parameter point, a larger non-zero $v_1(0)$ will minimize the potential since the overall contribution $-m^2_{12}v_1(0)v_2(0)$ is negative.
 
\subsubsection{Impact of the wall's width on the region of EWSR}

One important consequence of electroweak symmetry restoration inside the wall is the enhancement of the sphaleron transitions compared to their rate in the region far from the wall. This can lead to the possibility of baryon number violating processes inside the wall that would be protected from being washed out once the wall moves away due to the sphaleron rate being highly suppressed in the broken phase. However, for such a mechanism to be efficient, the EW symmetry restoration region inside the wall should be large enough to fit a sphaleron. 
At high temperatures $T$, the radius of a sphaleron of the weak interactions is proportional to \cite{Cline:1998rc}: $$R_{sph} \propto g^2T^{-1},$$ where $g$ denotes the coupling constant of $SU(2)_L$, while the sphaleron radius at $T=0$ is on the order of the inverse of the W boson mass \cite{Brihaye:1993ud}. In this subsection, we discuss the dependence of the wall's width $\delta_s$ on the parameters of the model as well as how $\delta_s$ influences the size of the region where electroweak symmetry restoration of the Higgs doublets occurs. 

\begin{figure}[h]
     \centering
      \begin{subfigure}[b]{0.49\textwidth}
         \centering
         \includegraphics[width=\textwidth]{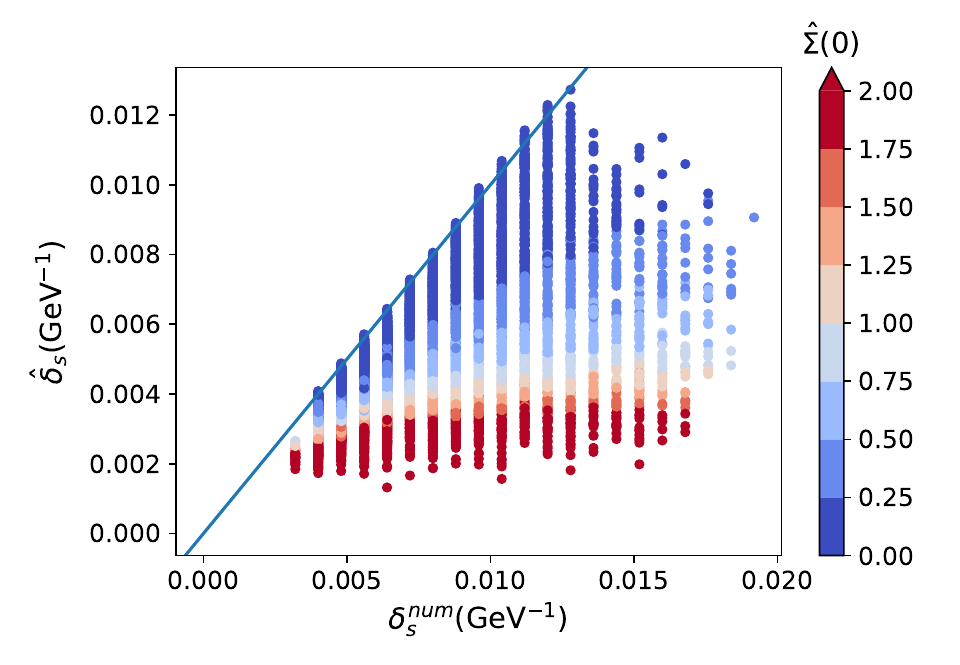}
       \subcaption{}  \label{subfig:generalwidthcorrelation}
        \end{subfigure}
     \begin{subfigure}[b]{0.49\textwidth}
         \centering
         \includegraphics[width=\textwidth]{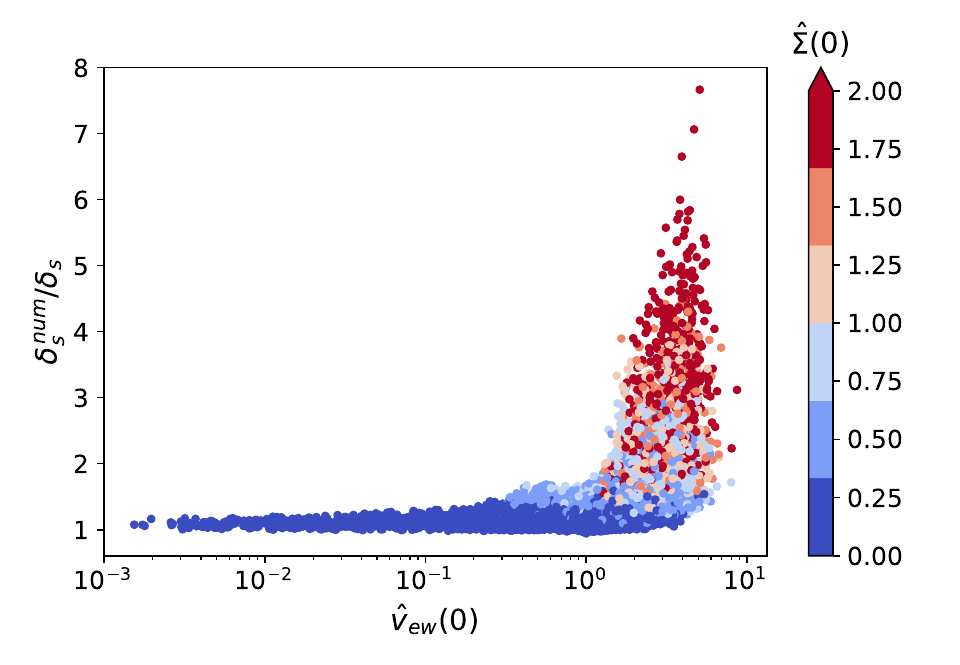}
        \subcaption{} \label{subfig:smallcouplingwidthcorrelation}
     \end{subfigure}
\caption{Correlations between the analytical formula for the domain wall width and the actual value calculated numerically for the parameter scan in Table \ref{tab:input_parameters_Scan} as a function of the normalized doublet vacuum expectation values inside the wall ($\hat{v}_{ew}(0)$) and the normalized quantity $ \hat{\Sigma}(0)$. Note that a change in $\hat{\Sigma}(0)$ for fixed $\hat{v}_{ew}(0)$ corresponds to the variation of the couplings $\lambda_7$ and $\lambda_8$. The blue line in (a) represents $\hat{\delta}_s = \delta^{num}_s$.} 
\label{fig:widthcorrelation}
\end{figure}

In the case of a pure scalar singlet model with no interaction with other scalar fields (e.g. $\lambda_{7,8} = 0 $), the width of the domain wall solution reduces to \cite{vachaspati_2023}:
\begin{equation}
  \hat{\delta}_s = (\sqrt{\dfrac{\lambda_6}{4}}v_s)^{-1}.
\label{eq:widthfree} 
\end{equation}
Naively, one would expect that, for small values of $\lambda_7$ and $\lambda_8$, the width of the singlet wall can be well approximated by (\ref{eq:widthfree}), as the mixing between the singlet and the doublet would then be negligible. However, this is not correct in general, as the profiles of $v_1(x)$ and $v_2(x)$ can change considerably inside the wall for high values of $v_s$, which can lead to $\lambda_{7,8} \sim \mathcal{O}(10^{-4})$. Looking at the equation of motion governing the profile of $v_s(x)$: 
\begin{equation}
    \dfrac{\partial^2 v_s}{\partial x^2} = \dfrac{1}{2}\biggl( 2 m^2_S + \lambda_7v^2_1(x) + \lambda_8v^2_2(x) \biggr)v_s(x) + \dfrac{\lambda_6}{4}v^3_s(x), 
 \label{eq:vseom}   
\end{equation}
one then expects that a sizable change in the doublets contribution to the singlet effective mass $ \lambda_7v^2_1(x) + \lambda_8v^2_2(x)$, which we define as:
\begin{equation}
\Sigma(x) = \lambda_7(v^2_1(x)-v^2_1(\infty))  + \lambda_8(v^2_2(x)-v^2_2(\infty)), 
\end{equation}
would lead to a considerable change in the width of the wall\footnote{A change in this quantity corresponds to the variation of the effective mass of the singlet field inside the wall. This will then lead to a change in the potential of the singlet inside the wall and therefore modify the path in field space that minimizes the energy of the field configuration.}. 
We verify this hypothesis using the parameter scan from the previous section (see Table \ref{tab:input_parameters_Scan}). Figure \ref{subfig:generalwidthcorrelation} shows the numerical values of the wall's width $\delta^{num}_s$ compared with the width $\hat{\delta}_s$ obtained via (\ref{eq:widthfree}) for each parameter point. The numerical value of the wall's width is obtained by calculating the full width at half the maximum of the field's profile.

Qualitatively, we find that parameter points with small $v_{ew}(0)$ and $\Sigma(0)$ lead to smaller ratios $r_{\delta} = \delta^{num}_s/\hat{\delta}_s $.
In particular, we find that $\delta^{num}_s$ is well approximated by $\hat{\delta}_s$ for parameter points with very small $\Sigma(0)$. This is expected as the equation of motion for $v_s(x)$ (\ref{eq:vseom}) reduces to the pure singlet scalar equation of motion in the limit of vanishing $\Sigma(x)$. Some parameter points show a slightly smaller width than the analytical formula. This, however, is due to numerical precision and the result should be interpreted as 1 or slightly higher than 1. For high values of $v_{ew}(0)$ and $\Sigma(0)$, the calculated numerical width is much higher than $\delta_s$. 
These general correlations can be intuitively understood by considering the change in the second derivative of $v_s(x)$ in (\ref{eq:vseom}). As the profiles of $v_{1,2}(x)$ change across the wall, the quantity $\Sigma(x)$ grows, decreasing $\dfrac{\partial^2 v_s}{\partial x^2}$ on the left side of the wall (where $v_s(x)$ is taken to have a negative sign). Therefore the kinetic energy of the solution gets smaller and the wall becomes thicker. Another intuitive way to look at this is interpreting the change in $\Sigma(x)$ as the change in the effective mass of the singlet. For all parameter points, $\Sigma(x)$ always grows in the vicinity of the wall. Therefore the effective mass of the singlet is higher, decreasing the barrier of the potential in the direction of the singlet field between the minima $-v_s$ and $v_s$. Consequently, the potential energy contribution to the domain wall solution gets smaller. According to the Bogomolnyi method for finding static kink solutions, the kinetic energy $(\dfrac{\partial \Phi_s}{\partial x })^2$ of the solution gets smaller and the wall gets thicker.
\begin{figure}[h]
     \centering           \includegraphics[width=0.65\textwidth]{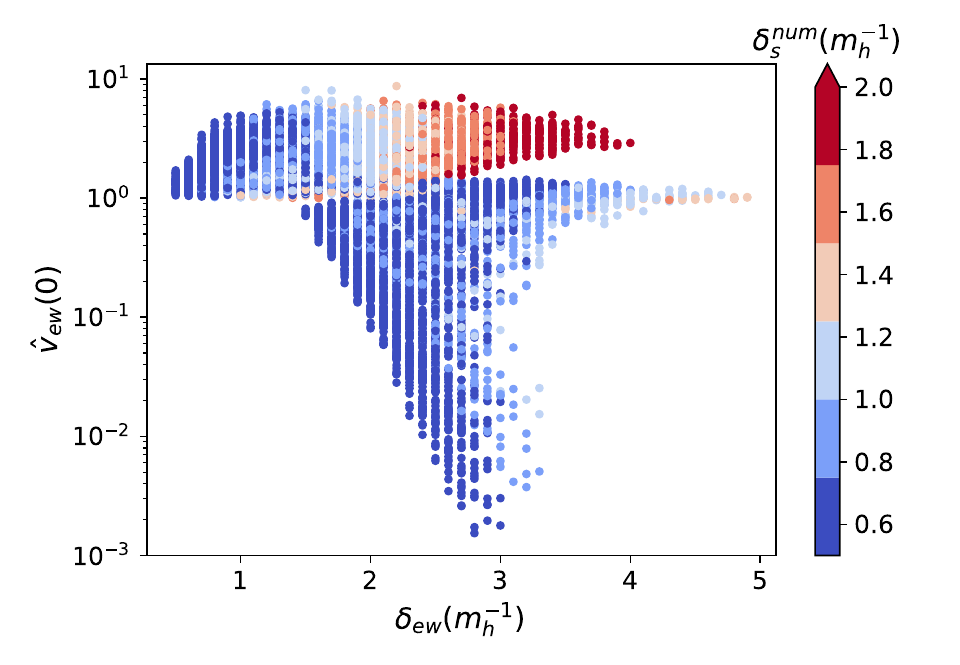}
\caption{Correlations between $\delta_{ew}$, the values of the electroweak vacuum inside the wall $v_{ew}(0)$ and the width of the singlet wall $\delta^{num}_s$. We observe that $\delta_{ew}$ increases with smaller $v_{ew}$ and larger $\delta^{num}_s$. } 
\label{fig:widthsinhgletewcorrelation}
\end{figure}

We now focus on the width of the region where the doublet scalar fields change their values, as this is a relevant quantity for electroweak baryogenesis mediated by domain walls. The change in the profile of $v_{1,2}(x)$ is related to the change in $v_s(x)$. Therefore, we expect that the width of the singlet domain wall influences the width of the region where $v_{1,2}(x)$ varies (which we define as $\delta_{ew}$) and as such the width of electroweak symmetry restoration region for small values of $r_{1,2}$. Figure \ref{fig:widthsinhgletewcorrelation} shows the correlation between $\delta_{ew}$, $v_{ew}(0)$ and $\delta^{num}_s$. For parameter points where the EW VEV gets smaller inside the wall ($\hat{v}_{ew}(0) < 1$) we find that the width $\delta_{ew}$ can be much bigger than the width of the domain wall. We also find that the width $\delta_{ew}$ increases with decreasing $v_{ew}(0)$ and can be very sizable even for small $\delta^{num}_s$.

For parameter points that have a higher EW VEV inside the wall ($\hat{v}_{ew}(0) > 1$), we also find that the width of the doublets increases with $\hat{\delta}^{num}_s$ as expected. Even though for this parameter scan one obtains higher values for $\delta^{num}_{s}$ when increasing $v_{ew}(0)$ (see Figure \ref{subfig:smallcouplingwidthcorrelation}), we observe that for a fixed value of $\delta^{num}_s$ and large $v_{ew}(0)$, the width $\delta_{ew}$ does not reach the high values observed for small values of $v_{ew}(0)$.

%% file: PhenoScenarios.tex
\section{Scenarios with EWSR in a large region around the wall}\label{section4}
As we discussed in the last sections, the VEV ratios inside the wall are dependent on the change in the effective mass terms inside the wall. As we are mainly interested in electroweak symmetry restoration inside the domain wall, we shall focus in this section on some possible scenarios where $v_1(x)$ and $v_2(x)$ become small for a large region in the vicinity of the wall. We saw in the previous random parameter scans that the electroweak symmetry is not completely restored inside the wall. Instead, the electroweak VEV $v_{ew}(0)$ gets rather small values when the effective mass terms of the doublets grow higher (and positive) inside the wall, forcing the 2HDM part of the potential (the first two lines of (\ref{eq:treepot})) to go into the symmetric phase. In order to make the profile of $v_1$ and $v_2$ vanish inside the wall, leading to complete electroweak symmetry restoration, we need the change in the effective masses $M_{1,2}$ to occur on a large region so that the doublet fields have enough space to converge to zero. 

As was shown in the last section, parameter points leading to smaller $v_{ew}(0)$ typically have singlet wall widths $\delta^{num}_s$ that can be well approximated by (\ref{eq:widthfree}) (see Figure \ref{subfig:smallcouplingwidthcorrelation}). As can be seen in Figure \ref{fig:widthsinhgletewcorrelation}, for a fixed value of $v_{ew}(0)$ the width $\delta_{ew}$ increases with $\delta^{num}_s$. Therefore, a larger region in the change of the effective masses is correlated with the quantity $\hat{\delta}_s = 2(\sqrt{\lambda_6}v_s)^{-1}$. Neglecting the contributions from $\lambda_{345}$, the change in the effective masses $M_{1,2}$ (see (\ref{eq:deltas})) is proportional to $-\lambda_{7,8}v^2_s$. One can then define the ratios $B_{1,2} = \lambda_{7,8}/\lambda_6 \propto (\Delta_{1,2}\times (\hat{\delta}_s)^2 )$ as dimensionless measures that provide a good correlation with parameter points leading to EWSR in a large region around the wall. This is the case namely when these ratios are big and negative. Therefore, one has to look for parameter points that lead to large values of $B_{1,2}$ with negative $\lambda_{7,8}$ (so that $\Delta_{1,2}$ are positive, leading to higher effective masses inside the wall). 

Using the expressions (\ref{eq:lambda7}) and (\ref{eq:lambda8}) for $\lambda_7$ and $\lambda_8$ in the physical mass basis of the N2HDM, we can write $B_{1,2}$ as:
\begin{align}
    B_1 & =  \biggl(\dfrac{v_s}{v_1}\biggr)\biggl(\dfrac{R_{13}R_{11}m^2_{h_1} + R_{23}R_{21}m^2_{h_2} + R_{33}R_{31}m^2_{h_3}  }{m^2_{h_1}R^2_{13} + m^2_{h_2}R^2_{23} + m^2_{h_3}R^2_{33}}\biggr), \label{eq:B1} \\
    B_2 & = \biggl(\dfrac{v_s}{v_2}\biggr)\biggl(\dfrac{R_{13}R_{12}m^2_{h_1} + R_{23}R_{22}m^2_{h_2} + R_{33}R_{32}m^2_{h_3}  }{m^2_{h_1}R^2_{13} + m^2_{h_2}R^2_{23} + m^2_{h_3}R^2_{33}}\biggr).
    \label{eq:B2}
\end{align}
Therefore, finding parameter points with large negative values for $B_{1,2}$ involves a rather complex interplay between the masses of the CP-even Higgs scalars and the components of the diagonalization matrix $R_{ij}$, which are functions of the mixing angles $\alpha_{1,2,3}$. 

In the following subsections, we discuss different scenarios for obtaining EWSR inside the wall. We generate parameter points using \texttt{ScannerS} satisfying all experimental and theoretical constraints discussed in the previous chapter. We also require that all parameter points restore the $Z'_2$ symmetry in the early universe to ensure the formation of the domain walls. Furthermore, for these sets of scenarios, we also require that the parameter points satisfy collider constraints (unless mentioned otherwise). This is done by using \texttt{HiggsBounds} \cite{Bechtle:2008jh,Bechtle:2011sb,Bechtle:2012lvg,Bechtle:2013wla,Bechtle:2015pma,Bechtle:2020pkv} and \texttt{HiggsSignals} \cite{Bechtle:2013xfa,Bechtle:2020uwn} implemented in \texttt{ScannerS}.
\subsection{Scenario 1: Small CP-even Higgs masses}\mbox{} 
\begin{table}[H]
\centering
{\renewcommand{\arraystretch}{1.0}
\footnotesize
\begin{tabular}{cccccc}
$m_{h_{a}}$  & $m_{h_{b}}$  & $m_{h_{c}}$ & $m_{A}$ &
$m_{H^{\pm}}$  & $\text{tan}\beta$ \\
\hline
\hline
$125.09$  & $[94,98]$  & $[200,300]$ & $[630, 750]$ &
$[650,750]$  & $[0.6,10]$ \\
\hline
\hline
$C^{2}_{h_{a}t\bar{t}}$ & $C^{2}_{h_{a}VV}$  & $R_{b3}$  & $m_{12}^{2}$ & $v_{S}$ & $type$ \\
\hline
\hline
$[0.6, 1.2]$ & $ [0.6, 1]$ & $[-1,1]$ & $[2 \times 10^4, 1.8 \times 10^5 ]$ & $[100,10000]$ & $1-4$ 
\end{tabular}
}
\caption{\small Set of input parameters for \texttt{ScannerS} scan of scenario 1. The masses and vacuum expectation values are given in $GeV$, while $m^2_{12}$ is given in $GeV^2$. $C_{h_{a}t\bar{t}}$ and $C_{h_{a}VV}$ are defined respectively as the coupling factors of the CP-even Higgs boson $h_a$ to the SM gauge bosons and the top quark and are defined as $C_{h_aVV} = \cos{(\beta)}R_{a1} + \sin{(\beta)}R_{a2} $ and $C_{h_at\bar{t}}=R_{a2}/\sin{(\beta)}$ (see \cite{Muhlleitner:2020wwk}). For these parameter points the $Z'_2$ is unbroken at very high temperatures.} 
\label{tab:input_parameters_Scan_scenario1}
\end{table}
The simplest way to get large values for the ratios $B_{1,2}$ is to use parameter points with large $v_s$. The second factor in (\ref{eq:B1}) and (\ref{eq:B2}) are complicated functions of the masses and mixing angles (that enter into $R_{ij}$) between the CP-even Higgs scalars and, as such, we expect some interplay between the values of the masses and mixing angles that lead to large $B_{1,2}$. For this first scenario, we focus on parameter points with small masses for the CP-even Higgs particles with one scalar fixed to be the SM-like Higgs with mass $m_{h_2} = 125.09 \text{ GeV}$. The lightest CP-even scalar is taken to be one with a mass in the vicinity of $95 \text{ GeV}$ depicting a particle in the same mass range of some recent excesses observed in ATLAS \cite{inproceedings} and CMS \cite{CMS:2024yhz}. We vary the singlet VEV $v_s$ between 100 GeV and 10 TeV as well as the mixing angles between the 3 CP-even Higgs states (see Table \ref{tab:input_parameters_Scan_scenario1}). The generated parameter points\footnote{Note that collider searches heavily constrain this particular scenario and therefore we start our discussion using a set of parameter points where these constraints were neglected. We later show the results for a parameter scan of 5000 points, where the collider constraints were considered.} satisfying the theoretical and experimental constraints discussed in earlier chapters as well as the requirement of $Z'_2$ symmetry restoration is in the range shown in Table \ref{tab:input_parameters_Scan_scenario1}. 

We start with a parameter scan where we do not require collider constraints in the parameter points search done by \texttt{ScannerS}. We later show the effects of these constraints by using a parameter scan where these constraints are imposed. This is done for pedagogical reasons in order to discuss how collider constraints affect the results.

\begin{figure}[t]
     \centering
     \begin{subfigure}[b]{0.49\textwidth}
         \centering
         \includegraphics[width=\textwidth]{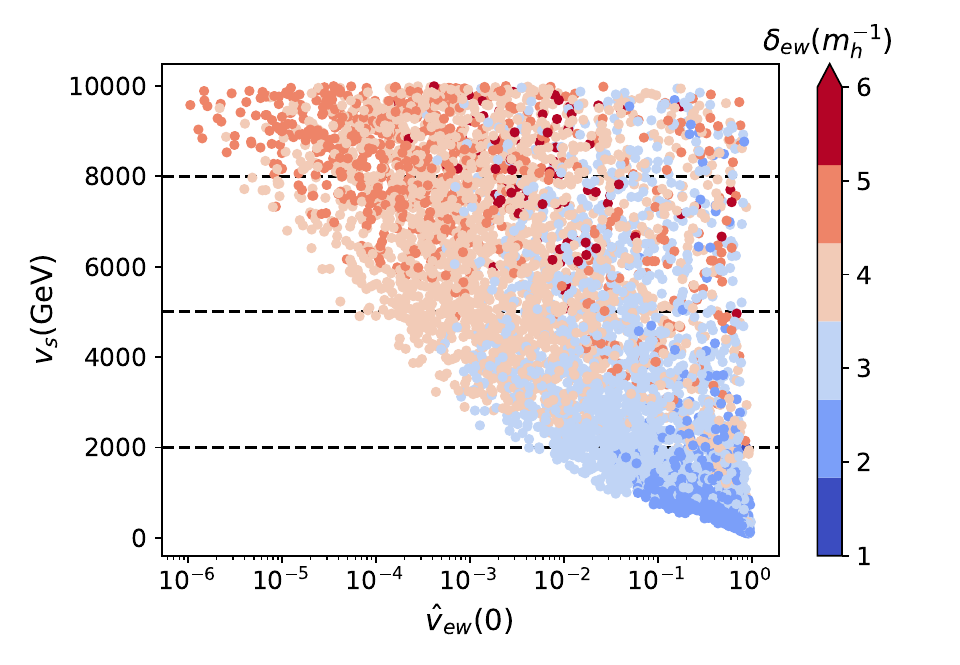}
        \subcaption{} \label{subfig:scenario1}
     \end{subfigure}
     \begin{subfigure}[b]{0.49\textwidth}
         \centering
         \includegraphics[width=\textwidth]{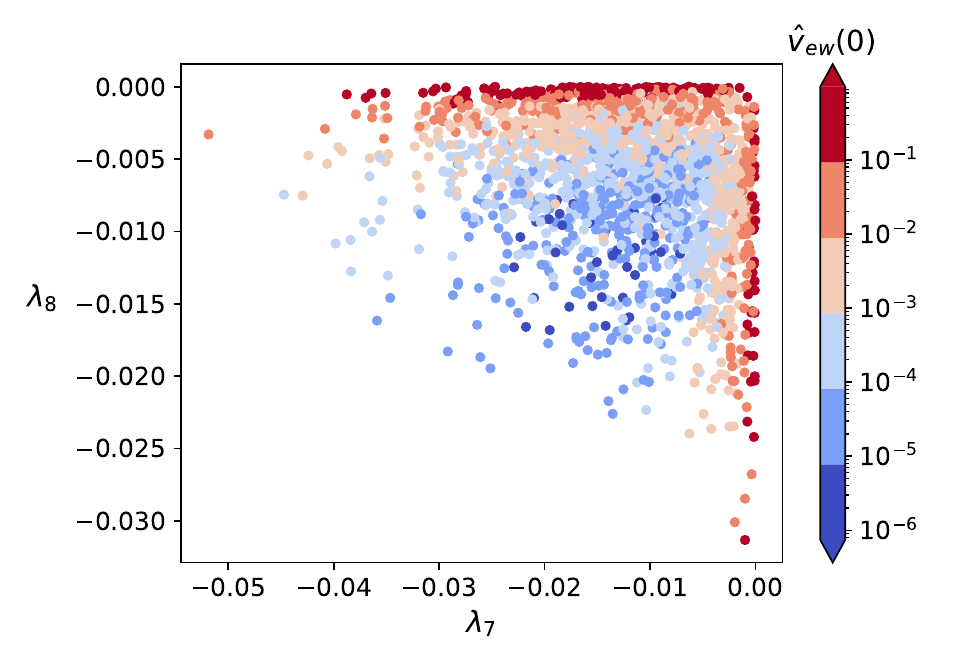}
       \subcaption{}  \label{subfig:scenario1lambdas}
        \end{subfigure}
\caption{Results of the parameter scan for scenario 1. (a) The amount of EWSR inside the domain wall $\hat{v}_{ew}(0)$ as a function of the singlet VEV $v_s$ and the width $\delta_{ew}$ of the symmetry restoration region $v_{ew}(x)$ around the wall. (b) The amount of EWSR inside the wall as a function of the couplings $\lambda_7$ and $\lambda_8$ for parameter points with $v_s > 6000 \text{ GeV}$.} 
\label{fig:scenario1}
\end{figure}
The results of the scan in terms of the effective EW VEV inside the wall $v_{ew}(0)$ as a function of $v_s$ and the width $\delta_{ew}$ are shown in Figure \ref{subfig:scenario1}. We find that for the lower range of $v_s < 2000 \text{ GeV}$, the EW symmetry restoration measure $\hat{v}_{ew}(0)$ is at least above $0.01$. We also find that for several parameter points with large $v_s > 6 \text{ TeV}$, the doublet VEVs inside the wall are not suppressed. These parameter points correspond to the red and orange points in Figure \ref{subfig:scenario1lambdas} where one of the couplings between the singlet and doublet scalar fields $\lambda_7$ or $\lambda_8$ is small compared to the other one. This leads to one of the doublets having its VEV highly suppressed inside the wall, while the other doublet is only slightly affected by the domain wall, leading to an overall effect where $\hat{v}_{ew}(0)$ is slightly smaller than 1.

Concerning the width $\delta_{ew}$, the observed indirect dependence on $v_s$ is due to the change in $\hat{v}_{ew}(0)$ which decreases with higher $v_s$ and leads to higher values for $\delta_{ew}$. For the chosen interval in the masses of the CP-even Higgses (see Table \ref{tab:input_parameters_Scan_scenario1}), we verified that the width of the singlet wall $\delta^{num}_s$ is independent of $v_s$. 

\begin{figure}[h]
     \centering
     \begin{subfigure}[b]{0.49\textwidth}
         \centering
         \includegraphics[width=\textwidth]{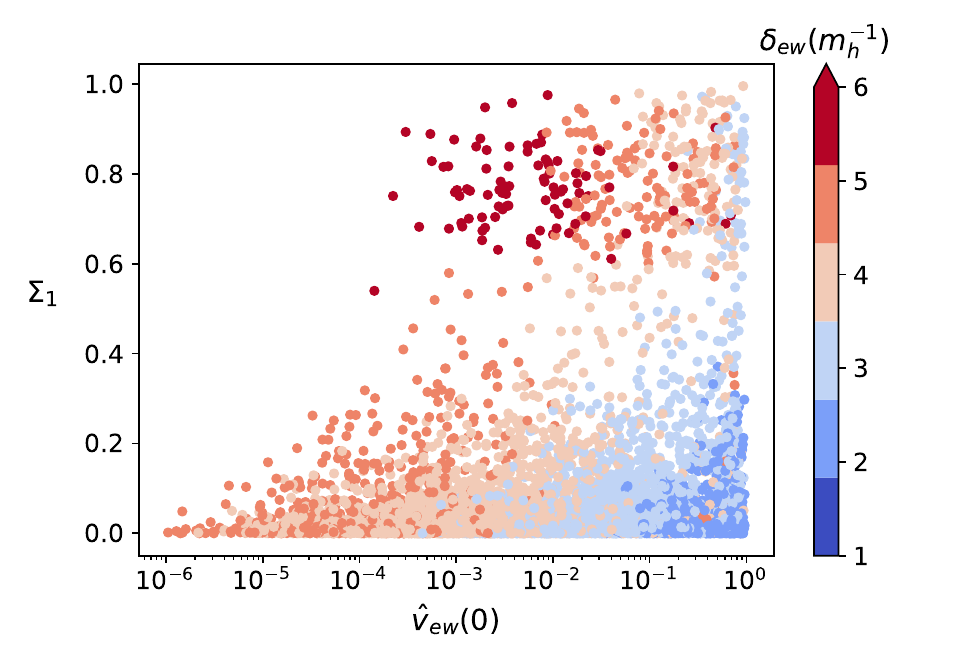}
        \subcaption{} \label{subfig:scenario1sigma1}
     \end{subfigure}
     \begin{subfigure}[b]{0.49\textwidth}
         \centering
         \includegraphics[width=\textwidth]{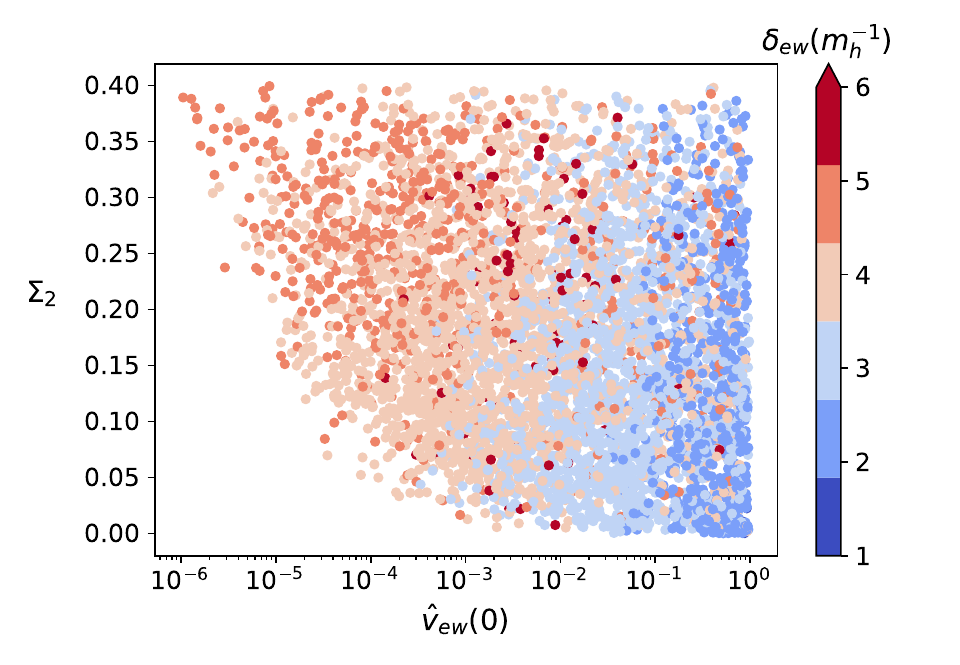}
        \subcaption{} \label{subfig:scenario1sigma2}
     \end{subfigure}
     \begin{subfigure}[b]{0.49\textwidth}
         \centering
         \includegraphics[width=\textwidth]{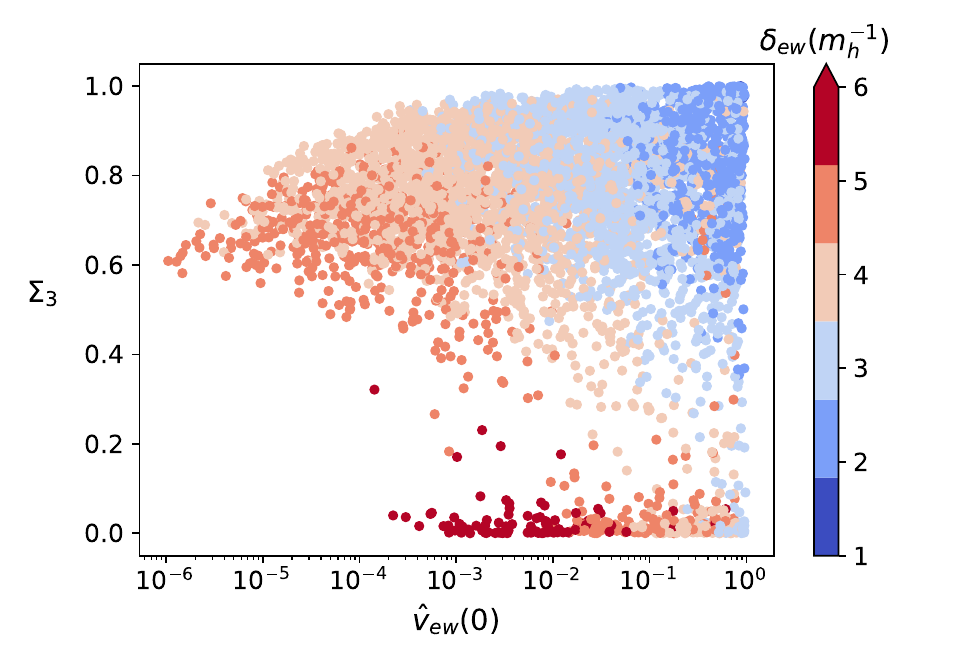}
       \subcaption{}  \label{subfig:scenario1sigma3}
        \end{subfigure}
\caption{Singlet admixture $\Sigma_i$ of the CP-even Higgs states $h_i$ in the physical mass basis. For this particular parameter scan, $\Sigma_2$ corresponds to the singlet admixture in the SM Higgs state $h_2$ with mass $m_{h_2} = 125.09 \text{ GeV}$, while $\Sigma_1$ corresponds to the singlet admixture of the CP-even Higgs boson with mass around $95 \text{ GeV}$.  } 
\label{fig:scenario1sigmas}
\end{figure}
We now discuss the effects of the mixing angles $\alpha_i$ on the results. In order to do this, we look at the singlet admixture $\Sigma_i$ of the CP-even Higgs states $h_i$ defined as: 
\begin{align}
\Sigma_1 = R^2_{13}, && \Sigma_2 = R^2_{23}, && \Sigma_3 = R^2_{33},
\label{eq:singletadmixture}
\end{align}
where $R_{ij}$ corresponds to the entries of the diagonalizing matrix R defined in (\ref{eq:Rmatrix}) and are functions of the mixing angles $\alpha_i$. 
The correlations between $\Sigma_i$ and the parameters $v_{ew}(0)$ and $\delta_{ew}$ are shown in Figure \ref{fig:scenario1sigmas}.
We find that the smallest values for $v_{ew}(0)$ are obtained when the singlet admixture in the SM Higgs scalar $h_2$ is the highest. Such a correlation puts constraints on this parameter region, as a large singlet admixture in the SM Higgs scalar is not allowed experimentally. 

\begin{figure}[t]
     \centering
     \begin{subfigure}[b]{0.49\textwidth}
         \centering
         \includegraphics[width=\textwidth]{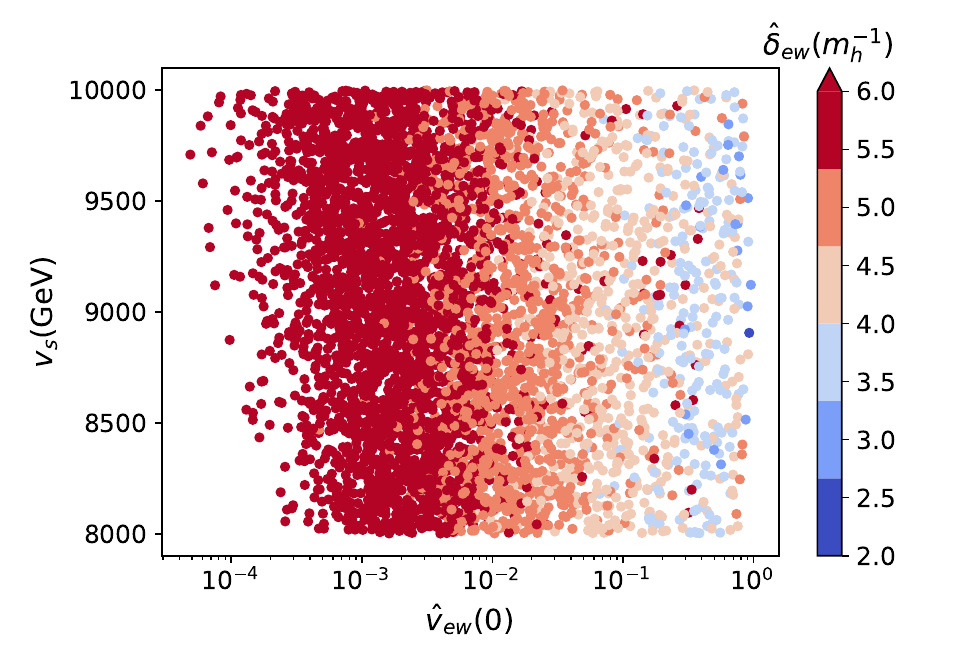}
        \subcaption{} \label{subfig:singletspecial}
     \end{subfigure}
     \begin{subfigure}[b]{0.49\textwidth}
         \centering
         \includegraphics[width=\textwidth]{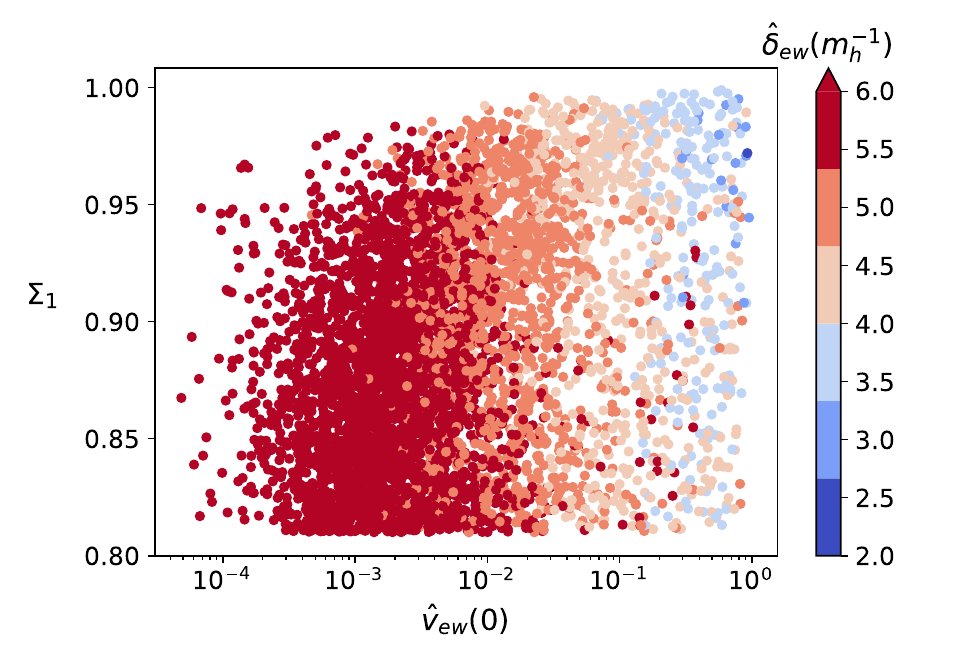}
       \subcaption{}  \label{subfig:sigma1special}
        \end{subfigure}
\caption{EW VEV inside the wall $v_{ew}(0)$ and the EW width $\delta_{ew}$ for the parameter scan where the singlet admixture of the lightest Higgs boson in scenario 1 is close to 1, while the singlet admixture of the SM Higgs boson is small, (a) dependent on $v_s$ and (b) dependent on $\Sigma_1$.} 
\label{fig:specialscanscenario1}
\end{figure}

We also perform a parameter scan for 5000 points where the singlet admixture is dominant in the lightest Higgs scalar ($94 \text{ GeV} < m_{h_1} < 98 \text{ GeV}$) while it is negligible in the SM Higgs. We choose $v_s$ to vary between $8 \text{ TeV}$ and $10 \text{ TeV}$ in order to get larger negative values for $B_{1,2}$ and therefore obtain parameter points that are more favorable to lead to EWSR inside the wall. The results are shown in Figure \ref{fig:specialscanscenario1}. We find that, for this parameter scan, $v_{ew}(0)$ is lower than the values it can reach when the singlet admixture in the SM-like Higgs boson $\Sigma_2$ is higher. We therefore conclude that a larger effect of EWSR inside the wall correlates with parameter points that have higher singlet admixture $\Sigma_2$ in the SM-like Higgs boson. Such a scenario is, however, strongly constrained by collider results as we discuss next.

\begin{figure}[H]
     \centering
     \begin{subfigure}[b]{0.49\textwidth}
         \centering
         \includegraphics[width=\textwidth]{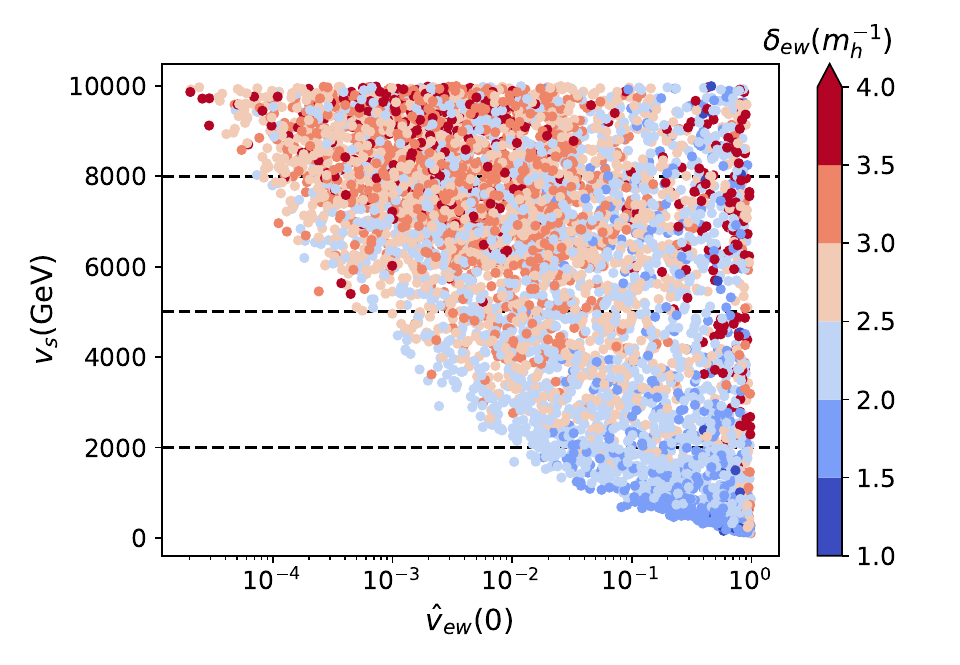}
        \subcaption{} \label{subfig:resultcollider1}
     \end{subfigure}
     \begin{subfigure}[b]{0.49\textwidth}
         \centering
         \includegraphics[width=\textwidth]{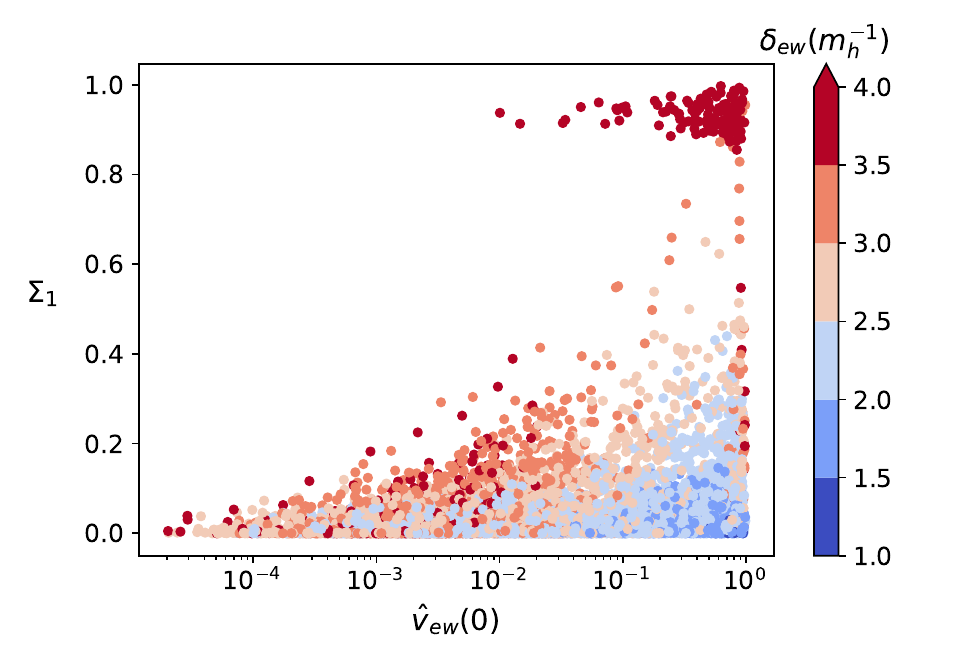}
       \subcaption{}  \label{subfig:sigma1collider}
        \end{subfigure}
     \begin{subfigure}[b]{0.49\textwidth}
         \centering
         \includegraphics[width=\textwidth]{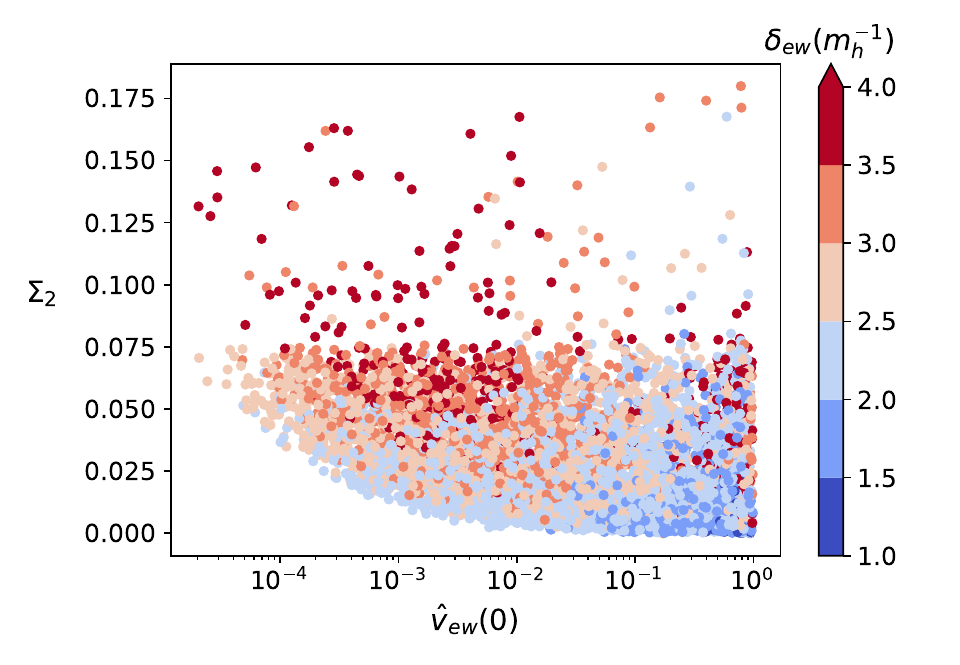}
        \subcaption{} \label{subfig:sigma2collider}
     \end{subfigure}
     \begin{subfigure}[b]{0.49\textwidth}
         \centering
         \includegraphics[width=\textwidth]{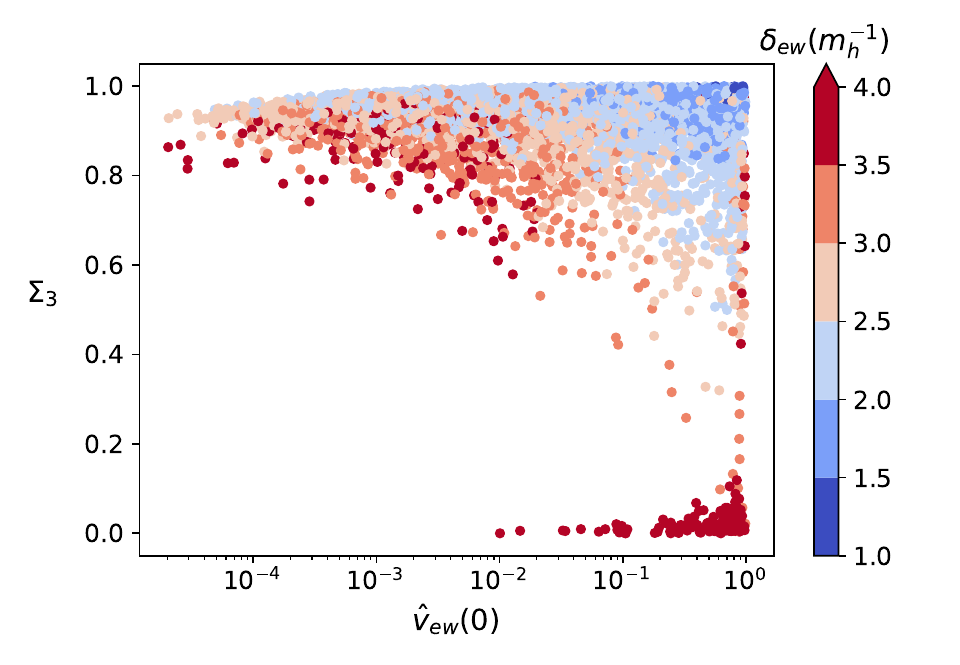}
       \subcaption{}  \label{subfig:sigma3collider}
        \end{subfigure}
\caption{Results of the parameter scan satisfying all theoretical and experimental constraints including collider searches: (a) $\hat{v}_{ew}(0)$ as a function of $v_s$ and the electroweak width; (b), (c) and (d) $\hat{v}_{ew}(0)$ as a function of the singlet admixtures in the CP-even Higgs bosons $\Sigma_1$, $\Sigma_2$ and $\Sigma_3$ respectively. We find that collider constraints impose that the singlet admixture in the SM Higgs boson should be rather small, which leads to higher values for $\hat{v}_{ew}(0)$ than in the previous scan.} 
\label{fig:scenario1collider}
\end{figure}

To demonstrate how collider constraints affect the results, we use \texttt{ScannerS} to generate a set of parameter points in the same range as the one shown in Table \ref{tab:input_parameters_Scan_scenario1}, but where also collider constraints are imposed. The results are shown in Figure \ref{fig:scenario1collider}. We find that imposing collider constraints reduces the maximum singlet admixture $\Sigma_2$ in the SM-like Higgs boson to values lower than $20\%$, which in turn increases the minimal values obtained for $v_{ew}(0)$ by approximately one order of magnitude. For most parameter points of this scan, we find that the heaviest CP-even Higgs boson is allowed to have the highest singlet admixture $\Sigma_3$ (see Figure \ref{subfig:sigma3collider}). For parameter points where the singlet admixture in the lightest CP-even Higgs boson with a mass around 95 GeV is close to one $\Sigma_1 \approx 1$ (see Figure \ref{subfig:sigma1collider}), we find that $v_{ew}(0)$ is large and therefore such a case will not yield electroweak symmetry restoration\footnote{Note that the values for $v_s$ for the parameter points where $\Sigma_1 \approx 1$ varied between 100 GeV and 10 TeV. Therefore, the obtained high values for $v_{ew}(0)$ cannot be explained by a small $v_s$. }.  
\subsection{Scenario 2: Intermediate CP-even masses}\mbox{} 
\begin{table}[H]
\centering
{\renewcommand{\arraystretch}{1.0}
\footnotesize
\begin{tabular}{cccccc}
$m_{h_{a}}$  & $m_{h_{b}}$  & $m_{h_{c}}$ & $m_{A}$ &
$m_{H^{\pm}}$  & $\text{tan}\beta$ \\
\hline
\hline
$125.09$  & $[300,700]$  & $[400,700]$ & $[500, 900]$ &
$[650,900]$  & $[0.5,8]$ \\
\hline
\hline
$C^{2}_{h_{a}t\bar{t}}$ & $C^{2}_{h_{a}VV}$  & $R_{b3}$  & $m_{12}^{2}$ & $v_{S}$ & $type$ \\
\hline
\hline
$[0.6, 1.2]$ & $ [0.6, 1]$ & $[-1,1]$ & $[2 \times 10^4, 2.2 \times 10^5 ]$ & $[100,10000]$ & $1-4$ 
\end{tabular}
}
\caption{\small Set of input parameters for \texttt{ScannerS} scan for \texttt{scenario 2}. The masses and vacuum expectation values are given in $GeV$, while $m^2_{12}$ is given in $GeV^2$. $C_{h_{a}t\bar{t}}$ and $C_{h_{a}VV}$ are defined respectively as the coupling factors of the CP-even Higgs boson $h_a$ to the SM gauge bosons and the top quark and are defined as $C_{h_aVV} = \cos{(\beta)}R_{a1} + \sin{(\beta)}R_{a2} $ and $C_{h_at\bar{t}}=R_{a2}/\sin{(\beta)}$ (see \cite{Muhlleitner:2020wwk}).} 
\label{tab:input_parameters_Scan_scenario2}
\end{table}
\noindent
In this case, we consider the scenario where the masses of the extra CP-even Higgs Bosons $m_{h_2}$ and $m_{h_3}$ are in the range $300 \text{ GeV} < m_{h_2} < 700 \text{ GeV}$ and $400 \text{ GeV} < m_{h_3} < 700 \text{ GeV}$, while $m_{h_1}$ depicts the SM-like Higgs boson. For the singlet VEV $v_s$, the 15000 generated parameter points are again chosen in the range $100 \text{ GeV} < v_s < 10 \text{ TeV}$. All parameter points in this scan also satisfy collider constraints.   

\begin{figure}[h]
     \centering
     \begin{subfigure}[b]{0.49\textwidth}
         \centering
         \includegraphics[width=\textwidth]{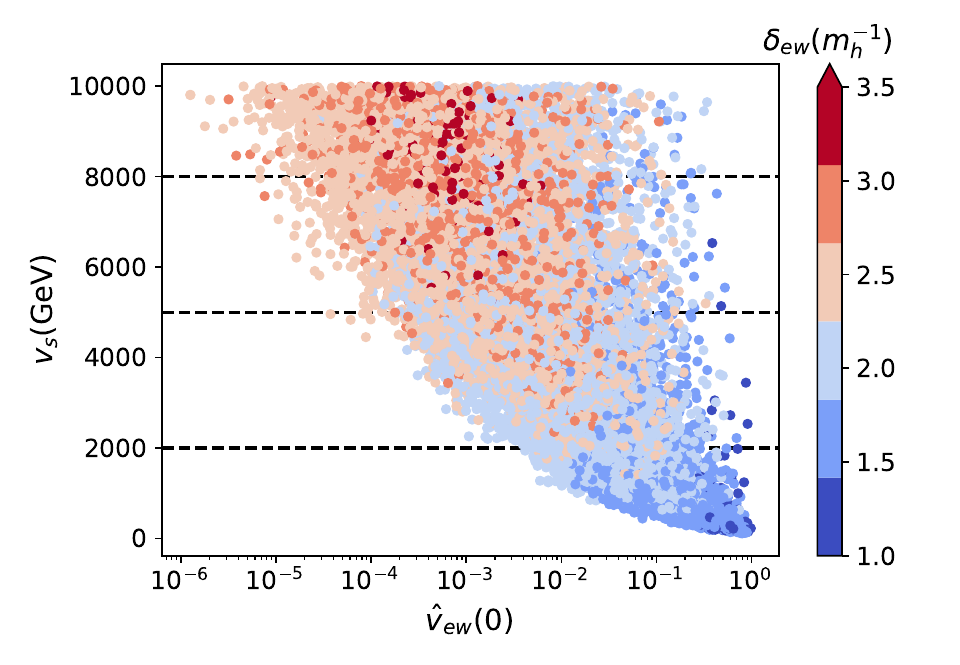}
        \subcaption{} \label{subfig:scenario2}
     \end{subfigure}
     \begin{subfigure}[b]{0.49\textwidth}
         \centering
         \includegraphics[width=\textwidth]{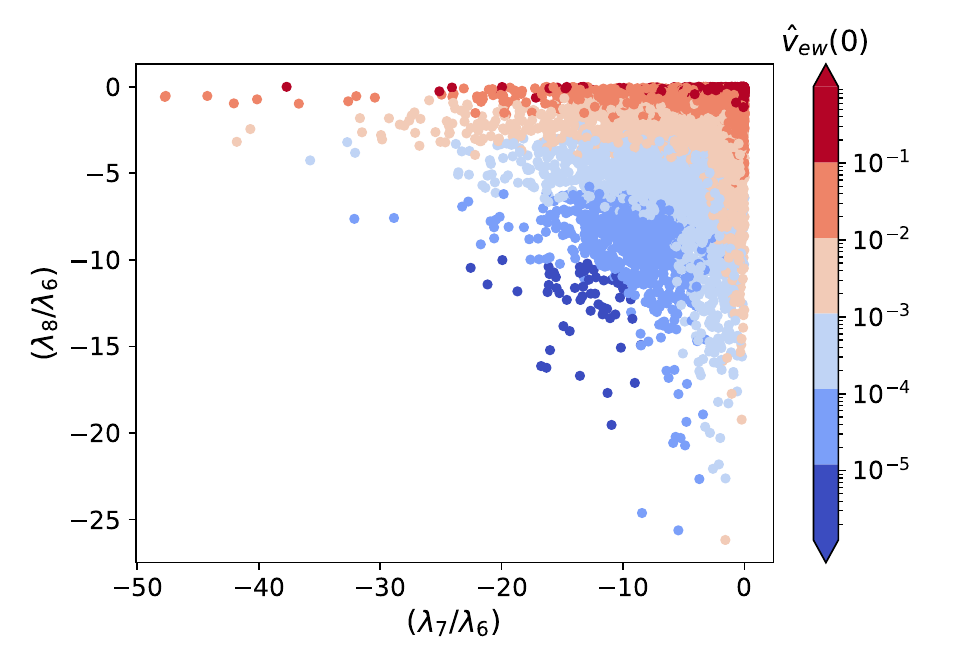}
       \subcaption{}  \label{subfig:scenario2lambdas}
        \end{subfigure}
\caption{Results of the parameter scan for scenario 2. (a) The amount of EWSR inside the domain wall $\hat{v}_{ew}(0)$ as a function of the singlet VEV $v_s$ and the width $\delta_{ew}$ of the symmetry restoration region $v_{ew}(x)$ around the wall. (b) The amount of EWSR inside the wall as a function of the couplings ratios $(\lambda_7/\lambda_6)$ and $(\lambda_8/\lambda_6)$.} 
\end{figure}
\begin{figure}[h]
\centering
\includegraphics[width=0.65\textwidth]{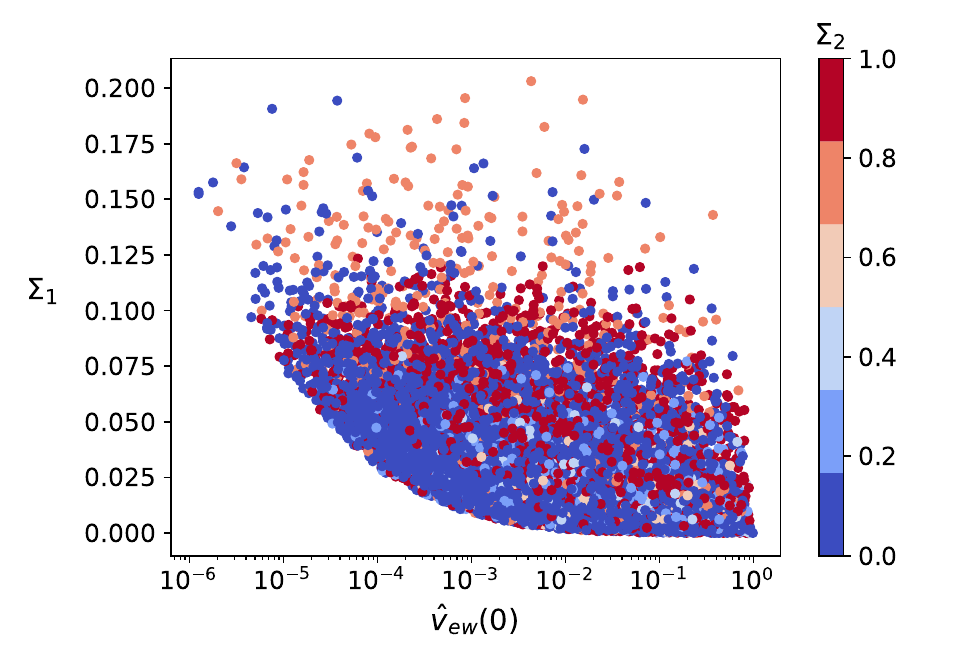}
\caption{Electroweak symmetry restoration measure $\hat{v}_{ew}(0)$ as a function of the singlet admixture in the CP-even Higgs bosons for scenario 2. In this case, the SM Higgs boson is the lightest particle. We find, similar to the previous case, that the smallest $\hat{v}_{ew}(0)$ correlates with higher singlet admixture in the SM Higgs boson state. }
\label{fig:scenario2singletadmixture}
\end{figure}

\noindent
The results of this parameter scan are shown in Figure \ref{subfig:scenario2}. We find that fewer parameter points are leading to large values of $\hat{v}_{ew}(0)$ for large $v_s$ compared to scenario 1 (see Figure \ref{subfig:resultcollider1}). We also find that $\delta_{ew}$ is overall smaller in this scenario. This is due to $\delta^{num}_s$ being smaller for higher masses $m_{h_{2,3}}$.

We show in Figure \ref{subfig:scenario2lambdas} the dependence of the EWSR measure $\hat{v}_{ew}(0)$ on the ratios $\lambda_{7,8}/\lambda_6$. In this case, we observe that the electroweak VEV inside the wall $\hat{v}_{ew}(0)$ decreases as the absolute value of these ratios increases. Again, this is interpreted as a larger change in the effective mass term alongside the wall occurring on a larger region in space. Therefore, the profiles of the doublet Higgs fields have enough space to reach their minimal values inside the wall.
The correlations between the singlet admixtures $\Sigma_i$ in Eq. (\ref{eq:singletadmixture}) and the measures $v_{ew}(0)$ and $\delta_{ew}$ are similar to the previous scenario. We find that EWSR is preferred for a larger singlet admixture in the SM Higgs boson as can be seen in Figure \ref{fig:scenario2singletadmixture}. Note that the collider constraints are fulfilled in the whole range of parameter points.

\subsection{Scenario 3: Heavy CP-even Higgs masses}\mbox{} 
\\
\begin{table}[h]
\centering
{\renewcommand{\arraystretch}{1.0}
\footnotesize
\begin{tabular}{cccccc}
$m_{h_{a}}$  & $m_{h_{b}}$  & $m_{h_{c}}$ & $m_{A}$ &
$m_{H^{\pm}}$  & $\text{tan}\beta$ \\
\hline
\hline
$125.09$  & $[700,1200]$  & $[700,3000]$ & $[500, 1000]$ &
$[650,1200]$  & $[0.5,10]]$ \\
\hline
\hline
$C^{2}_{h_{a}t\bar{t}}$ & $C^{2}_{h_{a}VV}$  & $R_{b3}$  & $m_{12}^{2}$ & $v_{S}$ & $type$ \\
\hline
\hline
$[0.6, 1.2]$ & $ [0.6, 1]$ & $[-1,1]$ & $[5 \times 10^4, 5 \times 10^5 ]$ & $[100,10000]$ & $1-4$ 
\end{tabular}
}
\caption{\small Set of input parameters for \texttt{ScannerS} scan of \texttt{scenario 3}. The masses and vacuum expectation values are given in $GeV$, while $m^2_{12}$ is given in $GeV^2$. $C_{h_{a}t\bar{t}}$ and $C_{h_{a}VV}$ are defined respectively as the coupling factors of the CP-even Higgs boson $h_a$ to the SM gauge bosons and the top quark and are defined as $C_{h_aVV} = \cos{(\beta)}R_{a1} + \sin{(\beta)}R_{a2} $ and $C_{h_at\bar{t}}=R_{a2}/\sin{(\beta)}$ (see \cite{Muhlleitner:2020wwk}).} 
\label{tab:input_parameters_Scan_scenario3}
\end{table}
In this scenario, we investigate the case where the extra CP-even Higgs bosons can be very heavy. We fix the SM Higgs to be $m_{h_1} = 125.09 \text{ GeV}$ and vary the heavier masses $700 \text{ GeV} < m_{h_2} < 3000 \text{ GeV}$ and $700 \text{ GeV} < m_{h_3} < 3000 \text{ GeV}$. The value for the singlet VEV $v_s$ varies again between 100 GeV and 10 TeV. The other parameter ranges are shown in Table \ref{tab:input_parameters_Scan_scenario3}.

\begin{figure}[h]
     \centering
     \begin{subfigure}[b]{0.49\textwidth}
         \centering
         \includegraphics[width=\textwidth]{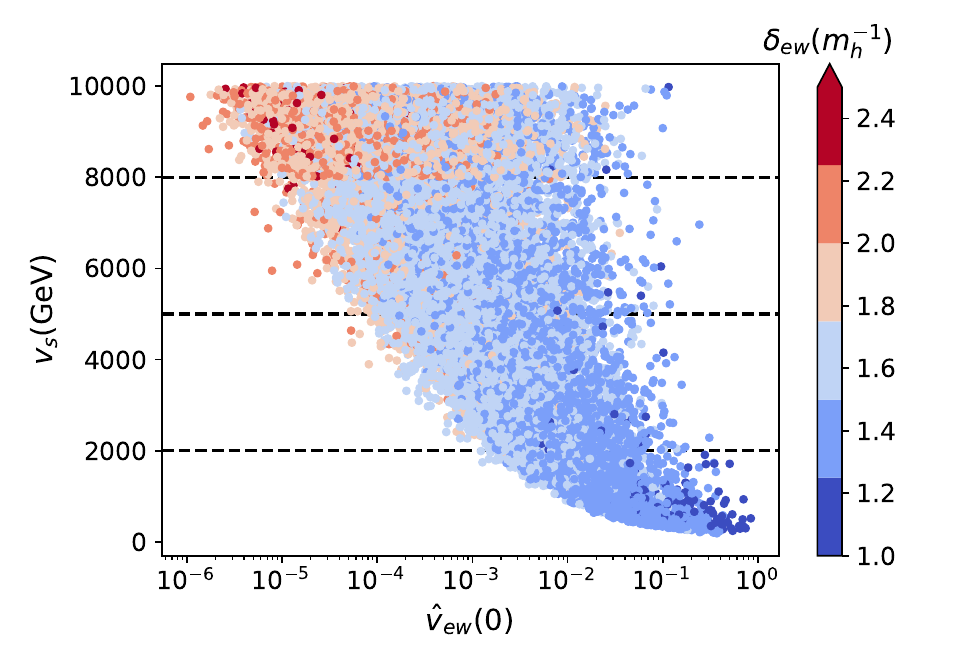}
        \subcaption{} \label{subfig:scenario3}
     \end{subfigure}
     \begin{subfigure}[b]{0.49\textwidth}
         \centering
         \includegraphics[width=\textwidth]{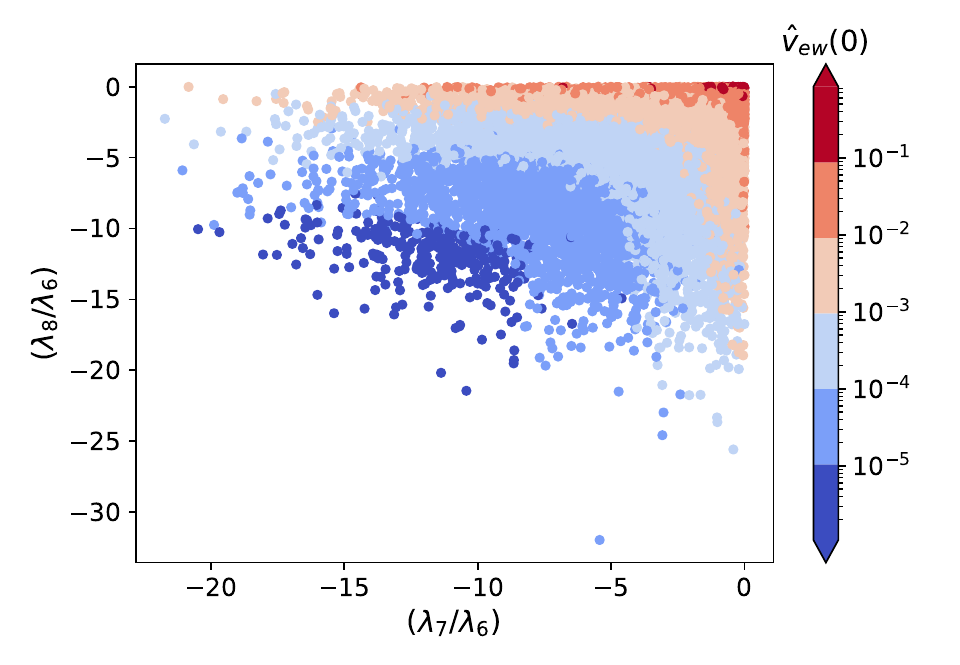}
       \subcaption{}  \label{subfig:scenario3lambdas}
        \end{subfigure}
\caption{Results of the parameter scan for scenario 3. (a) The amount of EWSR inside the domain wall $\hat{v}_{ew}(0)$ as a function of the singlet VEV $v_s$ and the width $\delta_{ew}$ of the symmetry restoration region $v_{ew}(x)$ around the wall. We find that this scenario leads to smaller minimal values for $v_{ew}(0)$ than the previous scenario. (b) The amount of EWSR inside the wall as a function of the ratios $(\lambda_7/\lambda_6)$ and $(\lambda_8/\lambda_6)$. We find that the lowest values for $v_{ew}(0)$ are obtained when both ratios are large and negative. } 
\end{figure}

\begin{figure}[H]
     \centering
     \begin{subfigure}[b]{0.49\textwidth}
         \centering
         \includegraphics[width=\textwidth]{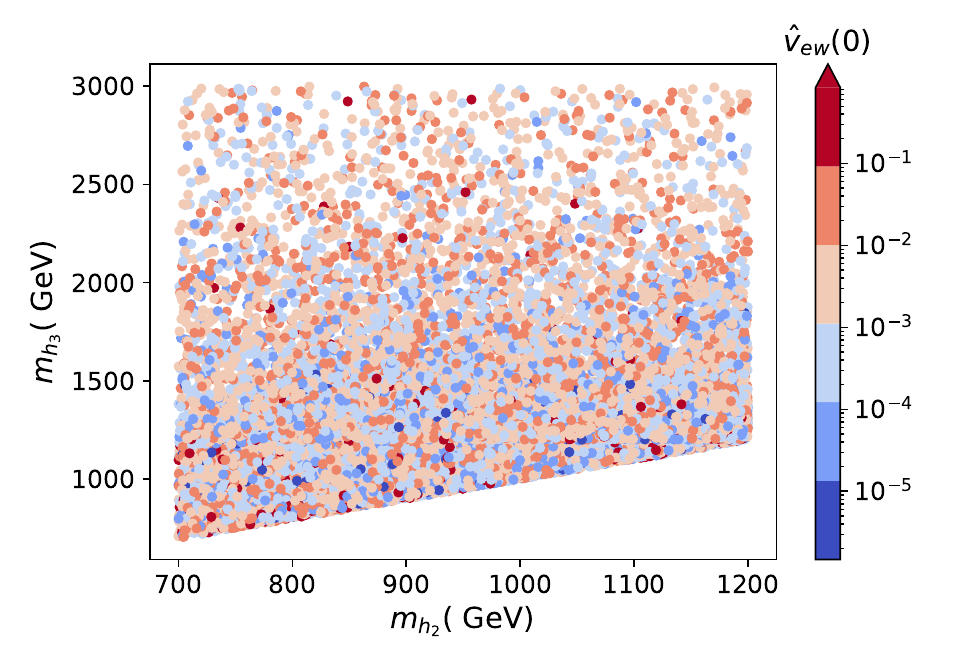}
        \subcaption{} \label{subfig:scenario3ew}
     \end{subfigure}
     \begin{subfigure}[b]{0.49\textwidth}
         \centering
         \includegraphics[width=\textwidth]{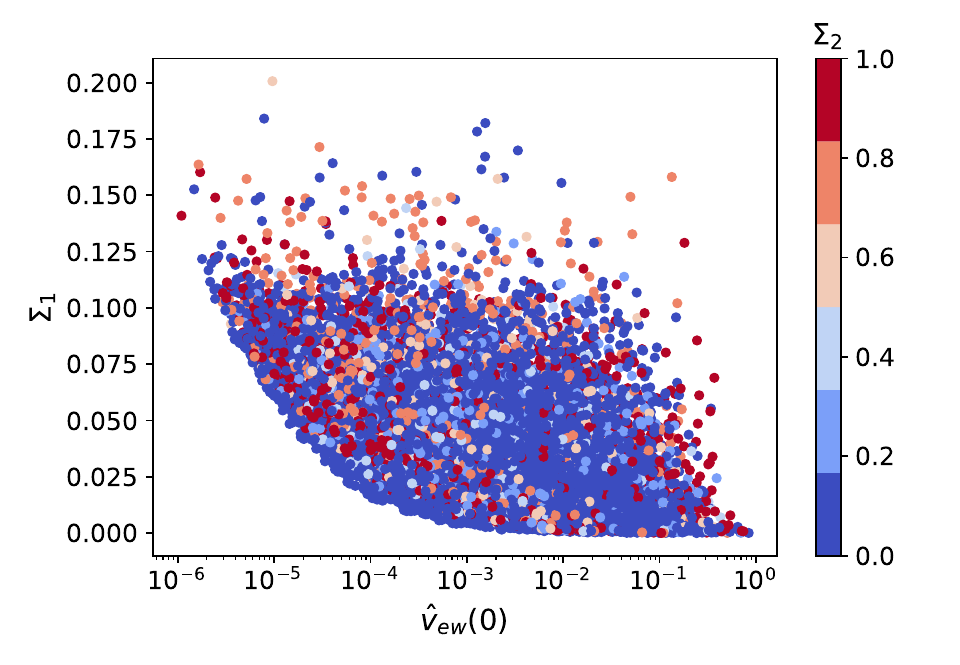}
       \subcaption{}  \label{subfig:scenario3singletadmixture}
        \end{subfigure}
\caption{Results of parameter scan for scenario 3. (a) The amount of EWSR inside the domain wall $\hat{v}_{ew}(0)$ as a function of the masses $m_{h_2}$ and $m_{h_3}$ (b) The amount of EWSR inside the wall as a function of the singlet admixtures $\Sigma_1$ related to the SM-like Higgs boson $h_1$ and $\Sigma_2$ related to $h_2$.  } 
\end{figure}

The results for the electroweak VEV inside the wall (see Figure \ref{subfig:scenario3}) are overall similar to the previous case, i.e. smaller $v_{ew}(0)$ inside the wall correlate with a higher singlet VEV $v_s$. The major difference can be seen in the decrease in the value of the width $\delta_{ew}$ due to the increase in the masses of the CP-even Higgses leading to smaller $\delta^{num}_s$ and therefore, to an overall smaller $\delta_{ew}$. We do not find a correlation between the values of the masses and the electroweak symmetry restoration measure $\hat{v}_{ew}(0)$ (see Figure \ref{subfig:scenario3ew}). The correlations are, however, dependent on the mixing angles and therefore the singlet admixtures $\Sigma_i$. These correlations are shown in Figure \ref{subfig:scenario3singletadmixture}, where the smallest values for $\hat{v}_{ew}(0)$ are obtained, again, for a higher singlet admixture $\Sigma_1$ in the SM-like Higgs boson state.

\begin{figure}[H]
     \centering
     \begin{subfigure}[b]{0.49\textwidth}
         \centering
         \includegraphics[width=\textwidth]{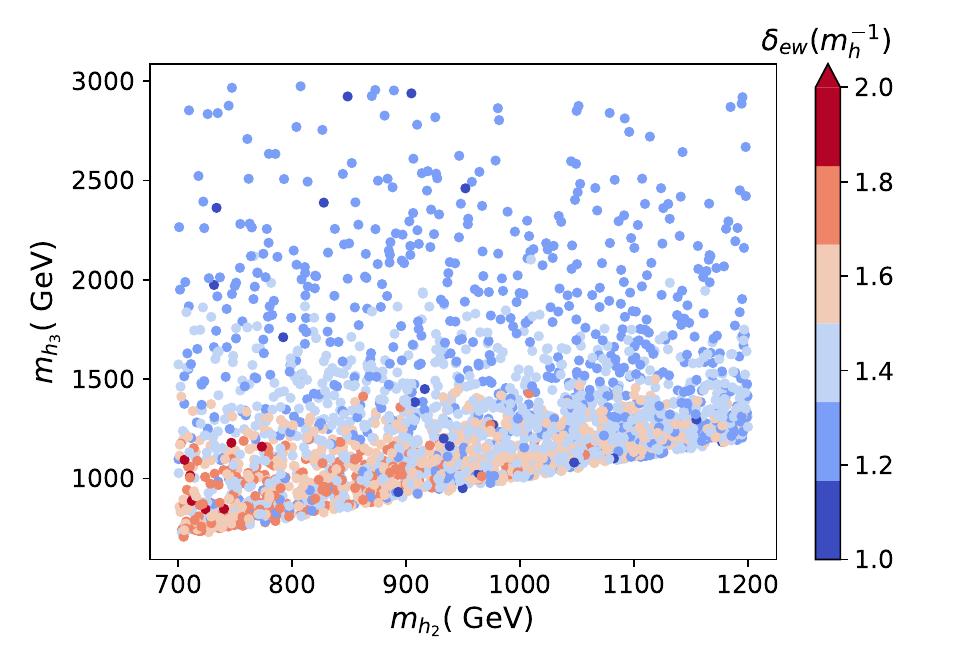}
        \subcaption{$ 100 \text{ GeV} <v_s < 2 \text{ TeV}$} \label{subfig:region1}
     \end{subfigure}
     \begin{subfigure}[b]{0.49\textwidth}
         \centering
         \includegraphics[width=\textwidth]{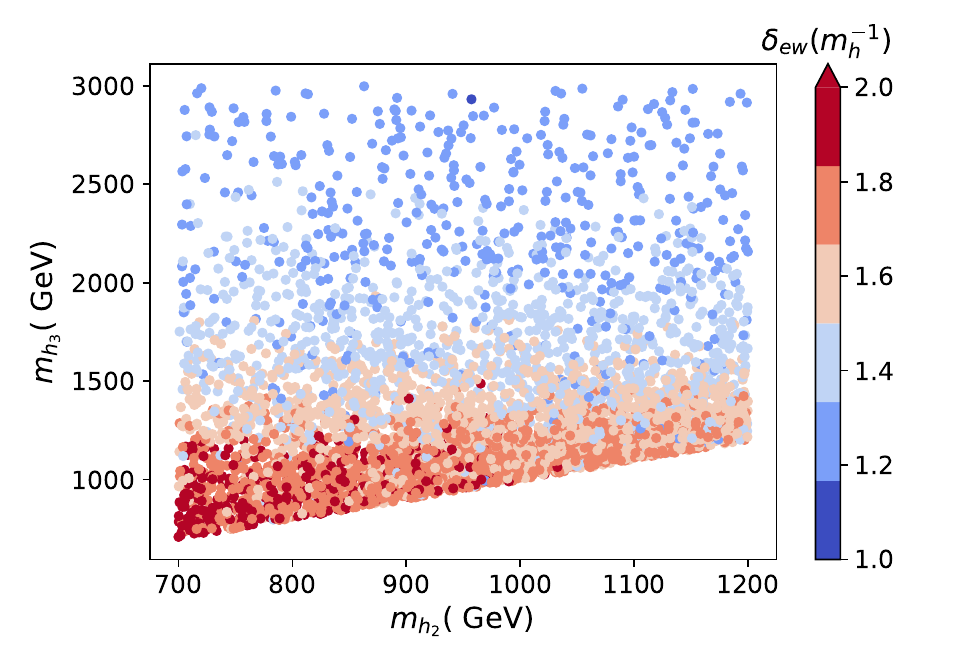}
       \subcaption{$ 2 \text{ TeV} <v_s < 5 \text{ TeV}$}  \label{subfig:region2}
        \end{subfigure}
        \begin{subfigure}[b]{0.49\textwidth}
         \centering
         \includegraphics[width=\textwidth]{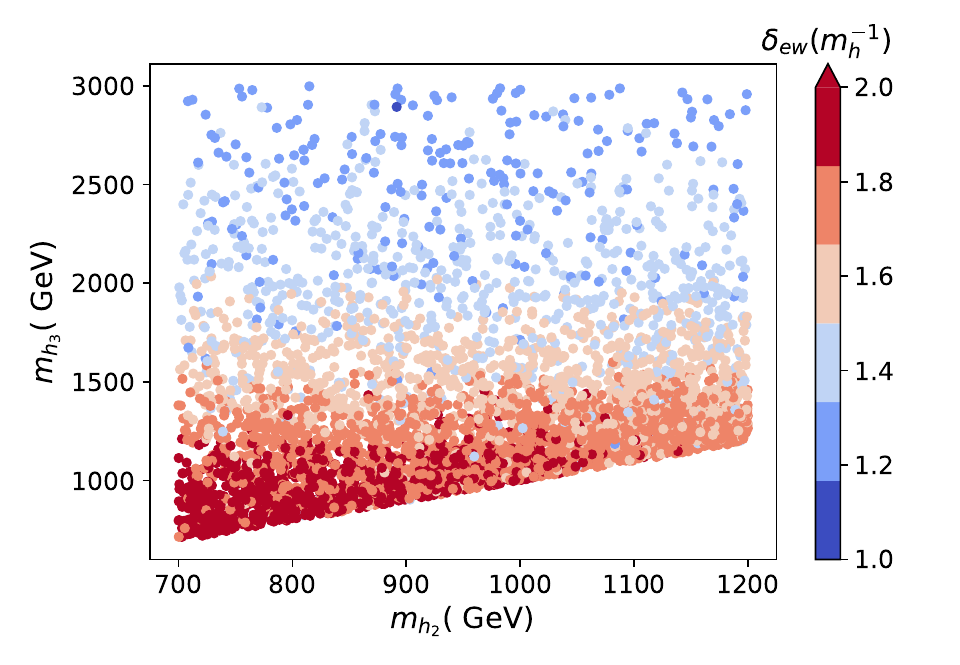}
        \subcaption{$ 5 \text{ TeV} <v_s < 8 \text{ TeV}$} \label{subfig:region3}
     \end{subfigure}
     \begin{subfigure}[b]{0.49\textwidth}
         \centering
         \includegraphics[width=\textwidth]{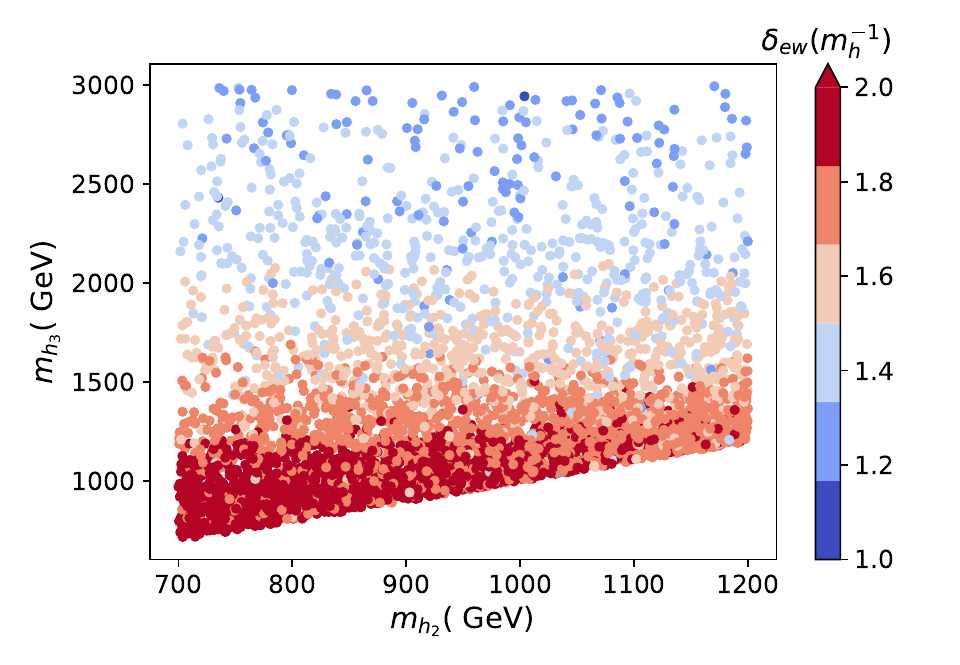}
       \subcaption{$ 8 \text{ TeV} <v_s < 10 \text{ TeV}$}  \label{subfig:region4}
        \end{subfigure}
\caption{Width $\delta_{ew}$ of the Higgs doublet variation inside the wall for different ranges of $v_s$. We find that the behavior of $\delta_{ew}$ is independent of the range of $v_s$ and is mostly determined by the mass of $m_{h_3}$.} 
\label{fig:scenario3differentrange}
\end{figure}

Concerning the width $\delta_{ew}$, we plot the results that we get for different ranges of $v_s$ as shown in Figure \ref{fig:scenario3differentrange}. We find that $\delta_{ew}$ is mostly independent of the range of $v_s$. The width is, however, largely dependent on the mass $m_{h_3}$ and we obtain the largest values of $\delta_{ew}$ for smaller $m_{h_3}$. 

As a summary of these three scenarios, we found that the singlet vev $v_s$ is the most important parameter in determining the lowest values for the electroweak symmetry restoration measure $\hat{v}_{ew}(0)$ and that higher $v_s$ leads to smaller $\hat{v}_{ew}(0)$. We found that the masses of the CP-even Higgs bosons influence the width $\delta_{ew}$ and that smaller masses lead, overall, to a higher $\delta_{ew}$. The singlet admixture in the CP-even Higgs states also plays a major role: a higher singlet admixture in the SM Higgs boson leads to the smallest values for $\hat{v}_{ew}(0)$. This correlation obviously puts rather strong experimental constraints on the feasibility of inducing electroweak symmetry restoration inside the singlet domain wall in the N2HDM.
\subsection{Scenarios with fixed $\mathbf{v_s}$}\mbox{} 
\begin{figure}[h]
     \centering
     \begin{subfigure}[b]{0.49\textwidth}
         \centering
         \includegraphics[width=\textwidth]{PhenoScenarios/mh2mh3vevscn4coolider.pdf}
        \subcaption{$\hat{v}_{ew}(0)$ for $v_s = 800 \text{ GeV}$} \label{subfig:scenario4vev}
     \end{subfigure}
     \begin{subfigure}[b]{0.49\textwidth}
         \centering
         \includegraphics[width=\textwidth]{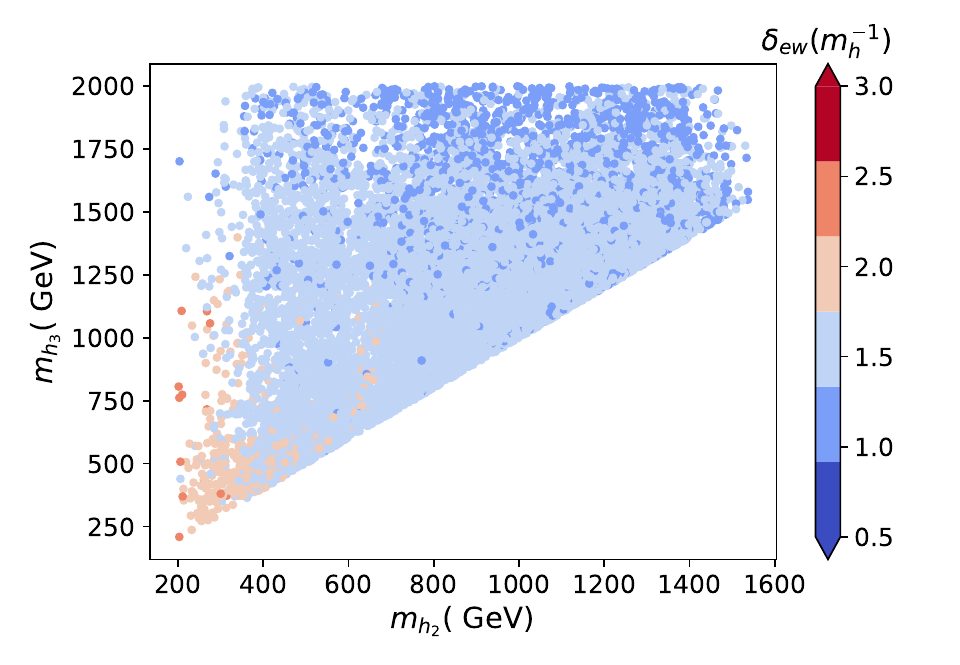}
       \subcaption{$\delta_{ew}$ for $v_s = 800 \text{ GeV}$}  \label{subfig:scenario4width}
        \end{subfigure}
     \centering
     \begin{subfigure}[b]{0.49\textwidth}
         \centering
         \includegraphics[width=\textwidth]{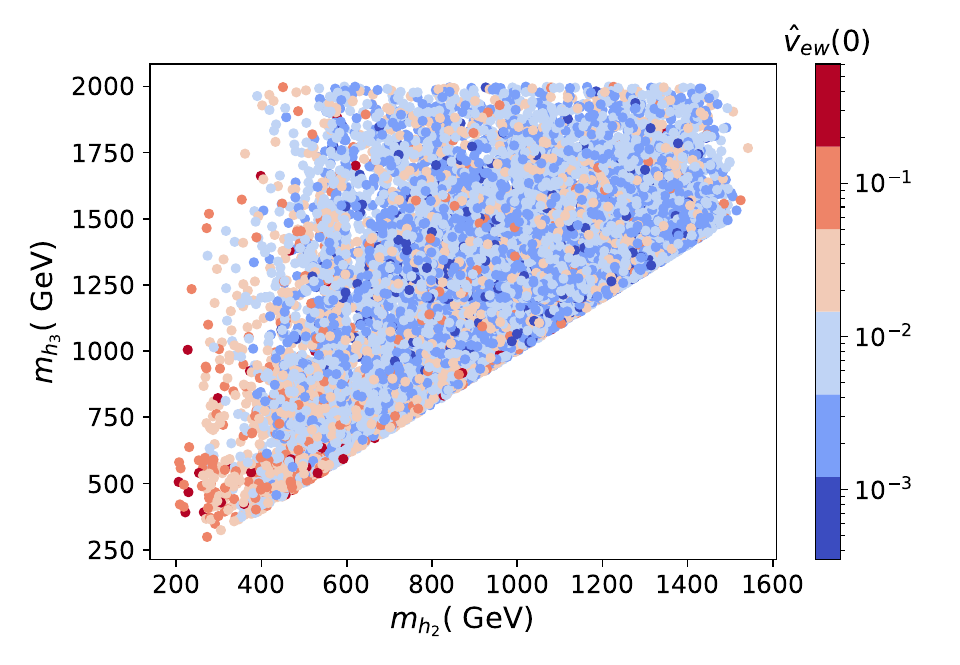}
        \subcaption{$\hat{v}_{ew}(0)$ for $v_s = 2500 \text{ GeV}$} \label{subfig:scenario5vev}
     \end{subfigure}
     \begin{subfigure}[b]{0.49\textwidth}
         \centering
         \includegraphics[width=\textwidth]{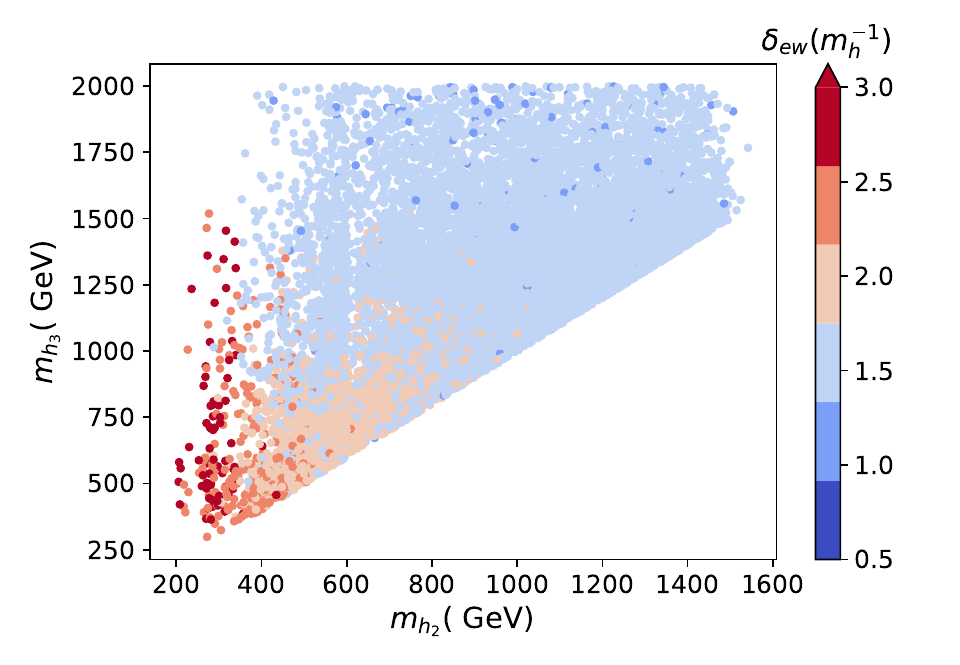}
       \subcaption{$\delta_{ew}$ for $v_s = 2500 \text{ GeV}$}  \label{subfig:scenario5width}
        \end{subfigure}
     \centering
     \begin{subfigure}[b]{0.49\textwidth}
         \centering
         \includegraphics[width=\textwidth]{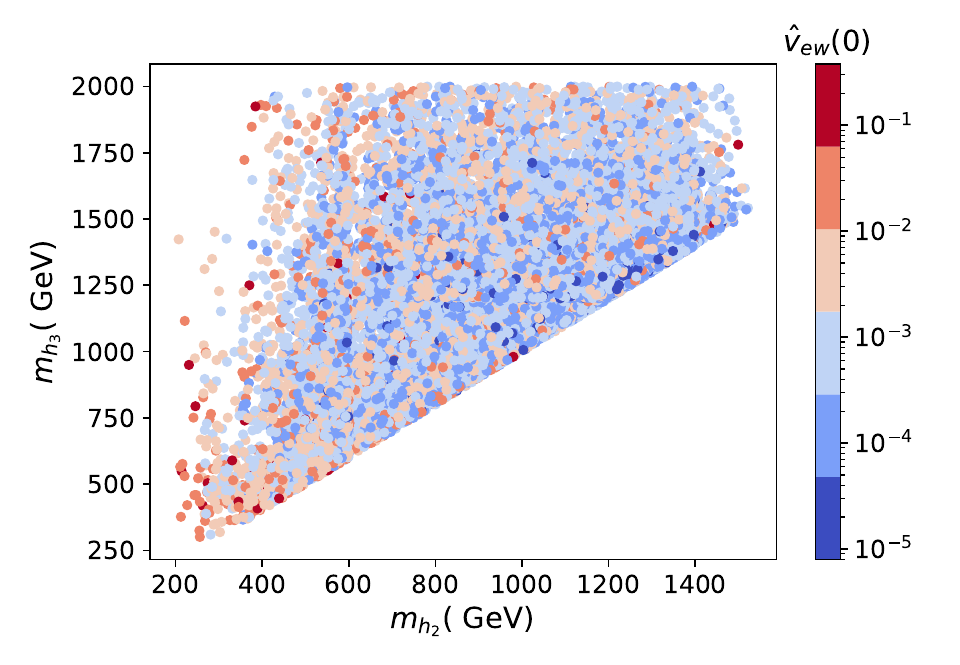}
        \subcaption{$\hat{v}_{ew}(0)$ for $v_s = 6000 \text{ GeV}$} \label{subfig:scenario6vev}
     \end{subfigure}
     \begin{subfigure}[b]{0.49\textwidth}
         \centering
         \includegraphics[width=\textwidth]{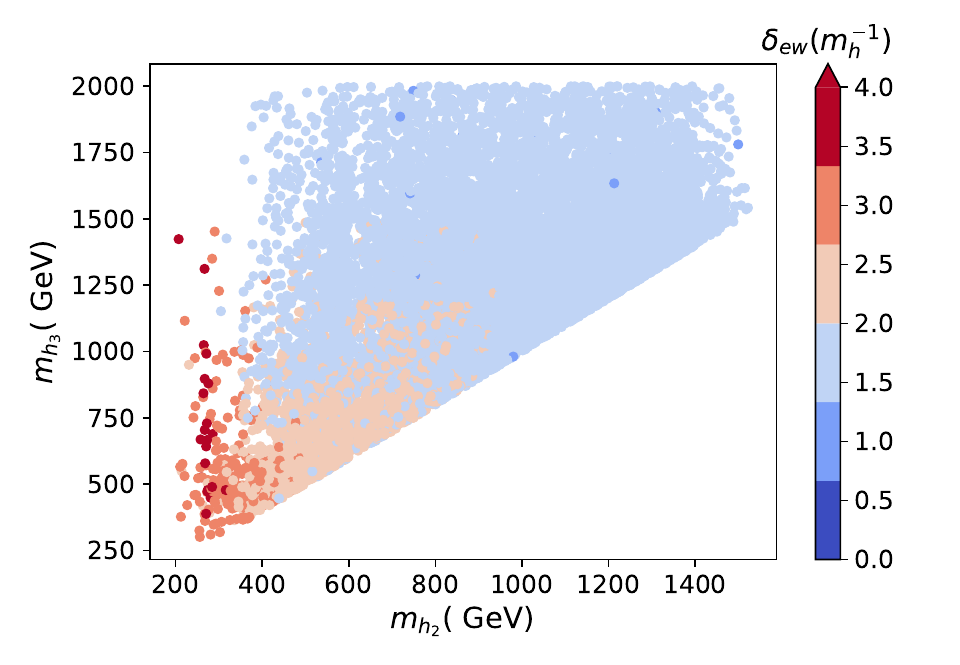}
       \subcaption{$\delta_{ew}$ for $v_s = 6000 \text{ GeV}$}  \label{subfig:scenario6width}
        \end{subfigure}
\caption{$\hat{v}_{ew}(0)$ and $\delta_{ew}$ for different fixed values for $v_s$ (800, 2500 and 6000 GeV) as a function of the masses $m_{h_2}$ and $m_{h_3}$ and varying mixing angles. For all these scenarios, the SM-like Higgs boson $h_1$ has the mass $m_{h_1} = 125.09 \text{ GeV}$. The generated parameter points satisfy all theoretical and experimental constraints, including collider constraints.} 
\label{fig:fixedvs}
\end{figure}
\begin{table}[t]
\centering
{\renewcommand{\arraystretch}{1.0}
\footnotesize
\begin{tabular}{cccccc}
 $v_{S} \text{(GeV)}$ & $m_{h_{a}} \text{(GeV)}$  & $m_{h_{b}} \text{(GeV)}$  & $m_{h_{c}} \text{(GeV)}$  & $\text{tan}\beta$ & $m_{12}^{2} \text{(GeV²)}$ \\
\hline
\hline
800 & $ 125.09 $ & $200-1500$ & $ 200-2000$ & $0.5-10$ & $0-10^6$  \\
\hline
2500 & $ 125.09 $ & $200-1500$ & $ 200-2000$ & $0.5-8$ & $0-10^6$\\ 
\hline
6000 & $ 125.09 $ & $200-1500$ & $ 200-2000$ & $0.5-7$ & $0-10^6$ \\ 
\end{tabular}
}
\caption{\small Range of the parameter points generated by \texttt{ScannerS} for the different scenarios. For the variables $C^2_{h_at\bar{t}}$, $C^2_{h_aVV}$ and $R_{b3}$ relevant for determining the mixing angles, we use the same range as in the previous scans in Tables \ref{tab:input_parameters_Scan_scenario1}, \ref{tab:input_parameters_Scan_scenario2} and \ref{tab:input_parameters_Scan_scenario3}.} 
\label{tab:input_parameters_Scan_scenario456}
\end{table}
In contrast to the previous scenarios where the singlet VEV $v_s$ was the primary variable determining the amount of EWSR inside the wall for different mass hierarchies, we focus here on the effects of varying the CP-even Higgs masses $h_2$ and $h_3$ as well as their mixing angles while fixing $v_s$ and the mass of $h_1$ to be the SM-like Higgs boson $m_{h_1} = 125.09 \text{ GeV}$ (see Table \ref{tab:input_parameters_Scan_scenario456}).

We show the results of the scans for different values of $v_s$ (800, 2500 and 6000 GeV) in Figure \ref{fig:fixedvs}.
In terms of $\hat{v}_{ew}(0)$ and for the three different values of $v_s$, we observe that smaller $\hat{v}_{ew}(0)$ are obtained for mostly big and intermediate values of the masses $m_{h_2}$ and $m_{h_3}$. However, those ranges of masses do not guarantee the possibility of having EWSR inside the wall as they can also lead to high values for $\hat{v}_{ew}(0)$. We observe a large dependence of the minimal obtained values for $\hat{v}_{ew}$ on $v_s$. As for the width $\delta_{ew}$, we find a strong correlation between the masses and $\delta_{ew}$. We observe that smaller masses lead, in general, to larger values as can be seen in Figures \ref{subfig:scenario4width}, \ref{subfig:scenario5width}, \ref{subfig:scenario6width}. These results are in good agreement with the general behavior found in the previous scans of scenarios 1, 2 and 3.   

For the case when $v_s = 800 \text{ GeV}$, we obtain larger values for $\hat{v}_{ew}(0)$ compared to the scenarios with higher $v_s$. This is because the generated parameter points for low $v_s$ (satisfying all theoretical and experimental constraints) lead mostly to small values for $\lambda_{7,8}/\lambda_6$. In order to further study this scenario of low $v_s$, we look for parameter points where the chosen masses and mixing angles lead to large negative $\lambda_{7,8}/\lambda_6$. We generated parameter points where we impose a limit on $\lambda_{7,8}/\lambda_6 < -8$ to obtain rather small $v_{ew}(0)$. We found that most parameter points of this scan have a large singlet admixture $\Sigma_1$ in the SM Higgs boson state $m_{h_1} = 125.09 \text{ GeV}$. However, this set of parameter points is experimentally ruled out and is even incompatible with the constraint of perturbative unitarity. We therefore can conclude that, at zero or low temperatures, achieving electroweak symmetry restoration using domain walls with rather low $v_s$ is ruled out.

%% file: DoubleDWSol.tex
\section{Different Goldstone modes and CP-violation in the vicinity of the wall}\label{section5}

Until now we focused on the trivial case when the vacua for the Higgs doublets have the same Goldstone modes on both domains. In a realistic electroweak phase transition, however, one expects causally disconnected domains of the universe to end up in vacua with different values for the Goldstone modes (\ref{eq:vacuumform}) given that they lead to degenerate minima of the potential. Recall that the VEVs of the Higgs doublets can be written in the general form (\ref{eq:vacuumform}):
\begin{align}
   \langle \Phi_1 \rangle = \text{U} \langle \tilde{\Phi}_1 \rangle = \text{U} \dfrac{1}{\sqrt{2}}
    \begin{pmatrix}
          0 \\      v_1
     \end{pmatrix},      
&& \langle \Phi_2 \rangle = \text{U} \langle \tilde{\Phi}_2 \rangle = \text{U} \dfrac{1}{\sqrt{2}}
      \begin{pmatrix}
     v_+ \\
     v_2e^{i\xi}
      \end{pmatrix} , && \text{U} = e^{i\theta} \text{exp}\biggl(i\dfrac{\tilde{g}_i\sigma_i}{2v_{sm}}\biggl),
\end{align}
where U is an element of the $\text{SU(2)}_L\times\text{U(1)}_Y$ symmetry group. The possibility of having different Goldstone modes ($\theta, \tilde{g}_i$) on different domains was found to have profound consequences on the solutions related to the $Z_2$-symmetry domain walls in the 2HDM \cite{Law:2021ing,Viatic:2020yme, Sassi:2023cqp}. In general, one obtains several classes of domain wall solutions with different properties such as CP-violating or electric charge-breaking condensates localized inside or in the vicinity of the wall. 

We now study these effects in the N2HDM. We consider two different cases of electroweak symmetry breaking. The first case is the breaking of the electroweak symmetry at the same time as the breaking of $Z'_2$. This is a one-step phase transition according to: $$ (0,0,0) \rightarrow (v_1,v_2, \pm v_s). $$ In such a case the two domains related by the $Z'_2$ symmetry will also have different Goldstone modes. The second case is when the electroweak and the $Z'_2$ symmetries are broken at different times:
$$(0,0,0) \rightarrow (0,0,\pm v'_s) \rightarrow (v_1,v_2, \pm v_s), $$ where $v_s$ and $v'_s$ can be equal or have different values. For this case, we assume that the $Z'_2$ symmetry is spontaneously broken before the electroweak symmetry in order to form the domain walls that will modify the doublet VEVs. Therefore, a domain with a given sign of $v_s$ can include several smaller domains where the doublet VEVs have different Goldstone modes. 

We start with the first case. This scenario requires that both the singlet and the doublets acquire their vacuum expectation values at the same time in the early universe. Checking whether such a one-step phase transition is the correct evolution in the early universe would require a substantial finite-temperature numerical analysis for every considered parameter point and is beyond the scope of the current analysis. Therefore, we assume for simplicity and pedagogical reasons that this requirement is fulfilled and restrict ourselves to the discussion of the extra domain walls properties that can occur in such a case. We postpone a complete discussion of this requirement for a future comprehensive work discussing the electroweak baryogenesis generated by the domain walls in the N2HDM. \\

\begin{figure}[h]
     \centering
     \begin{subfigure}[b]{0.49\textwidth}
         \centering
         \includegraphics[width=\textwidth]{Goldstones/ThetaSol2.pdf}
        \subcaption{} \label{subfig:thetasol}
     \end{subfigure}
     \begin{subfigure}[b]{0.5\textwidth}
         \centering
         \includegraphics[width=\textwidth]{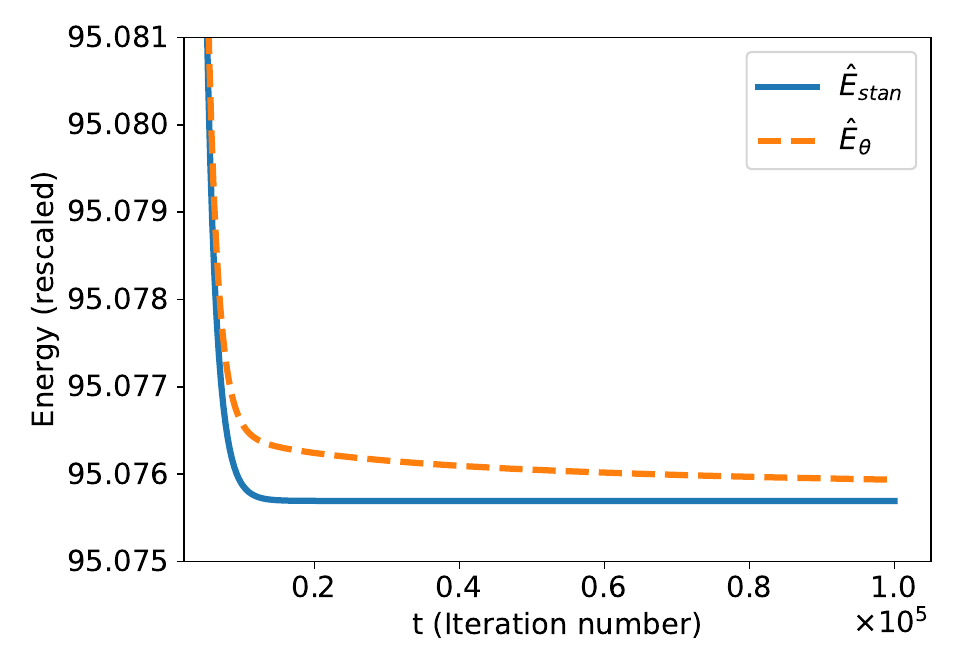}
       \subcaption{}  \label{subfig:energytheta}
        \end{subfigure}
\caption{(a) Rescaled profile of the fields $\hat{v}_i = v_i/v_{sm}$ for a domain wall solution with different Goldstone modes $\theta$ on both domains. (b) The domain wall energy of the standard solution (blue) and the CP-violating solution (dashed orange). The used parameter point corresponds to the variables in Table \ref{tab:thetapp1}} 
\label{fig:thetaresult}
\end{figure}

We discuss the domain wall solution in the case when the Goldstone mode $\theta$ related to the $U(1)_Y$ symmetry is different on both domains. In order to get the domain wall solution of this scenario, we need to find the solution that minimizes the energy functional $\mathcal{E_{\theta}}(x)$:
\begin{align}
    \notag & \mathcal{E_{\theta}}(x) = \dfrac{1}{2}\biggl(\dfrac{dv_1}{dx}\biggr)^2 + \dfrac{1}{2}\biggl(\dfrac{dv_2}{dx}\biggr)^2 + \dfrac{1}{2}\biggl(\dfrac{dv_+}{dx}\biggr)^2 + \dfrac{1}{2}v^2_2(x)\biggl(\dfrac{d\xi}{dx}\biggr)^2  +
  \dfrac{1}{2}v^2_1(x)\biggl(\dfrac{d\theta}{dx}\biggr)^2  \\ & + \dfrac{1}{2}v^2_2(x)\biggl[  \biggl(\dfrac{d\theta}{dx}\biggr)^2 + 2\dfrac{d\theta}{dx}\dfrac{d\xi}{dx} \biggr]  + \dfrac{1}{2}v^2_+(x)  \biggl(\dfrac{d\theta}{dx}\biggr)^2  +  V_{N2HDM}(x). 
\label{eq:energytheta}  
\end{align}
The gradient flow equations of motion of the domain wall solution are given by:
\begin{align}
   & \dfrac{\dd v_1}{\dd t} = \dfrac{\dd^2 v_1}{\dd x^2} - \dfrac{ \dd \mathcal{E_{\theta}}}{\dd v_1}  , \label{eq:eomv1} \\
   & \dfrac{\dd v_2}{\dd t} = \dfrac{\dd^2 v_2}{\dd x^2} - \dfrac{ \dd \mathcal{E_{\theta}}}{\dd v_2} ,\\
    & \dfrac{\dd v_s}{\dd t} = \dfrac{\dd^2 v_s}{\dd x^2} - \dfrac{ \dd \mathcal{E_{\theta}}}{\dd v_s}  ,\\
   & \dfrac{\dd \xi}{\dd t} = \dfrac{\dd \mathcal{E_{\theta}}}{ \dd ( \dd \xi/ \dd x)} - \dfrac{ \dd \mathcal{E_{\theta}}}{\dd \xi},\\
    & \dfrac{\dd \theta}{\dd t} = \dfrac{\dd \mathcal{E_{\theta}}}{ \dd ( \dd \theta/ \dd x)}.
\label{eq:eomtheta}    
\end{align}
We numerically solve this system of differential equations using a gradient flow algorithm \cite{Law:2021ing, Viatic:2020yme} and take the boundary conditions for the Goldstone mode $\theta$ to be $0$ at $-\infty$ and $\pi/2$ at $+\infty$ using von Neumann boundary conditions. The chosen parameter point is given in table \ref{tab:thetapp1}
\begin{table}
    \centering
    \begin{tabular}{ccccc}
        $m_{h_1}\text{ (GeV)}$ & $m_{h_2}\text{ (GeV)}$  & $m_{h_3}\text{ (GeV)}$ & $v_s \text{ (GeV)}$ & $\tan(\beta)$ \\
         \hline
         \hline
        125.09 & 483.50 & 567.65 & 1340 & 3.14 \\
        \hline
        \hline
        $m^2_{12}\text{ (GeV)}^2$ & $\alpha_1$ & $\alpha_2$ & $\alpha_3$ & type \\
        \hline
        \hline
         65316 & 1.29 & 0.51 & 0.33 & 1 \\
    \end{tabular}
    \caption{Parameter point used to calculate the CP-violating solution in Figure \ref{fig:thetaresult}.}
    \label{tab:thetapp1}
\end{table}
and the results are shown in Figure \ref{subfig:thetasol}. We find that, in the vicinity of the wall, $\xi(x)$ is non-zero leading to a non-zero imaginary mass in the Yukawa sector and therefore to CP-violating phenomena \cite{Sassi:2023cqp}. This condensate vanishes in the core of the wall given that $v_2(0) = 0$ for this parameter point. We also obtain a kink-like profile for $\theta(x)$ interpolating between 0 and $\pi/2$. 
The profile of $\xi(x)$ can be explained using the equation of motion for the Goldstone mode $\theta$. 
It was found in \cite{Law:2021ing} that a change in $\theta$ across the wall will also induce a change in $\xi(x)$ according to the formula:
\begin{equation}
    \dfrac{\dd \theta}{\dd x} = \dfrac{-v^2_2(x)}{v^2_1(x) + v^2_2(x) + v^2_+(x)}\dfrac{\dd \xi}{\dd x}.
\end{equation}
Accordingly, a change in $\theta(x)$ in the vicinity of the wall will lead to a change in $\xi(x)$ at the same point of space. 
To determine whether this solution is stable or not, we compute its energy $\sigma_{\theta} = \int \dd x \, \, \mathcal{E}_{\theta}(x)$ and compare it to the energy of the standard solution as shown in Figure \ref{subfig:energytheta}. Numerically, we find that the CP-violating solution has a slightly higher energy causing such a solution to be unstable and to decay to the standard domain wall solution with $\xi(x) = 0$ everywhere. This decay process occurs due to $\theta(x)$ varying with time in such a way as to make both domains have the same Goldstone mode values ($\theta(-\infty) = \theta(+\infty)$).

In the case of electroweak symmetry restoration in a large region around the wall, we found that the decay of this CP-violating domain wall solution takes a longer iteration time. Looking at $\mathcal{E}_{\theta}(x)$ (\ref{eq:energytheta}), we see that all terms with a $\theta$ and $\xi$ contributions are dependent on $v_1(x)$ and $v_2(x)$. In the case of a total electroweak symmetry restoration inside the wall, these terms vanish and therefore a CP-violating solution has almost the same energy as the stable standard domain wall field configuration, leading the CP-violating vacua around the wall to be long-lived. However, the imaginary mass proportional to $\text{Im}(v_2(x)e^{i\xi(x)})=\sin (\xi(x))v_{2}(x)$ providing the CP-violation effects for fermions will, in such a case, be small.
\begin{table}
    \centering
    \begin{tabular}{ccccc}
        $m_{h_1}\text{ (GeV)}$ & $m_{h_2}\text{ (GeV)}$  & $m_{h_3}\text{ (GeV)}$ & $v_s \text{ (GeV)}$ & $\tan(\beta)$ \\
         \hline
         \hline
        125.09 & 589.5 & 697.5 & 9635 & 1.28 \\
        \hline
        \hline
        $m^2_{12}\text{ (GeV)}^2$ & $\alpha_1$ & $\alpha_2$ & $\alpha_3$ & type \\
        \hline
        \hline
         208249 & 0.94 & 0.25 & -1.37 & 4 \\
    \end{tabular}
    \caption{Parameter point used to calculate the CP-violating solution in Figure \ref{fig:asythetaresult}.}
    \label{tab:thetapp2}
\end{table}

\begin{figure}[h]
     \centering
     \begin{subfigure}[b]{0.49\textwidth}
         \centering
         \includegraphics[width=\textwidth]{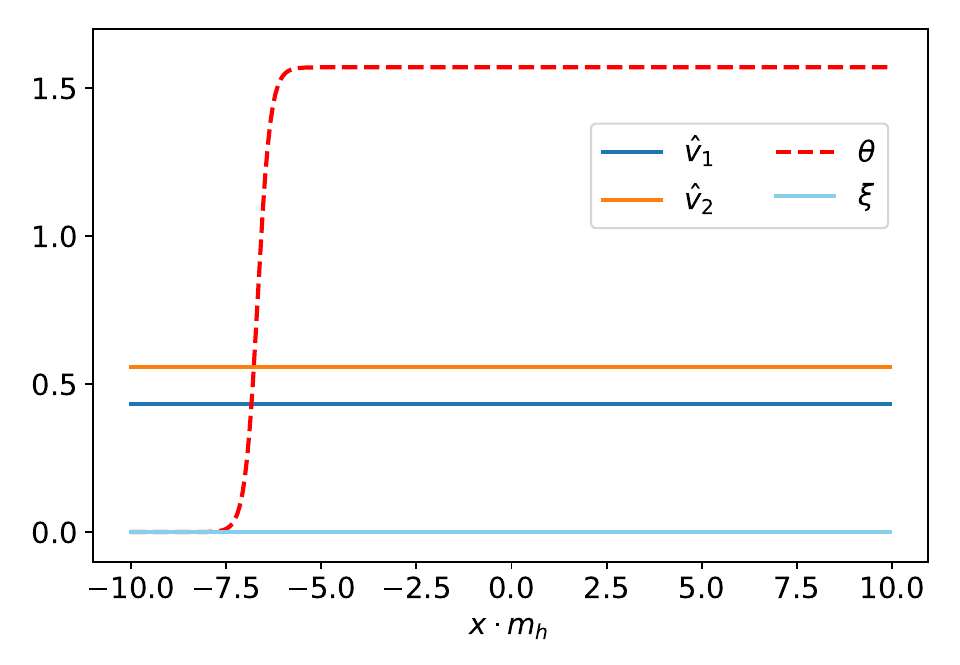}
        \subcaption{} \label{subfig:initheta}
     \end{subfigure}
     \begin{subfigure}[b]{0.5\textwidth}
         \centering
         \includegraphics[width=\textwidth]{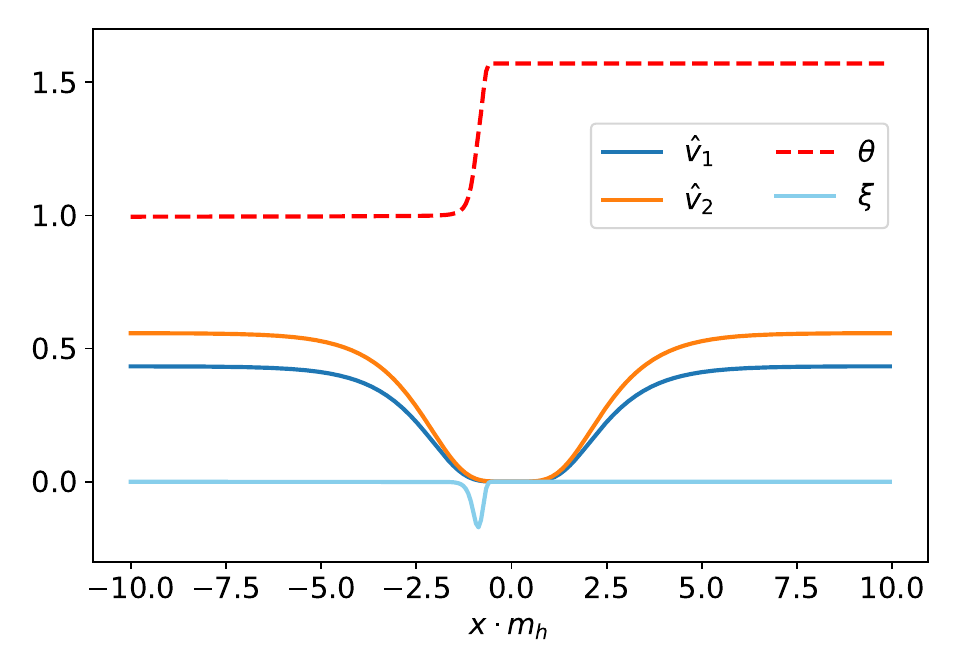}
       \subcaption{}  \label{subfig:intertheta}
        \end{subfigure}
\caption{Scenario where the EWSB occurs after the formation of the walls. The domain wall solution for $v_s$ is centered at $x=0$.  (a) Initial field configuration, where the value $\pi/2$ for $\theta$ is obtained for the domain $x>0$ corresponding to $v_s>0$ and part of the region $x<0$ corresponding to $v_s<0$, while the value $\theta=0$ is obtained for the rest of the $x<0$ domain. (b) Intermediate solution showing CP-violating vacua in the vicinity of the wall. The parameter point used for this scenario is shown in table \ref{tab:thetapp2}.} 
\label{fig:asythetaresult}
\end{figure}

We now look at the second scenario where the electroweak symmetry gets broken after the formation of the walls. In this case, one expects that a single domain of $v_s$ can have multiple different values of Goldstone modes. \\
We model this scenario as follows. After EWSB, vacua with the same sign of $v_{1,2}$ start expanding in the region of the false vacuum $v_{1,2} = 0$. Assuming that one of these expanding vacuum bubbles (in case of a first-order phase transition) collides with the other bubbles that have a different Goldstone mode only after it crosses the singlet domain wall, we then obtain the usual solution discussed in the previous chapter in the vicinity of the wall and a kink-like solution for the Goldstone modes interpolating between the two different values that are obtained far from the wall (see the initial 
vacuum field configuration in Figure \ref{subfig:initheta}). 

We solve the system of equations (\ref{eq:eomv1})-(\ref{eq:eomtheta}) using the initial boundary conditions and the initial guess solution shown in Figure \ref{subfig:initheta}. We find that the kink-like solution for $\theta(x)$ evolves and moves to the vicinity of the wall, where it keeps its shape after some iteration time, indicating that the solution minimizes the energy. This behavior is, however, only possible in the case of electroweak symmetry restoration in a large region around the wall: this is due to the extra contributions from a non-zero $\theta(x)$ and $\xi(x)$ to the energy $\mathcal{E}_{\theta}(x)$ (\ref{eq:energytheta}) being vanishingly small for $v_{1,2}(x) \rightarrow 0$ as discussed earlier. Notice that in such a case, we obtain a CP-violating vacuum configuration only in the vicinity of the region where the Goldstone mode changes.

%% file: Summary_and_conclusions.tex
\section{Summary and conclusions}\label{section6}
In this article, we investigated domain walls that are related to the singlet scalar field of the N2HDM arising after spontaneous symmetry breaking of the $Z'_2$ symmetry in the early universe. We numerically calculated the equations of motion of the scalar fields present in the N2HDM in order to determine the profiles of the doublet scalar fields in the background of the singlet domain wall. We found that the profile of the doublet fields can change considerably in the vicinity and inside of the wall, making either the VEVs $v_{1,2}$ smaller or larger inside the wall. We focused, in particular, on the scenario where $v_{1,2}(x)$ become very small inside the wall, leading to electroweak symmetry restoration. 

The presence of the domain wall solution effectively renders the 2HDM part of the scalar potential x-dependent. This has the effect that the 2HDM potential in the vicinity and inside the wall can be in the symmetric phase where the minima of the potential are $v_{1,2} = 0$. We showed that this case is mostly related to the sign of the effective mass terms of the doublets $M_{1,2}$ which can turn positive inside the wall where the contribution $\lambda_{7,8}v^2_s(x)$ vanishes. We discussed in detail the different behaviors of the doublet fields inside the wall and showed that most parameter points where the effective mass terms get larger inside the wall lead to smaller values for the doublet VEVs inside and in the vicinity of the wall, while smaller (more negative) values for the effective mass term lead to higher values of $v_{1,2}(0)$. We also discussed the different possible anomalous behaviors for some particular parameter points.

We showed, in particular, that positive effective mass terms inside the wall are not sufficient to force the doublet VEVs to become zero even though the potential of the Higgs doublets is in the symmetric phase inside the wall. To achieve EWSR, it was crucial to have a large change in the effective mass terms occurring in a large region of space in order for the doublet VEVs to converge to zero inside the wall. We found that parameter points that can satisfy this requirement have large and negative ratios ($\lambda_{7,8}/\lambda_6$) and we found that they lead to very small $v_{1,2}$ in a large region around the wall. 

To find parameter points with large ($\lambda_{7,8}/\lambda_6$), we looked at different scenarios that satisfy all theoretical and experimental constraints including collider searches and showed that the vacuum expectation value of the singlet scalar as well as the masses of the CP-even Higgs bosons are the most important model parameters. In particular, we find that larger $v_s$ mostly lead to smaller doublet VEVs inside the wall, while lower masses of the CP-even Higgs bosons mostly lead to a larger width for the EWSR region. Effects from the mixing angles between the different CP-even Higgs bosons also play an important role, as we observed that parameter points with higher singlet admixture in the SM-like Higgs boson state tend to have a higher electroweak symmetry restoration effect inside the wall. This, however, already puts constraints on the amount of EWSR that can be achieved inside the wall given that collider constraints restrict the amount of singlet admixture in the SM-like Higgs boson state.

We also showed that it is possible to induce CP-violating vacua in the vicinity of the wall in the case when different regions of the universe acquire different values for the Goldstone modes after EWSB. In contrast to the analogous case in the 2HDM \cite{Law:2021ing, Sassi:2023cqp}, we found that the energy difference between CP-violating solutions and standard domain wall solutions is very small, especially for parameter points that lead to EWSR in a large region around the wall. One would expect that the CP-violating domain wall in such scenarios would be long-lived. Determining the lifetime of these CP-violating solutions is crucial for the calculation of the matter-antimatter asymmetry generated by the motion of the domain walls in the N2HDM. 

We performed in this study the first steps toward studying the viability of electroweak baryogenesis via domain walls in the N2HDM. This mechanism relies on the weak sphaleron rate inside the wall being less suppressed than outside of it. The sphaleron rate in the broken electroweak phase and at a temperature $T$ has an exponential suppression due to a non-zero doublets vacuum expectation value $v_{ew}$ and is proportional to $\Gamma_{sphaleron} \propto e^{(-4\pi (v_{ew}/gT))}$ \cite{Morrissey:2012db}, where g is the weak coupling. Therefore, for small or vanishing values for $v_{ew}$, the sphaleron rate is not suppressed and the rate of baryon-violating processes inside the wall will be significantly higher than that outside of it. Another important requirement is that the region of symmetry restoration needs to be large enough to fit a sphaleron. Given that the weak sphaleron only couples to left-handed fermions, CP-violation in the vicinity of the wall is required in order to create a chiral asymmetry that is subsequently transformed into a baryon asymmetry inside the wall \cite{Schroder:2024gsi, Brandenberger:1994mq}. A complete and detailed calculation of the amount of baryogenesis generated via this mechanism in the case of an annihilating singlet domain wall network in the N2HDM is the subject of future work, where we incorporate thermal effects as well as the calculation of the transmission and reflection rates of fermions that scatter off the CP-violating walls discussed in this article.